\documentclass[aps,prx,reprint,floatfix,amsmath,amssymb,nofootinbib,bibnotes,longbibliography,superscriptaddress]{revtex4-2}

\usepackage[T1]{fontenc}
\usepackage{geometry}
\usepackage{color}
\usepackage{xcolor}
\usepackage{babel}
\usepackage{float}
\usepackage{amsmath}
\usepackage{amsfonts}
\usepackage{amssymb}
\usepackage{graphicx}
\usepackage[caption=false,position=top]{subfig}
\usepackage[normalem]{ulem} 
\usepackage{enumitem}

\usepackage{titlesec}
\usepackage[dont-mess-around]{fnpct}

\usepackage{siunitx}
\usepackage[version=4,textfontname=sffamily,mathfontname=mathsf]{mhchem}
\usepackage{nicefrac}

\usepackage{mathtools}

\usepackage{algpseudocode}

\usepackage{empheq}

\usepackage[braket, qm]{qcircuit}
\usepackage{ragged2e}
\usepackage{chngcntr}
\usepackage{changes}
\usepackage{etoolbox}
\usepackage{soul} 
\usepackage[normalem]{ulem}

\makeatletter
\newcommand{\appendixtocext}{atoc}

\newcommand{\startappendixtoc}{%
  \let\orig@addcontentsline\addcontentsline
  \renewcommand{\addcontentsline}[3]{%
    \ifstrequal{##1}{toc}
      {\orig@addcontentsline{\appendixtocext}{##2}{##3}}
      {\orig@addcontentsline{##1}{##2}{##3}}%
  }%
}

\newcommand{\printappendixtoc}{%
  \@starttoc{\appendixtocext}%
}
\makeatother

\newif\ifshownotes
\shownotesfalse   

\ifshownotes
  \newcommand{\authorcomment}[3]{{\color{#1}[#2: #3]}}
  \newcommand{\customauthorcomment}[4]{{%
    \color[#1]{#2}[#3: #4]%
  }}
\else
  \newcommand{\authorcomment}[3]{}
  \newcommand{\customauthorcomment}[4]{}
\fi

\newcommand{\cupdot}{\mathbin{\mathaccent\cdot\cup}}

\usepackage[unicode=true,pdfusetitle,bookmarks=true,bookmarksnumbered=false,bookmarksopen=false,breaklinks=false,pdfborder={0 0 1},backref=false,colorlinks=true]{hyperref}

 \hypersetup{
     colorlinks   = true,
     linkcolor    = blue,
     urlcolor     = blue,
     citecolor    = blue
}

\newcounter{algorithm}
\renewcommand{\thealgorithm}{\arabic{algorithm}}

\newenvironment{algorithm}[1][]{%
  \par\medskip
  \refstepcounter{algorithm}%
  \phantomsection
  \noindent\begin{minipage}{\linewidth}%
  \def\caption##1{\noindent\textbf{Algorithm~\thealgorithm:} ##1\par\medskip}%
}{%
  \end{minipage}%
  \par\medskip
}

\usepackage{xcolor}
\usepackage{comment}

\algrenewcommand\algorithmicrequire{\textbf{Input:}}
\algrenewcommand\algorithmicensure{\textbf{Output:}}
\algnewcommand\Input{\item[\algorithmicrequire]}
\algnewcommand\Output{\item[\algorithmicensure]}
\algrenewcommand\algorithmiccomment[1]{\hfill{\color{gray}\texttt{\textbackslash\textbackslash}~#1}}
\algnewcommand\LComment[1]{{\color{gray}\texttt{\textbackslash*}~#1~\texttt{*\textbackslash}}}

\global\long\def\ket#1{\left|#1\right\rangle }%
\global\long\def\braket#1#2{\left\langle #1\right|\left.#2\right\rangle }%
\DeclarePairedDelimiterX\braketExp[1]{\langle}{\rangle}{#1}%

\makeatletter
\DeclareFontFamily{OMX}{MnSymbolE}{}
\DeclareSymbolFont{MnLargeSymbols}{OMX}{MnSymbolE}{m}{n}
\SetSymbolFont{MnLargeSymbols}{bold}{OMX}{MnSymbolE}{b}{n}
\DeclareFontShape{OMX}{MnSymbolE}{m}{n}{
    <-6>  MnSymbolE5
   <6-7>  MnSymbolE6
   <7-8>  MnSymbolE7
   <8-9>  MnSymbolE8
   <9-10> MnSymbolE9
  <10-12> MnSymbolE10
  <12->   MnSymbolE12
}{}
\DeclareFontShape{OMX}{MnSymbolE}{b}{n}{
    <-6>  MnSymbolE-Bold5
   <6-7>  MnSymbolE-Bold6
   <7-8>  MnSymbolE-Bold7
   <8-9>  MnSymbolE-Bold8
   <9-10> MnSymbolE-Bold9
  <10-12> MnSymbolE-Bold10
  <12->   MnSymbolE-Bold12
}{}
\let\llangle\@undefined
\let\rrangle\@undefined
\DeclareMathDelimiter{\llangle}{\mathopen}%
                     {MnLargeSymbols}{'164}{MnLargeSymbols}{'164}
\DeclareMathDelimiter{\rrangle}{\mathclose}%
                     {MnLargeSymbols}{'171}{MnLargeSymbols}{'171}
\DeclarePairedDelimiterX\kket[1]{\lvert}{\rrangle}{#1}
\DeclarePairedDelimiterX\bbra[1]{\llangle}{\rvert}{#1}
\DeclarePairedDelimiterX\bbrakket[2]{\llangle}{\rrangle}{#1\delimsize\vert\mathopen{}#2}%
\DeclarePairedDelimiterX\bbrakketExp[1]{\llangle}{\rrangle}{#1}
\DeclarePairedDelimiterX\bbrakketOP[3]{\llangle}{\rrangle}{#1\,\delimsize\vert\,\mathopen{}#2\,\delimsize\vert\,\mathopen{}#3}%
\newcommand{\ignore}[1]{}
\makeatother

\makeatletter
\providecommand*{\cupdot}{%
  \mathbin{%
    \mathpalette\@cupdot{}%
  }%
}
\newcommand*{\@cupdot}[2]{%
  \ooalign{%
    $\m@th#1\cup$\cr
    \hidewidth$\m@th#1\cdot$\hidewidth
  }%
}
\makeatother

\makeatletter
\long\def\@makecaption#1#2{%
  \vskip\abovecaptionskip
  \begingroup
    \justifying
    \small
    \noindent
    \textbf{#1:} #2\par
  \endgroup
  \vskip\belowcaptionskip
}
\makeatother

\begin{document}

\title{Resolving Structure in Prethermal Floquet Dynamics with Precision Quantum Computation}

\author{Eyal Leviatan}\affiliation{Qedma Quantum Computing, Tel Aviv, Israel}

\author{Tasneem Watad}\affiliation{Qedma Quantum Computing, Tel Aviv, Israel}

\author{Roy Perry}\affiliation{Qedma Quantum Computing, Tel Aviv, Israel}

\author{Lukas Broers}\affiliation{RIKEN Center for Computational Science (R-CCS), Hyogo 650-0047, Japan}

\author{Mohammed Zuhair Mullath}\affiliation{BlueQubit, San Francisco, CA 94105, USA}

\author{Ori Alberton}\affiliation{Qedma Quantum Computing, Tel Aviv, Israel}

\author{Itai Arad}\affiliation{Qedma Quantum Computing, Tel Aviv, Israel}

\author{Yosi Atia}\affiliation{Qedma Quantum Computing, Tel Aviv, Israel}

\author{Eyal Bairey}\affiliation{Qedma Quantum Computing, Tel Aviv, Israel}

\author{Shaul Barkan}\affiliation{Qedma Quantum Computing, Tel Aviv, Israel}

\author{Matan Ben Dov}\affiliation{Qedma Quantum Computing, Tel Aviv, Israel}

\author{Asaf Berkovitch}\affiliation{Qedma Quantum Computing, Tel Aviv, Israel}

\author{Ewout van den Berg}\affiliation{IBM Quantum, IBM T.J. Watson Research Center, Yorktown Heights, NY 10598, USA}

\author{Itsik Cohen}\affiliation{Qedma Quantum Computing, Tel Aviv, Israel}

\author{Omri Golan}\affiliation{Qedma Quantum Computing, Tel Aviv, Israel}

\author{Ilya Gurwich}\affiliation{Qedma Quantum Computing, Tel Aviv, Israel}

\author{Avieli Haber}\affiliation{Qedma Quantum Computing, Tel Aviv, Israel}

\author{Barak A. Katzir}\affiliation{Qedma Quantum Computing, Tel Aviv, Israel}

\author{Oded Kenneth}\affiliation{Qedma Quantum Computing, Tel Aviv, Israel}

\author{Roei Levi}\affiliation{Qedma Quantum Computing, Tel Aviv, Israel}

\author{Yotam Y. Lifshitz}\affiliation{Qedma Quantum Computing, Tel Aviv, Israel}

\author{Yaron Lukovsky}\affiliation{Qedma Quantum Computing, Tel Aviv, Israel}

\author{Ron Melcer}\affiliation{Qedma Quantum Computing, Tel Aviv, Israel}

\author{Adiel Meyer}\affiliation{Qedma Quantum Computing, Tel Aviv, Israel}

\author{Boris Muratov}\affiliation{Qedma Quantum Computing, Tel Aviv, Israel}

\author{Aviad Panahi}\affiliation{Qedma Quantum Computing, Tel Aviv, Israel}

\author{Gili Schul}\affiliation{Qedma Quantum Computing, Tel Aviv, Israel}

\author{Tali Shnaider}\affiliation{Qedma Quantum Computing, Tel Aviv, Israel}

\author{Maor Shutman}\affiliation{Qedma Quantum Computing, Tel Aviv, Israel}

\author{Alireza Seif}\affiliation{IBM Quantum, IBM T.J. Watson Research Center, Yorktown Heights, NY 10598, USA}

\author{Tomonori Shirakawa}\affiliation{RIKEN Center for Computational Science (R-CCS), Hyogo 650-0047, Japan}\affiliation{RIKEN Center for Quantum Computing (RQC), Saitama 351-0198, Japan}\affiliation{RIKEN Pioneering Research Institute (PRI), Saitama 351-0198, Japan}\affiliation{RIKEN Center for Interdisciplinary Theoretical and Mathematical Sciences (iTHEMS), Saitama 351-0198, Japan}

\author{Asif Sinay}\affiliation{Qedma Quantum Computing, Tel Aviv, Israel}

\author{Vincent P. Su}\affiliation{BlueQubit, San Francisco, CA 94105, USA}

\author{Hayk Tepanyan}\affiliation{BlueQubit, San Francisco, CA 94105, USA}

\author{Omri Trebitch}\affiliation{Qedma Quantum Computing, Tel Aviv, Israel}

\author{Assaf Zubida}\affiliation{Qedma Quantum Computing, Tel Aviv, Israel}

\author{Dorit Aharonov}\affiliation{Qedma Quantum Computing, Tel Aviv, Israel}
\affiliation{The Benin School of Computer Science and Engineering, Hebrew University, Jerusalem, Israel}

\author{Hrant Gharibyan}\affiliation{BlueQubit, San Francisco, CA 94105, USA}

\author{Abhinav Kandala}\affiliation{IBM Quantum, IBM T.J. Watson Research Center, Yorktown Heights, NY 10598, USA}

\author{Seiji Yunoki}\affiliation{RIKEN Center for Computational Science (R-CCS), Hyogo 650-0047, Japan}\affiliation{RIKEN Center for Quantum Computing (RQC), Saitama 351-0198, Japan}\affiliation{RIKEN Pioneering Research Institute (PRI), Saitama 351-0198, Japan}\affiliation{RIKEN Center for Emergent Matter Science (CEMS), Saitama 351-0198, Japan}

\author{Netanel H. Lindner}\affiliation{Qedma Quantum Computing, Tel Aviv, Israel}\affiliation{Department of Physics, Technion, Haifa 320003, Israel }

\date{\today{}}

\begin{abstract} 
Periodically driven interacting quantum many-body systems can exhibit long-lived prethermal dynamics, where local observables retain coherent structure even as entanglement and operator complexity grow. Accessing this regime at the system sizes and times needed to determine physical properties of the prethermal state remains a central challenge: state-of-the-art classical methods become unreliable, while noise in quantum hardware degrades observable expectation values. Here we overcome these limitations for a Floquet Ising magnet realized on a heavy-hex lattice. Using the advanced error mitigation software QESEM on an IBM Heron r3 superconducting quantum processor, we measure magnetization dynamics with percent-level precision and resolve long-lived subharmonic prethermal oscillations in systems of up to 74 qubits. These experiments reach regimes for which leading tensor-network simulations fail to converge, while sparse Pauli-path simulations remain strongly truncation dependent despite extensive computations on advanced GPUs and the Fugaku supercomputer. Leveraging this quantum-accessible regime, we extend finite-size scaling to larger systems and find an unexpectedly slow decrease of the oscillation amplitude with system size, providing strong evidence that this oscillatory response persists in the thermodynamic limit of heavy-hex ladders. A hierarchy of mitigation and validation tests, including unbiased error mitigation, agreement between independent mitigation estimators, noise-model validation on the superconducting hardware, and cross-platform corroboration at selected Floquet cycles on Quantinuum System Model H2 and Quantinuum Helios trapped-ion hardware, supports the reliability of these findings. Our work establishes error-mitigated quantum processors as quantitative scientific instruments for discovering new physics in non-equilibrium quantum matter.
\end{abstract}

\maketitle
\section{Introduction}
\label{sec.introduction}

\begin{figure*}[tb]
\centering
\begin{minipage}[t]{0.42\textwidth}
\centering
\subfloat[]{
    \includegraphics[width=\linewidth]{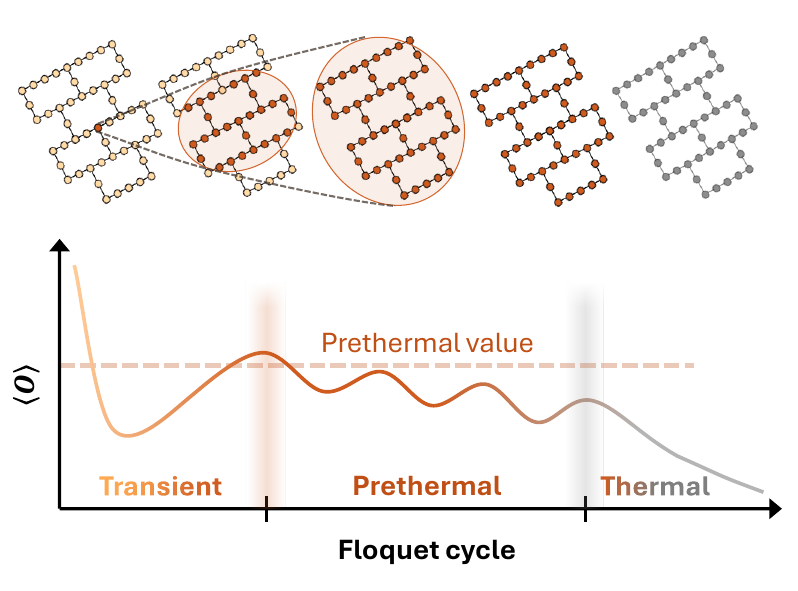}
    \label{fig:phase_diagram_floquet}
}
\end{minipage}
\hfill
\begin{minipage}[t]{0.57\textwidth}
\centering
\subfloat[]{
    \includegraphics[width=\linewidth]{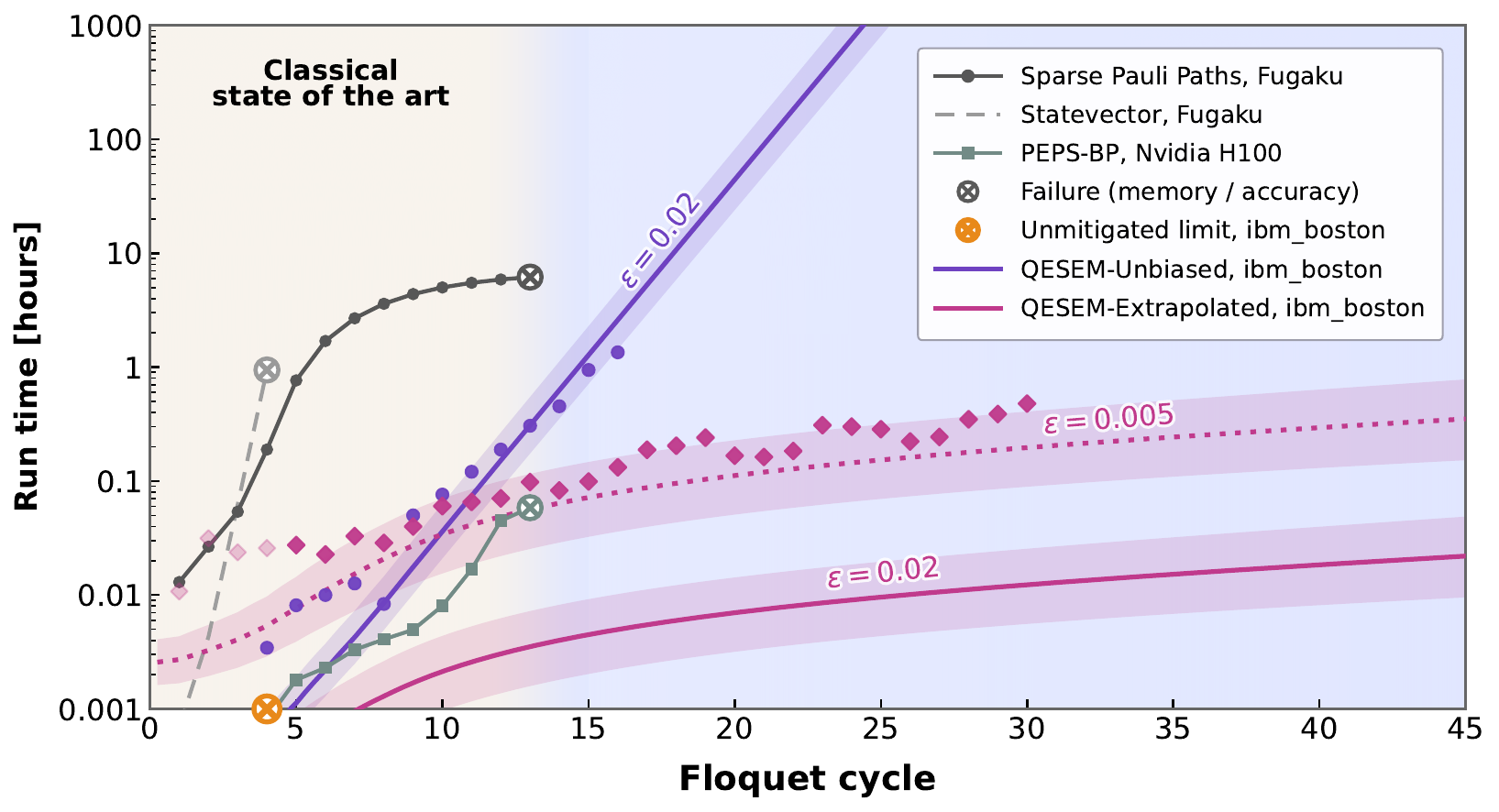}
    \label{fig:fig1b}
}
\end{minipage}
\caption{
    \textbf{\protect\subref{fig:phase_diagram_floquet}} Schematic of Floquet prethermalization. A generic local operator $O$ spreads and develops many-body complexity across a $51$-qubit heavy-hex patch, one of the geometries used in the experiments (top). After an initial transient, its expectation value $ \left\langle O \right\rangle $ exhibits long-lived structured oscillations about a prethermal value, before eventual heating drives the system to a featureless thermal state (bottom).
    \textbf{\protect\subref{fig:fig1b}} Runtime and accuracy frontier for estimating the magnetization of the $51$-qubit Floquet circuits shown in \hyperref[fig:magnetization]{Fig.~\ref*{fig:magnetization}}. Unmitigated execution loses accuracy by cycle $4$ (yellow marker), whereas QESEM extends the accessible depth by trading additional QPU sampling time for reduced bias. The classical curves show measured sparse Pauli-path (SPP) runtimes on the supercomputer Fugaku, statevector runtime estimates on Fugaku that account for the growing observable light-cone and assume a correspondingly increasing allocation of parallel resources, and measured runtimes for projected entangled-pair-state simulations with belief propagation (PEPS-BP) on an NVIDIA H100 GPU. Crossed open circles mark the endpoints reached with the computational resources used: the statevector resource limit on Fugaku, the point at which the SPP calculations shown lose accuracy despite retaining $\mathcal{O}(10^{12})$ Pauli strings over $12888$ Fugaku nodes, and the point at which the largest PEPS-BP bond dimension fitting in the available GPU memory no longer meets the accuracy target. The shading indicates the classical frontier represented by these calculations. QESEM markers show measured mitigation QPU runtimes on \texttt{ibm\_boston}, normalized to absolute magnetization precisions of $\epsilon=0.02$ for QESEM-Unbiased (using probabilistic error cancellation) and $\epsilon=0.005$ for QESEM-Extrapolated (using zero-noise extrapolation); curves and bands show the corresponding runtime estimates, with QESEM-Extrapolated also shown at $\epsilon=0.02$. The quantum runtime model assumes an effective two-qubit-gate
    infidelity $I_F=2.4\times10^{-3}$, representative of the
    \texttt{ibm\_boston} experiments, while the bands span the
    observed ranges of the circuit-dependent model parameters. The accuracy limits and classical convergence breakdowns are discussed in
    \hyperref[sec.results]{Sec.~\ref*{sec.results}} and
    \hyperref[sec.classical]{Sec.~\ref*{sec.classical}}. Detailed analyses of SPP,
    PEPS-BP, and complementary classical methods, including convergence and
    projected resource requirements beyond the computational resources used in this work, are
    given in Apps.~\ref{app.spp} and \ref{app.tn}; the quantum and classical
    runtime models are detailed in
    \hyperref[app.runtimes]{App.~\ref*{app.runtimes}}.
}
\end{figure*}

The dynamics of non-equilibrium quantum systems remain a largely unexplored frontier. They pose some of the most fundamental and challenging questions in modern physics, including how universal dynamical behaviour emerges from microscopic quantum evolution, what mechanisms govern or obstruct relaxation, and which collective phases and long-lived states can arise far from equilibrium~\cite{polkovnikov2011nonequilibrium,dalessio2016eth,abanin2019mbl}. Addressing these questions requires probing large quantum systems over long evolution times, where transient and finite-size effects recede, revealing robust collective behaviour. Yet this is precisely the regime in which the growth of entanglement and operator complexity severely limits controlled classical simulation~\cite{schollwoeck2011dmrg,orus2019tensornetworks,begusic2025realtime,shao2026characterizing}.

Quantum systems subjected to a time-periodic drive, or Floquet systems, provide a particularly rich setting for studying non-equilibrium dynamics. In particular, they exhibit phenomena with no equilibrium analogue, including anomalous Floquet topological phases
~\cite{rudner2013anomalous,titum2016anomalous,nathan2019anomalous} and discrete time-crystalline order~\cite{else2016floquettimecrystals,khemani2016floquetphases,yao2017dtc,zhang2017dtc,choi2017dtc}. However, stabilizing and observing such phenomena in this non-equilibrium setting is hindered by the tendency of generic interacting quantum systems to rapidly absorb energy from the drive until local observables approach the featureless values of an effectively infinite-temperature state~\cite{dalessio2014floquetheating,lazarides2014floquetheating}.

Observing the unique phenomena that can be realized by Floquet systems therefore requires mechanisms that prevent or delay thermalization. Heating can be avoided by local conservation laws, as in integrable systems~\cite{Lazaridis2014}, or by disorder-induced many-body localization~\cite{abanin2019mbl}.
Even in the absence of such conservation laws, under certain conditions thermalization can be significantly delayed, whereby after an initial rapid relaxation, the system enters a long-lived quasi-equilibrium regime governed by an approximately conserved effective Hamiltonian~\cite{abanin2017prethermal,mori2016floquetheating,fleckenstein2021thermalization,Lindner2017}. This prethermal regime can persist for many drive cycles before the system ultimately heats toward an effectively infinite-temperature state. In this time window, local observables remain close to thermal values of the effective Hamiltonian before eventually approaching their infinite-temperature values (Fig.~\ref{fig:phase_diagram_floquet}).

Known mechanisms leading to long prethermal time windows include high-frequency driving~\cite{abanin2017prethermal},  weak breaking of integrability~\cite{Roy2026integrability}, and kinetic constraints or bottlenecks~\cite{Lindner2017,Galitski2024, Ghosh2025DestructiveInterference,Lan2018QuantumSlowRelaxation,Joshi2026TunableFloquetSelectionRules,Sala2020ErgodicityBreaking,Birnkammer2022Prethermalization}, where the redistribution of energy or information is dynamically suppressed. The prethermal time window is especially interesting when observables retain well-structured time dependence while the dynamics also develop substantial entanglement and operator complexity, distinguishing this regime from both simple static equilibrium and rapidly featureless heating.

Quantum processors offer a route to studying quantum dynamics at time scales that are difficult to access with classical simulations~\cite{feynman1982simulating,lloyd1996universal,kim2023utility,CiracPEVP,fischer2026dynamical,haghshenas2026digitalmagnetism,shinjo2026unveiling,alam2025programmable,alam2025fermionic,granet2025superconducting,hartnett2026fast}. Floquet dynamics are especially natural in this setting. For a time-periodic Hamiltonian, the stroboscopic dynamics are generated by repeated applications of a Floquet unitary~\cite{bukov2015floquet,goldman2014floquet, eckardt2017floquet}, a structure naturally implemented by quantum circuits.  However, the large-system, long-time regime that makes these dynamics physically interesting requires large-volume quantum circuits, which are highly sensitive to hardware errors. 

On present-day quantum computers, accumulated hardware errors bias expectation values and can obscure the quantum dynamics of interest. Error mitigation methods are designed to reduce this bias, extending the range of circuit volumes over which accurate expectation values can be extracted in exchange for sampling and runtime overhead~\cite{Cai2023, YEMpaper,temme2017qem,Endo2018}. In particular, given accurate noise characterization, quasi-probability (QP) representations can be used to perform probabilistic error cancellation (PEC), producing unbiased estimators of ideal expectation values~\cite{temme2017qem,Endo2018,vandenberg2023pec,QESEM}. Alternatively, QP distributions can be used to construct noise-reduced or noise-amplified estimators, enabling heuristic but often lower-overhead zero-noise extrapolation (ZNE)~\cite{mari2021noisescaling,kim2023utility,haghshenas2026digitalmagnetism}. 

In this work, we empirically show that cutting-edge error-mitigated quantum processors can be used to investigate prethermal dynamics with the precision required to uncover new physical phenomena in a regime beyond the reach of classical simulations. Our quantum computing stack combines IBM Heron r3 superconducting processors with QESEM~\cite{QESEM}, a software framework for quantum error suppression and error mitigation, supporting efficient and reliable QP-based  implementations of both PEC (QESEM-Unbiased) and ZNE (QESEM-Extrapolated); see Fig.~\ref{fig:fig1b} for empirical runtimes. We establish the accuracy of our results through a hierarchy of validation tests. These include validation of the QESEM-characterized noise model on IBM hardware; agreement between the unbiased and extrapolated estimators over their common time window; consistency between extrapolated estimators obtained using different noise amplification factors; and comparisons with exact or converged classical results where available. Finally, we perform an additional cross-platform test, using characterization-free extrapolated QESEM estimators obtained with Quantinuum H2 and Helios trapped-ion processors for selected Floquet cycles, demonstrating consistency with the corresponding results obtained on IBM.

We focus on a Floquet mixed-field Ising quantum magnet on the heavy-hex lattice, and on circuits that do not admit a description in terms of either high-frequency driving or a small-step Suzuki--Trotter expansion of a static Hamiltonian. These circuits therefore generate intrinsically Floquet dynamics and can lead to phenomena with no equilibrium analogue. At the same time, generically they have no obvious mechanism for stabilizing a long prethermal time window. Using exact statevector simulations of small systems, we nevertheless find circuit parameters for which the system rapidly relaxes into long-lived prethermal dynamics. Intriguingly, the prethermalized system exhibits magnetization oscillations with a period approximately four times that of the underlying Floquet drive. These simulations indicate that the oscillation amplitude decreases as the system size increases. However, finite-size scaling based only on the classically simulated systems does not distinguish whether this oscillatory response is a robust property of the thermodynamic system or ultimately becomes too small to resolve. 

To address this question, we simulate dynamics in the prethermal time window in systems of up to $74$ qubits using an IBM Heron r3 processor. Using QESEM, the error-mitigated quantum data for the large-system dynamics show clear oscillations, enabling an accurate extraction of their amplitudes, which decay surprisingly slowly with system size. An extended finite-size scaling analysis then strongly suggests that the observed oscillations retain a significant, measurable value in the thermodynamic limit of heavy-hex ladders. 

Finally, we address whether classical simulations can investigate the same prethermal dynamics with the precision obtained using our quantum computing stack, and assess the unique contribution of error-mitigated quantum computation to this physical problem. To this end, we apply leading approximate classical simulation methods based on one- and two-dimensional tensor networks (TNs)~\cite{verstraete2004renormalization, paeckel2019time,alkabetz2021tensor,tindall2023gauging, liao2023simulation,begusic2023fastconverged} and sparse Pauli-path representations~\cite{begusic2023fastconverged,begusic2025sparsepauli,gharibyan2025practicalguideusingpauli,rudolph2026paulipropagationcomputationalframework, Loizeau25paulistrings,broers2025scalablesimulationquantummanybody,Broers_2026}, implemented on H100 GPUs and the Fugaku supercomputer. We additionally consider a ``noisy'' variant of the Pauli-path method with zero-noise extrapolation, and a recently proposed heuristic correction of one-dimensional tensor-network estimates~\cite{mandra2025heuristic}, which reproduced the error-mitigated quantum data of Ref.~\cite{haghshenas2026digitalmagnetism}. All of these methods fail to converge to controlled, accurate physical predictions for the relevant large-system, late-cycle dynamics (see Fig.~\ref{fig:fig1b} for an overview). We conclude that error-mitigated quantum computation extends beyond  the reach of the state-of-the-art classical methods considered here.

\section{A prethermal Floquet magnet at the classical frontier}
\label{sec.setup}

\begin{figure*}[tb]
\begin{minipage}[t]{0.24\textwidth}
\vspace{0pt}
\centering
\subfloat[]{%
    \includegraphics[width=0.95\linewidth]{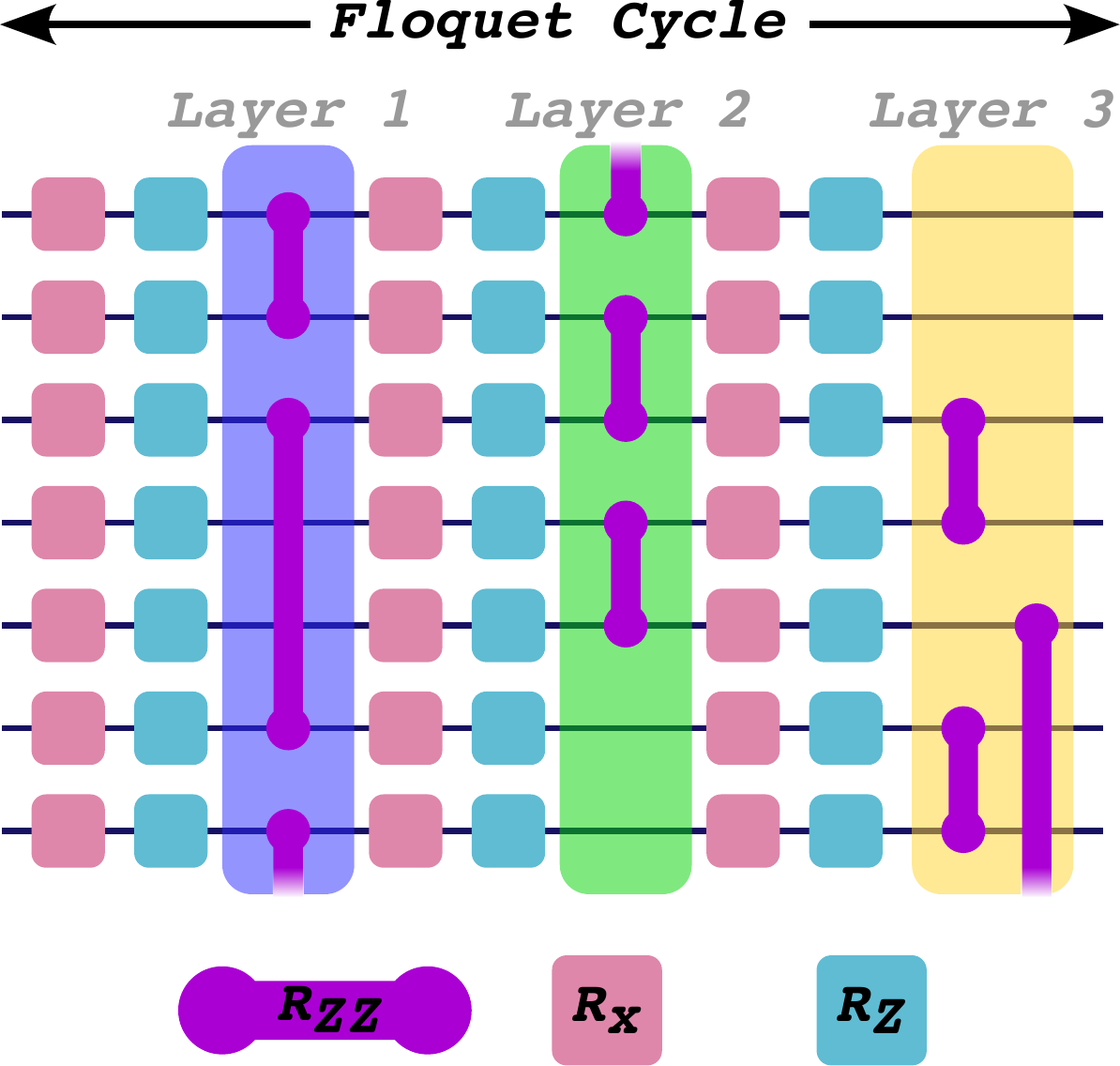}
    \label{fig:floquet_cycle}
}%
\vspace{1.6em}
\subfloat[]{%
    \includegraphics[width=0.98\linewidth]{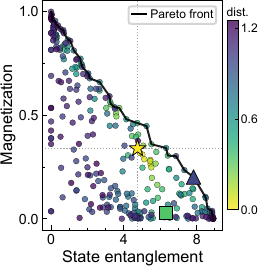}
    \label{fig:tradeoff_plane}
}
\end{minipage}%
\hfill
\begin{minipage}[t]{0.75\textwidth}
\vspace{0pt}
\centering
\includegraphics[width=\linewidth]{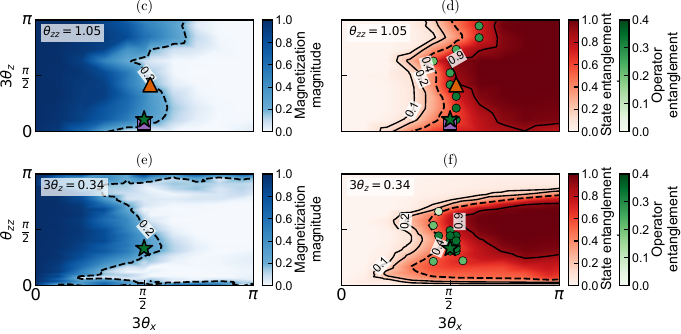}
\label{fig:phase_diagram_combined}
\vspace{-0.25em}
\includegraphics[width=\linewidth]{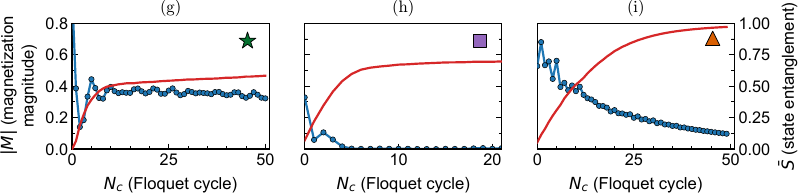}
\label{fig:dynamics_combined}
\end{minipage}
\caption{
    \textbf{\protect\subref{fig:floquet_cycle}} One Floquet cycle of the driven magnet. Each cycle consists of parallel single-qubit $X$ and $Z$ rotation layers on all qubits interleaved with three edge-coloured layers of nearest-neighbour fractional-angle $ZZ$ rotations.
    \textbf{\protect\subref{fig:tradeoff_plane}} Tradeoff between magnetization and state entanglement entropy in 21-qubit exact simulations after 30 Floquet cycles. Each point corresponds to one choice of Floquet parameters in the scan; colour indicates distance, in parameter space, from the chosen set of parameters, marked by the star and located near the Pareto front.
    \textbf{(c)} Retained magnetization after 30 Floquet cycles in the $(3\theta_x,3\theta_z)$ plane at fixed $\theta_{zz}=1.05$, from the same 21-qubit exact simulations as in \protect\subref{fig:tradeoff_plane}. The selected point is located near $3\theta_x \simeq \pi/2$, where a finite $\theta_z$ is needed to retain a measurable signal.
    \textbf{(d)} State and operator complexity for the same $(3\theta_x,3\theta_z)$ cut. The colour map shows the normalized half-system state entanglement entropy, while green markers show the normalized operator entanglement generated by evolving a single-site $Z$ observable at representative points along the high-state-entanglement ridge (App.~\ref{app.scan}).  
    \textbf{(e)} Retained magnetization after 30 Floquet cycles in the $(3\theta_x,\theta_{zz})$ plane at fixed $3\theta_z=0.34$. The selected point balances a long-lived magnetization signal with a non-negligible interaction strength. 
    \textbf{(f)} State and operator complexity for the same $(3\theta_x,\theta_{zz})$ cut. The colour map shows the normalized state entanglement entropy, while green markers show the corresponding normalized operator-entanglement diagnostic. The selected interaction angle $\theta_{zz}$ is large enough to generate substantial many-body and operator complexity while preserving a measurable magnetization signal.
    \textbf{(g)} Exact 28-qubit dynamics at the selected parameters, $(3\theta_x,3\theta_z,\theta_{zz})=(1.57,0.34,1.05)$, showing initial rapid state-entanglement growth followed by long-lived structured magnetization oscillations and a slower rate of entanglement growth.
    \textbf{(h)} Exact 28-qubit dynamics at $(3\theta_x,3\theta_z,\theta_{zz})=(1.57,0,1.05)$. The entanglement shows a similar initial growth profile, but the magnetization rapidly relaxes.
    \textbf{(i)} Exact 28-qubit dynamics at a more strongly entangling point, $(3\theta_x,3\theta_z,\theta_{zz})=(1.65,1.32,1.05)$. The magnetization lacks the long-lived structured oscillations seen at the selected point.
} 
\label{fig:param_scan_circ}
\end{figure*}
We consider a family of hardware-native Floquet circuits defined on heavy-hex lattices of $N_q$ qubits. One Floquet cycle consists of parallel single-qubit rotation layers applied uniformly across the lattice, interleaved with nearest-neighbour fractional-angle $ZZ$ rotations on the three edge-coloured bond layers shown in Fig.~\ref{fig:floquet_cycle}. The stroboscopic evolution is generated by
\begin{align}
\label{eq:unitary}
    U_F &= 
    \prod_{r=1}^3
    \left(
        e^{-i \frac{\theta_{zz}}{2} C_r}
        e^{-i \frac{\theta_z}{2} Z_\Sigma}
        e^{-i \frac{\theta_x}{2} X_\Sigma}
    \right),
\end{align}
where $X_\Sigma = \sum_{i=1}^{N_q} X_i$, $Z_\Sigma = \sum_{i=1}^{N_q} Z_i$, and $C_r = \sum_{\langle i,j\rangle \in E_r} Z_i Z_j$. The edge sets $E_1,E_2,E_3$ form a balanced three-colouring of the heavy-hex lattice, allowing the $ZZ$ rotations within each $C_r$ to be implemented in parallel (see Fig.~\ref{fig:ibm-boston} in App.~\ref{app.ibm_exp_detail} for an example). The unitary $U_F$ defines a Floquet mixed-field Ising model whose interactions naturally match the heavy-hex connectivity. 

The many-body dynamics generated by repeated applications of $U_F$ depend sensitively on the parameters $(\theta_x,\theta_z,\theta_{zz})$ and on the initial state. Generic choices either thermalize too rapidly, leading highly mixed local reduced density matrices, or generate entanglement too slowly, delaying the onset of possible prethermal behaviour. We use exact statevector simulations of a 21-qubit system to scan parameter space and identify an intermediate regime in which both local observables and entanglement entropy are substantial after $30$ Floquet cycles, using this as a proxy for relaxation toward a prethermal state. This corresponds to an intermediate physical time that allows us to monitor prethermal dynamics, while also producing circuit depths amenable to efficient, high-precision error-mitigated execution on quantum hardware. Specifically, we take the fully ferromagnetic state $\bigotimes_{i=1}^{N_q}|0\rangle$ as the initial state and record the magnetization
\begin{align}
\label{eq:magnetization}
    M = \frac{1}{N_q} \sum_{q=1}^{N_q} \left\langle Z_q \right\rangle,
\end{align}
 and the normalized half-system entanglement entropy of the state generated by the circuit (App.~\ref{app.scan}). We choose parameters that balance rapid entanglement buildup with long-lived measurable magnetization.

The resulting choice, $\theta_x \approx \pi/6$, $\theta_z \approx \pi/27$, and $\theta_{zz} = \pi/3$, is shown in the tradeoff plot of Fig.~\ref{fig:tradeoff_plane} and further resolved in the two-dimensional cuts of Figs.~\ref{fig:param_scan_circ}c--\ref{fig:param_scan_circ}f. In the tradeoff plot, the selected point is located near the Pareto front, meaning that within the scanned parameter set neither magnetization nor state entanglement can be substantially improved without degrading the other~\cite{deb2001multiobjective}. The cuts make explicit why this point is favourable: highly entangling regions typically have suppressed magnetization, while high-magnetization regions are often weakly entangling. Values near $3\theta_x\simeq\pi/2$ provide a useful balance between retained magnetization and generated entanglement, $\theta_{zz}$ must be of order unity to generate substantial many-body complexity, and a finite $\theta_z$ is needed to retain the desired signal after 30 Floquet cycles.

The parameter cuts shown in Figs.~\ref{fig:param_scan_circ}d and \ref{fig:param_scan_circ}f also include a diagnostic of operator complexity. The green markers show the normalized operator entanglement generated by evolving a single-site $Z$ observable at representative points along the high-state-entanglement ridge (App.~\ref{app.scan}). At the chosen parameters, the operator entanglement is substantial, showing that complexity growth is not limited to entanglement in the many-body state but also appears in the evolution of local observables.

We next examine the dynamics generated by this Floquet circuit in larger systems that remain accessible to direct exact simulation. At 28 qubits, the selected parameters show the combination of features targeted in this work: after an initial transient with rapid entanglement growth, the magnetization remains long-lived and develops structured oscillations, while the rate of entanglement growth decreases (Fig.~\ref{fig:param_scan_circ}g). This structured prethermal dynamics is notable because it occurs in a regime with substantial state entanglement: in the 28-qubit exact simulation, the normalized half-system entanglement entropy (App.~\ref{app.scan}) exceeds $S=1/2$ by $N_c=10$ Floquet cycles (Fig.~\ref{fig:param_scan_circ}g). The comparison points in Figs.~\ref{fig:param_scan_circ}h and \ref{fig:param_scan_circ}i illustrate why this parameter choice is nontrivial. At $\theta_z=0$ (Fig.~\ref{fig:param_scan_circ}h), the system also shows prethermal behaviour, with an initial rapid growth of entanglement followed by slower growth. However, its magnetization rapidly relaxes and therefore does not retain the long-lived oscillatory signal targeted here~\footnote{Other observables, such as the average $\langle X\rangle$, remain finite.}. By contrast, the more strongly entangling point in Fig.~\ref{fig:param_scan_circ}i shows thermalizing dynamics, with rapid entanglement growth that eventually approaches a near-maximal value and no long-lived structured magnetization oscillations.

The same qualitative behaviour persists at 35 qubits, albeit with a smaller oscillation amplitude; the magnetization at late Floquet-cycles is shown in Fig.~\ref{fig:fit_to_decay_sin}. Thus, this analysis leaves open whether the oscillatory response in the prethermal state remains measurable as the system size increases. Addressing this question calls for larger systems, such as the 51- and 74-qubit heavy-hex geometries used in the experiments below, which lie beyond direct exact statevector simulation~\cite{Willsch2020JUQCS,DeRaedt2026JUQCS50}. Moreover, the substantial state and operator entanglement seen at the selected parameters suggests that the dynamics should challenge both Schrödinger- and Heisenberg-picture classical simulation strategies. This motivates the error-mitigated quantum experiments of the next section.

\section{Error-mitigated Floquet dynamics beyond exact classical simulations}
\label{sec.results}

\begin{figure*}[tb]
\centering
\begin{minipage}[t]{0.59\textwidth}
\centering
\subfloat[]{
    \includegraphics[width=\linewidth]{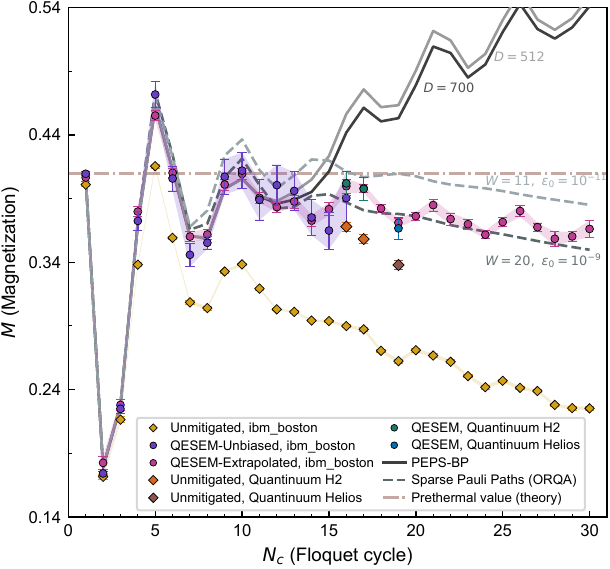}
    \label{fig:magnetization}
}
\end{minipage}
\hfill
\begin{minipage}[t]{0.39\textwidth}
\centering
\subfloat[]{
    \includegraphics[width=\linewidth]{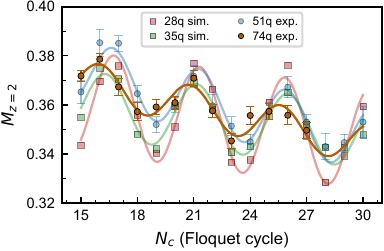}
    \label{fig:fit_to_decay_sin}
}

\subfloat[]{
    \includegraphics[width=\linewidth]{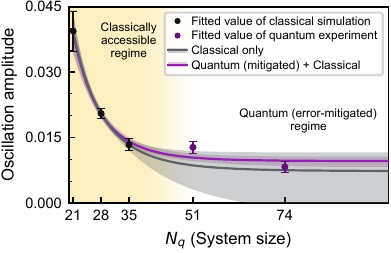}
    \label{fig:finite_size_scaling}
}

\end{minipage}
\caption{
    \textbf{\protect\subref{fig:magnetization}} Magnetization versus Floquet cycle for the $51$-qubit system. Vertical error bars denote one-standard-deviation statistical uncertainties, including sampling overhead for the error-mitigated data. The unmitigated results decay rapidly, while QESEM-Unbiased agrees with PEPS-BP at early times, where the $D=512$ and $D=700$ curves coincide, and continues to give stable mitigated estimates for several cycles after they separate. QESEM-Extrapolated extends the mitigated dynamics to later cycles, resolving slow relaxation and long-time oscillations. Sparse Pauli-path results from two truncation choices fail to provide a converged late-cycle reference and do not reproduce the late-cycle oscillatory response. The horizontal line marks the prethermal magnetization estimated from the fourth-order effective Hamiltonian (App.~\ref{app.effective_ham}).    
    \textbf{\protect\subref{fig:fit_to_decay_sin}} Late-cycle coordination-two magnetization, $M_{z=2}$, for exact $28$- and $35$-qubit simulations and error-mitigated $51$- and $74$-qubit experiments. Vertical error bars denote one-standard-deviation statistical uncertainties, as in \textbf{\protect\subref{fig:magnetization}}. The $74$-qubit data are highlighted, with the remaining system sizes shown in the background. Solid curves show fits to a slowly decaying background plus oscillations [Eq.~\eqref{eq:fit_sin}]. The overall relaxation and oscillation period are similar across system sizes, while the fitted amplitude decreases with increasing size.
    \textbf{\protect\subref{fig:finite_size_scaling}} Fitted oscillation amplitude versus system size, extracted from the late-cycle fits in \protect\subref{fig:fit_to_decay_sin}. Points show the fitted amplitudes; vertical error bars denote one-standard-deviation uncertainties from the weighted nonlinear least-squares fits of the late-cycle traces in \protect\subref{fig:fit_to_decay_sin}. For the error-mitigated data these propagate the sampling uncertainties of the magnetization estimates; for the exact statevector data they reflect only the uncertainty of extracting an amplitude from the chosen fit model and time window. The fit using only the exactly simulated small systems gives $c=7.3\times10^{-3}$ with a central $68\%$ interval of $(-3.7\times10^{-3},1.2\times10^{-2})$, whereas the fit using all available sizes gives $c=9.6\times10^{-3}$ with a central $68\%$ interval of $(8.1\times10^{-3},1.1\times10^{-2})$. The all-size fit also gives $\xi=0.7$ with a central $68\%$ interval of $(0.6,0.9)$, comparable to the independently extracted correlation length $\xi'=1.1\pm0.1$ (App.~\ref{app.correlation}).
} 
\end{figure*}

We therefore turn to experiments on IBM superconducting hardware, using $51$- and $74$-qubit heavy-hex geometries on the \texttt{ibm\_boston} Heron r3 processor. We begin with the $51$-qubit system and apply QESEM-Unbiased to recover the underlying ideal dynamics from noisy measurements. As described in Ref.~\cite{QESEM}, this protocol uses predictive characterization of the gate layers, including the non-Clifford fractional-angle $R_{ZZ}$ gate layers, to construct multi-type QP distributions, which are sampled efficiently using active-volume identification and QPU-time optimization, and stabilized against device drift. The resulting estimators are formally unbiased provided that the characterized noise model accurately describes the device. We therefore validate this model independently (App.~\ref{app.validation}) and compare the mitigated estimates with exact or converged classical calculations wherever these are available. The unbiased results agree with the classical data throughout their converged range, approximately up to $11$ Floquet cycles, and remain statistically feasible through cycle $16$ (Fig.~\ref{fig:magnetization}), thereby providing a quantitatively trusted reference window for the dynamics.

The same comparison also identifies where the available classical references cease to be controlled. The projected entangled pair states with belief propagation (PEPS-BP) results agree with QESEM-Unbiased over their converged early-cycle range, but the largest bond-dimension calculations, $D=512$ and $D=700$, separate at later cycles and develop a nonphysical upward drift (Fig.~\ref{fig:magnetization}; Sec.~\ref{sec.classical}). Sparse Pauli-path (SPP) simulations show strong truncation dependence and develop a clear phase mismatch with the mitigated oscillatory response around cycle $15$. By cycle $16$, neither method provides a converged classical reference consistent with the mitigated data.

QESEM-Unbiased becomes statistically demanding beyond cycle $16$. To reach later cycles, we use QESEM-Extrapolated, a lower-overhead QESEM protocol described in App.~\ref{app.methods}. This protocol relies on a heuristic extrapolation ansatz and can therefore introduce extrapolation bias. We benchmark the extrapolated protocol against QESEM-Unbiased over their common cycle window, finding agreement for the magnetization dynamics within statistical uncertainties (Fig.~\ref{fig:magnetization}). We further test QESEM-Extrapolated in smaller systems, where exact late-cycle dynamics remain available, and through internal consistency checks using different noise amplifications, together with validation of the characterized noise models used to construct the QESEM estimators (App.~\ref{app.validation}). Finally, we compare these results with independent data obtained on Quantinuum using a simplified, characterization-free version of QESEM based on unitary gate folding and extrapolation (See App.~\ref{app.quantinuum} and Ref.~\cite{qedmaForthcoming}). At the selected Floquet cycles evaluated on Quantinuum, the Quantinuum and IBM estimates agree within statistical uncertainties, providing cross-platform corroboration despite the use of different mitigation heuristics on hardware platforms with distinct noise characteristics. Together with the preceding validation tests, this agreement increases our confidence in the results beyond cycle $16$, where the overhead of QESEM-Unbiased becomes prohibitive.

Using QESEM-Extrapolated, we study the dynamics on \texttt{ibm\_boston} up to $30$ Floquet cycles, reaching circuit depths beyond the convergence regime of the leading controlled classical methods considered here, as discussed further in Sec.~\ref{sec.classical}. In the runtime model of Fig.~\ref{fig:fig1b} and App.~\ref{app.ibm_exp_detail}, this corresponds to a circuit-volume boost of approximately $B_V\simeq30$ at fixed target accuracy relative to unmitigated execution. The error-mitigated magnetization shows a slowly decaying background together with long-cycle oscillations at a period approximately four times that of the Floquet drive (Fig.~\ref{fig:magnetization}). In contrast, the SPP simulations fail to reproduce this late-cycle oscillatory response. The horizontal reference value denotes the magnetization predicted for the prethermal plateau from the effective static Hamiltonian obtained by a fourth-order Magnus expansion of the Floquet unitary (App.~\ref{app.effective_ham}); the mitigated dynamics approach this value after roughly ten Floquet cycles.

On the heavy-hex lattice, the local response is coordination number dependent, and the oscillatory component of the magnetization is most clearly resolved on sites with coordination number two, i.e., sites connected to two nearest neighbours. To quantify the system-size dependence of this oscillatory component, we extend the error-mitigated experiment to a $74$-qubit heavy-hex system. The late-cycle traces of the coordination-two magnetization, $M_{z=2}$, are shown in Fig.~\ref{fig:fit_to_decay_sin}. The $51$- and $74$-qubit results are qualitatively consistent with the exact dynamics of the smaller systems, but are resolved at system sizes beyond direct exact statevector simulation. Together, the exact simulations of the smaller systems and the error-mitigated $51$- and $74$-qubit experiments allow us to perform a finite-size analysis of the oscillation amplitude.

We fit the late-cycle dynamics to
\begin{align}
\label{eq:fit_sin}
    M_{z=2}(N_c) = C e^{-\gamma N_c} + A\cos\left(\frac{2\pi}{T}N_c+\phi\right),
\end{align}
where $A$ is the fitted oscillation amplitude, $T$ is the fitted period, $C$ and $\gamma$ set the scale and decay rate of the background, and $\phi$ is the oscillation phase\footnote{Physically, the oscillation amplitude should itself decay with Floquet cycle. We do not include this additional cycle dependence in the fit because, over the fitted cycle window, the corresponding decay rate is consistent with zero while increasing the uncertainties of the other fit parameters}. Applying this fitting procedure to the exact $21$-, $28$-, and $35$-qubit simulations and to the error-mitigated $51$- and $74$-qubit data gives a finite-size comparison of the oscillation amplitude across the available heavy-hex geometries (Fig.~\ref{fig:fit_to_decay_sin}).

We next examine how the extracted oscillation amplitude varies with system size. A priori, the appropriate finite-size scaling ansatz is not clear: exponential decay in $\sqrt{N_q}$, exponential decay in $N_q$, and algebraic decay are all plausible phenomenological forms. The exact $21$-, $28$-, and $35$-qubit amplitudes alone cannot distinguish among these forms (App.~\ref{app.scaling}) or determine whether a nonzero offset is required: with three data points and three fit parameters, the large-size trend is essentially unconstrained. The error-mitigated $51$- and $74$-qubit data provide the additional constraints needed to discriminate among these phenomenological trends over the accessible system sizes (App.~\ref{app.scaling}).

Figure~\ref{fig:finite_size_scaling} shows the exponential finite-size fit using $\sqrt{N_q}$ as the scaling variable appropriate for a two-dimensional geometry,
\begin{align}
\label{eq:exp_sqrt}
    A(N_q) = a e^{-\frac{\sqrt{N_q}}{\xi}} + c.
\end{align}
Fitting the small-system data alone gives an offset coefficient $c=7.3\times 10^{-3}$. Using parametric Monte Carlo uncertainty propagation (App.~\ref{app.scaling}), we obtain a central $68\%$ interval of $(-3.7\times10^{-3},1.2\times10^{-2})$, which includes $c=0$. Thus, the small-system data alone do not determine whether a nonzero offset is required. By contrast, when the larger error-mitigated points are included, the fit gives $c=9.6\times 10^{-3}$ with a central $68\%$ interval of $(8.1\times10^{-3},1.1\times10^{-2})$. 

The same all-size fit gives a decay scale $\xi=0.7$ with a central $68\%$ interval of $(0.6,0.9)$, comparable to the independently extracted correlation length $\xi'=1.1\pm0.1$ (App.~\ref{app.correlation}). This agreement supports the interpretation that the finite-size decay is governed by a physical length scale of order the correlation length. Within this phenomenological finite-size ansatz, the large-system error-mitigated data therefore provide strong evidence for a nonzero asymptotic oscillation amplitude in the thermodynamic limit of heavy-hex ladders that could not be inferred from the small-system simulations alone.
\section{Comparison with classical simulation methods}
\label{sec.classical}
We next compare the error-mitigated magnetization dynamics with leading classical simulation methods that approach the problem from complementary directions. Projected Entangled Pair States with Belief Propagation (PEPS-BP) \cite{verstraete2004renormalization, alkabetz2021tensor,tindall2023gauging} evolves an approximate two-dimensional tensor-network representation of the many-body state, matched to the heavy-hex connectivity, in the Schrödinger picture, while Sparse Pauli Path (SPP) methods~\cite{begusic2023fastconverged,begusic2025sparsepauli,gharibyan2025practicalguideusingpauli,Loizeau25paulistrings,rudolph2026paulipropagationcomputationalframework} propagate observables in the Heisenberg picture using a truncated Pauli expansion. These methods provide stringent tests because their convergence failures are controlled by different measures of complexity: state entanglement for PEPS-BP and operator spreading for SPP. Details of the implementations and convergence diagnostics are given in App.~\ref{app.peps_bp} and App.~\ref{app.spp}.

\begin{figure*}[tb]
\centering
\begin{minipage}[t]{0.96\textwidth}
\centering
\begin{minipage}[t]{0.492\linewidth}
\centering
\subfloat[]{
    \includegraphics[width=\linewidth]{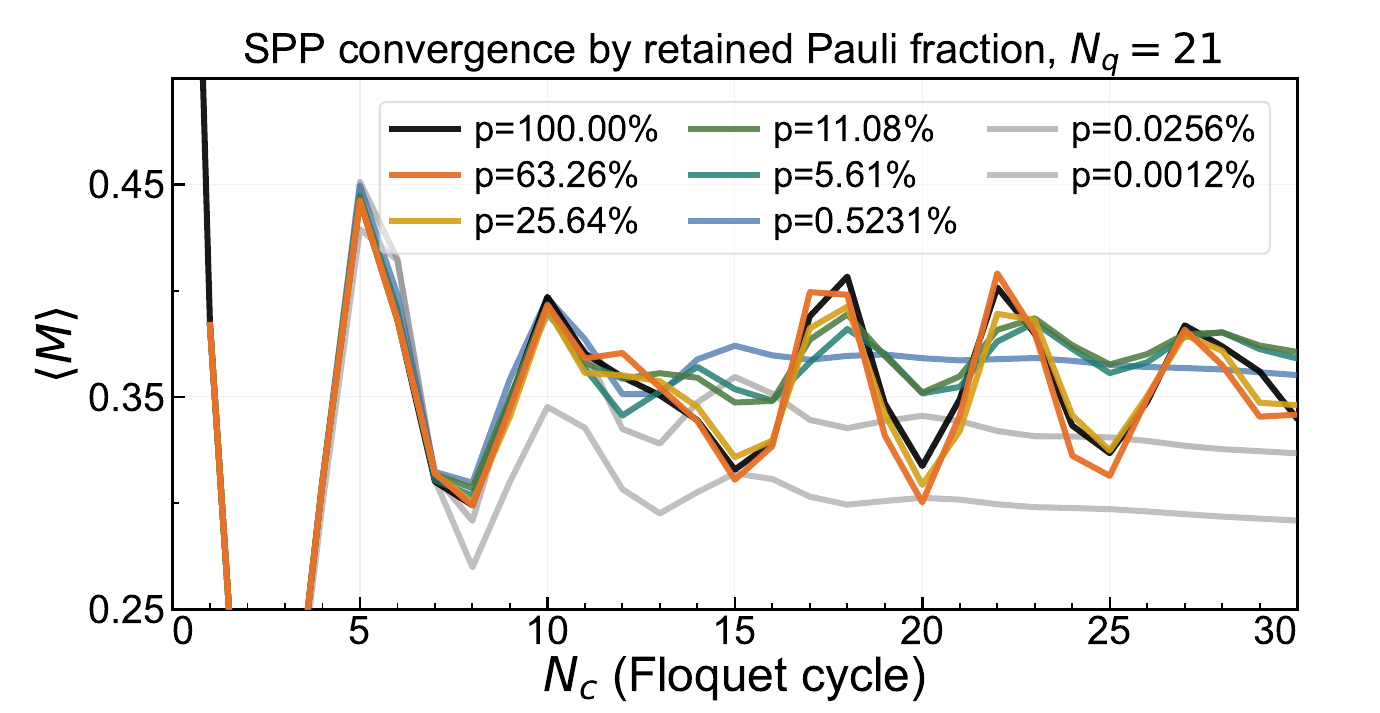}
    \label{fig:spp_21}
}
\end{minipage}
\begin{minipage}[t]{0.492\linewidth}
\centering
\subfloat[]{
    \includegraphics[width=\linewidth]{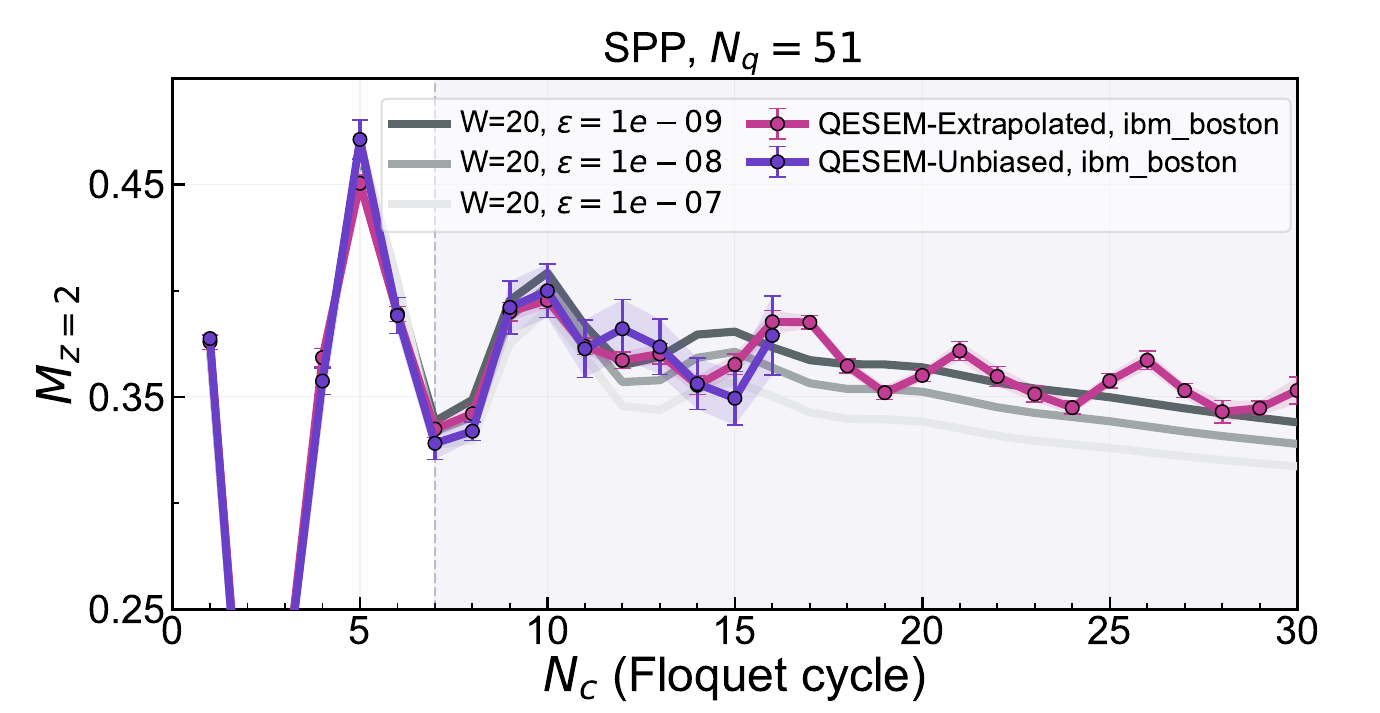}
    \label{fig:spp_51}
}
\end{minipage}
\vspace{0.5em}

\begin{minipage}[t]{0.325\linewidth}
\centering

\subfloat[]{
    \includegraphics[width=\linewidth, trim=0 0 0 25pt, clip]{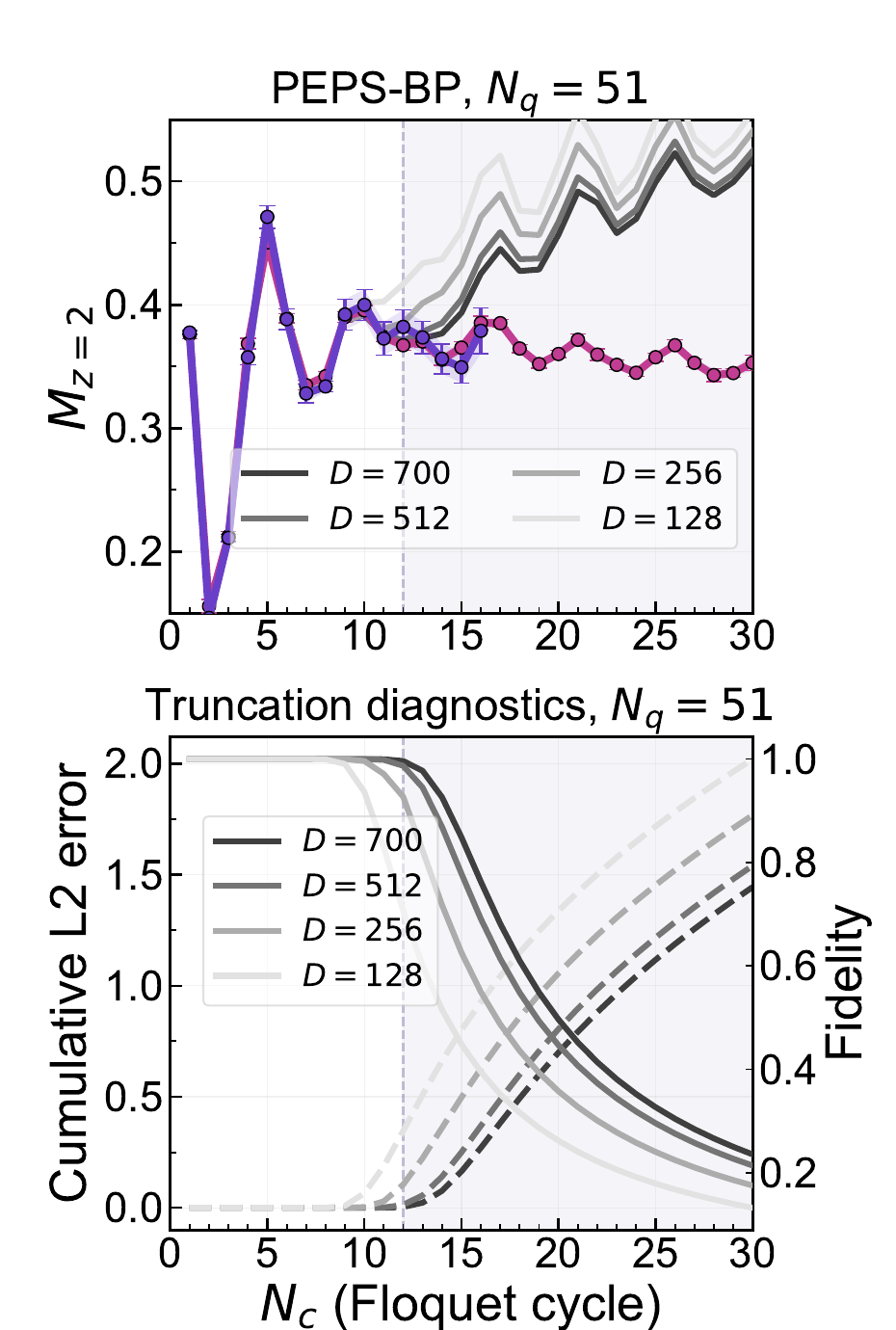}
    \label{fig:peps_bp}

}
\end{minipage}
\hfill
\begin{minipage}[t]{0.325\linewidth}
\centering
\subfloat[]{
    \includegraphics[width=\linewidth, trim=0 0 0 25pt, clip]{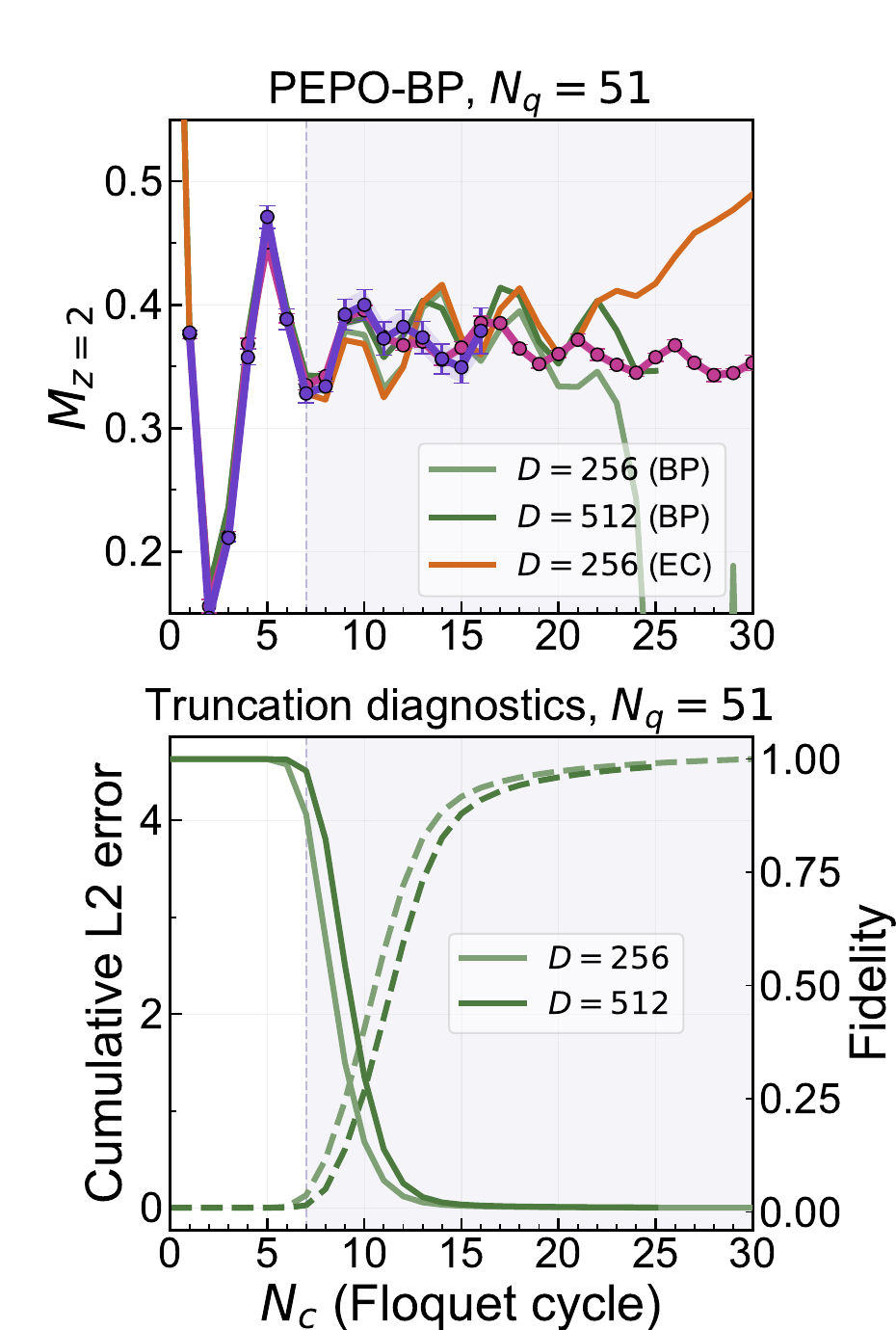}
    \label{fig:pepo_bp}
}
\end{minipage}
\hfill
\begin{minipage}[t]{0.325\linewidth}
\centering
\subfloat[]{
    \includegraphics[width=\linewidth, trim=0 0 0 25pt, clip]{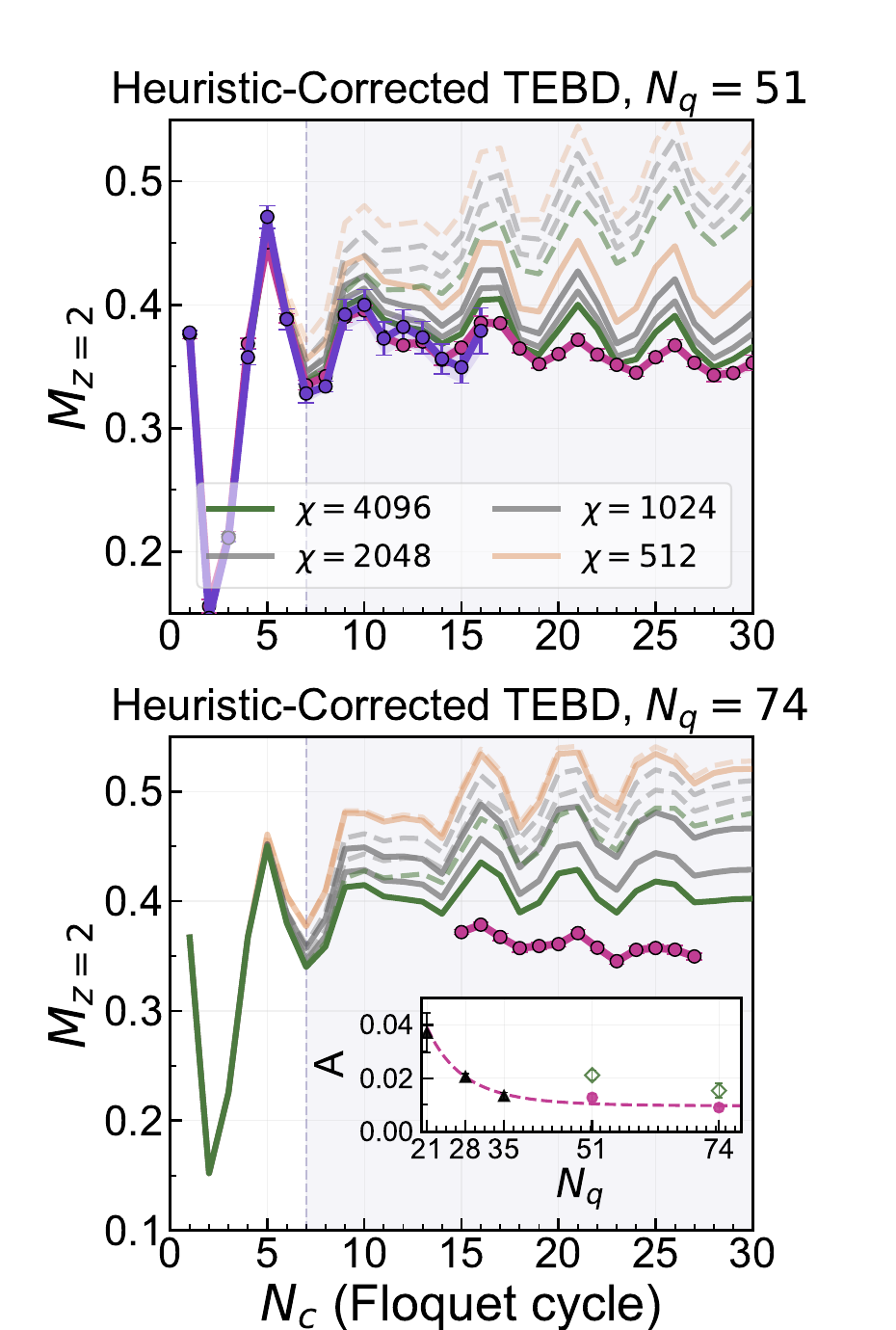}
    \label{fig:ibe_heuristic}
}
\end{minipage}
\end{minipage}
\caption{%
\textbf{\protect\subref{fig:spp_21}}
Magnetization dynamics for the $21$-qubit patch obtained with the ORQA SPP method on the
Fugaku supercomputer for different retained Pauli fractions, $p$, of the
complete Pauli operator space. The solid black curve shows the exact
statevector evolution.
\textbf{\protect\subref{fig:spp_51}}--\textbf{\protect\subref{fig:ibe_heuristic}}
Comparison of the coordination-two magnetization, $M_{z=2}$, as a
function of Floquet cycle, obtained using four classical simulation
methods and compared with the error-mitigated QESEM experiments
(purple: QESEM-Unbiased; magenta: QESEM-Extrapolated).
\textbf{\protect\subref{fig:spp_51}}
SPP results for the $51$-qubit patch. Gray curves show SPP simulations
performed with ORQA using $W=20$ and different coefficient-truncation
thresholds. Although SPP agrees with the quantum results at early times,
it deviates at later Floquet cycles where it fails to reproduce the oscillations of the magnetization.
\textbf{\protect\subref{fig:peps_bp},\protect\subref{fig:pepo_bp}}
Comparison of PEPS-BP (PEPO-BP) with the quantum results. Top: simulated
$M_{z=2}$ dynamics. Bottom: fidelity and cumulative $\ell_2$ truncation
error of the evolved PEPS (PEPO), indicating the onset of loss of
convergence at $N_c\!\approx\!12$ ($N_c\!\approx\!7$). For the PEPO-BP results, expectation values are evaluated using exact contraction (EC) and belief propagation (BP), whereas for the PEPS-BP we employ BP for the contraction part.
\textbf{\protect\subref{fig:ibe_heuristic}}
$M_{z=2}$ obtained using TEBD with different bond dimensions and the
Heuristic-Corrected TEBD procedure described in the main text and App.~\ref{app.mps}. Top:
$51$-qubit patch (Fig.~\ref{fig:ibm-boston}). Bottom: $74$-qubit patch
(Fig.~\ref{fig:ibm-boston}). Inset: late-cycle oscillation amplitude
versus system size, from statevector simulation (black), QESEM-Extrapolated (magenta),
and Heuristic-Corrected TEBD with $\chi=4096$ (green). Combined with the statevector results for the smaller systems, the Heuristic-Corrected TEBD oscillation amplitudes exhibit a non-monotonic system-size dependence, unlike the monotonic decrease observed in the combined statevector and QESEM results.
}
\label{fig:classical_simulation_methods_vs_quantum}
\end{figure*}

As shown in Fig.~\ref{fig:magnetization}, the PEPS-BP simulations obtained with the two largest PEPS bond dimensions $D=512$ and $D=700$, agree with each other through approximately $12$ Floquet cycles, accurately reproducing the quantum dynamics over this interval. Beyond this point, the two simulations begin to separate from one another and simultaneously deviate from the quantum results, signaling the onset of loss of convergence. Together with the truncation diagnostics shown in the bottom panel of Fig.~\ref{fig:peps_bp} and the small effect of belief-propagation loops on the magnetization observed for smaller systems (see App.~\ref{app.peps_bp} for details), this indicates the PEPS-BP results as reliable up to $N_c\approx 12$ cycles, with the $D=700$ result likely remaining accurate for one to two additional cycles. At later times however, the PEPS-BP simulation exhibits an upward drift, exceeding the prethermal magnetization predicted by the Gibbs state obtained from the effective Floquet Hamiltonian (see App.~\ref{app.effective_ham}) at $\sim 16$ Floquet cycles, and continuing to increase thereafter. This behavior is inconsistent with the expected Floquet dynamics, in which local observables are expected to relax toward a prethermal plateau before eventually drifting toward a featureless infinite-temperature state. We therefore interpret the late-cycle increase of the PEPS-BP magnetization as a numerical artifact that is associated with the breakdown of the PEPS-BP approximation rather than a physical feature of the dynamics.

SPP simulations likewise fail to provide converged late-cycle results. Unlike PEPS-BP, where convergence can be tested by increasing the bond dimension, SPP commonly involves at least two competing truncation controls: the maximum Pauli weight $W_0$ and the coefficient cutoff $\epsilon_0$. The two $51$-qubit SPP simulations shown in Fig.~\ref{fig:magnetization}, performed using the high-performance ORQA framework~\cite{broers2025scalablesimulationquantummanybody,Broers_2026} on the supercomputer Fugaku, use different values for these competing truncation parameters while retaining $\mathcal{O}(10^{12})$ Pauli strings over $12888$ Fugaku nodes. Nevertheless, the two trajectories already diverge after approximately seven Floquet cycles and neither reproduces the long-lived oscillatory dynamics observed experimentally. The breakdown becomes particularly evident around the $N_c=15$ Floquet cycle, where both simulation curves exhibit a clear phase mismatch with respect to the experimental quantum error-mitigated data, displaying a maximum where the quantum data shows a minimum, and consequently fail to reproduce the late-cycle oscillations. 

To diagnose this behavior, we study a $21$-qubit version of the same circuit, for which we can retain all $4^{21}$ Pauli strings. This smaller system shows the same qualitative failure under strong truncation, where the late-cycle oscillations are suppressed or shifted in phase (Fig.~\ref{fig:spp_21}). These oscillations are recovered only when a substantial fraction of the full Pauli basis is retained, roughly $10$--$20\%$ in this case. This provides a direct explanation for the $51$-qubit failure: the conventional truncation heuristics used here appear to discard the high-complexity Pauli components needed to reproduce the oscillatory response. Extrapolating this convergence behavior to $51$ qubits suggests that a converged SPP calculation would require resources far beyond the scale of present-day classical computation, with a direct retained-fraction estimate giving an order of $10^{30}$ Pauli strings (App.~\ref{app.spp}). We therefore, as in the case of PEPS-BP, regard the accessible late-cycle SPP trajectories as truncation-dependent numerical artifacts, not as converged predictions of the physical dynamics. The same behavior is observed for the coordination-two magnetization $M_{z=2}$, where both the SPP and PEPS-BP again agree only in the early-time regime before losing convergence (Figs.~\ref{fig:spp_51} and \ref{fig:peps_bp}).

The truncation dependence identified above also motivates a noisy variant of SPP, in which an artificial Pauli-noise channel suppresses small coefficients before truncation, followed by extrapolation to zero noise (App.~\ref{app.noisy_spp}). We find that this approach yields well below an order of magnitude reduction in the peak number of retained Pauli strings, while requiring several noisy simulations for the extrapolation. Consequently, it does not change our conclusions regarding SPP convergence.

We next consider a Heisenberg-picture tensor-network approach, Projected Entangled Pair Operators with Belief Propagation (PEPO-BP) \cite{liao2023simulation,begusic2023fastconverged}.  This method represents the evolving observable as a two-dimensional operator tensor network and back-propagates it through the Floquet circuit. As shown in Fig.~\ref{fig:pepo_bp} PEPO-BP loses reliability after approximately seven cycles, earlier than the Schrödinger-picture PEPS-BP calculation (Fig.~\ref{fig:pepo_bp}). This earlier breakdown is consistent with the rapid operator-complexity growth identified in the Floquet-parameter scan of Fig.~\ref{fig:param_scan_circ} as well as with the SPP results.
 
Finally, we examine Heuristic-Corrected Time-Evolving Block Decimation (TEBD), the heuristic Matrix Product State (MPS) based rescaling scheme introduced in Ref.~\cite{mandra2025heuristic}. Unlike PEPS-BP, which preserves the two-dimensional heavy-hex connectivity, this method maps the system onto a one-dimensional MPS ordering that snakes through the heavy-hex geometry (App.~\ref{app.mps}). It then produces the time evolved state using the TEBD and attempts to compensate for finite-bond-dimension bias using an estimate of the MPS fidelity (see App.~\ref{app.mps} for more details). The uncorrected MPS magnetization dynamics are rescaled cycle by cycle using a factor of the form $F(N_c)^{\gamma(\chi,N_q)}$, where $F(N_c)$ is the MPS simulation fidelity. The exponent is fitted using smaller systems, where exact statevector results are available. For the $51$-qubit system, some rescaled MPS curves visually resemble the error-mitigated oscillations. However, the rescaled trajectories have not yet converged with increasing bond dimension, and the same calibrated correction does not generalize consistently to the 74-qubit geometry (Fig.~\ref{fig:ibe_heuristic}), indicating that substantially larger exact benchmarks would be needed to calibrate such a correction reliably. In particular, the resulting finite-size amplitudes do not follow the monotonic decay expected from the exact small-system simulations and the error-mitigated data (inset of Fig.~\ref{fig:ibe_heuristic}). We therefore regard the MPS-corrected curves as a heuristic extrapolation with no evidence of predictive accuracy.

These comparisons indicate that the late-cycle oscillatory response lies beyond the convergence regime of the leading classical methods considered here. This conclusion is supported by independent failure modes across complementary approaches: bond-dimension truncation in PEPS-BP, truncation dependence in SPP, rapid operator-complexity growth in PEPO-BP, and inconsistent size generalization in the heuristic MPS correction. The error-mitigated quantum experiments therefore extend the finite-size analysis into a regime where no controlled classical calculation in this study provides a reliable late-cycle reference.

\section{Discussion}
\label{sec.disscussion}

In this work, we extended finite-size scaling using large-scale, high-precision quantum simulations, finding strong evidence for unexpected subharmonic prethermal oscillations in the thermodynamic limit of a periodically driven quantum magnet on the heavy-hex lattice. Geometrically, the systems we studied correspond to a series of ladders with maximal width of two heavy-hex plaquettes, and our thermodynamic-limit conclusion therefore applies specifically to this geometry. Further work on other geometries is needed in order to conclude whether this conclusion extends to genuinely two-dimensional systems. Moreover, our experiment reached only intermediate physical times. A theoretical understanding of the physical mechanism behind these oscillations, and in particular, the strong dependence of the dynamics of the qubits on their coordination number, would help guide future experiments. Such an understanding can also guide the construction of Floquet models exhibiting related phenomena with different temporal structures and on different lattice geometries. 

Our results demonstrate that today's commercially accessible quantum processors, when paired with advanced error suppression and mitigation software, can accurately resolve many-body dynamics in a regime beyond the reach of state-of-the-art classical simulations. Here, error-mitigated measurements performed on IBM’s superconducting quantum processors remained accurate across the large circuit depths required to resolve the subharmonic prethermal oscillations. On the other hand, we were not able to accurately observe this many-body phenomenon classically, despite using a variety of recent methodological advances and extensive classical compute resources. 

The above comparison suggests that error mitigation enables quantum computing to access a  regime that is not reliably reproduced classically for the problem at hand \cite{YEMpaper, zimboras2025myths, lanes2025frameworkquantumadvantage, eisert2025mind}. This observation should be distinguished from an asymptotic, complexity-theoretic separation based on error mitigation, which is essentially excluded on fundamental grounds \cite{YEMpaper,schuster2024polynomial,takagi2022fundamental,Takagi2023universal,Tsubouchi2023universal,quek2024exponentially}. Better classical algorithms or problem-specific strategies may nevertheless reproduce these results, and we invite the community to address this challenge through the Quantum Advantage Tracker \cite{quantumAdvantageTrackerObservableEstimations}. 

By uncovering a physical phenomenon that was not predicted by prior classical analyses, our results show that error-mitigated quantum computation is beginning to serve not only as a computational benchmark, but as a tool for scientific discovery, with near-term applications in condensed-matter and high-energy physics \cite{daley2022practical,bauer2023fundamental}. More broadly, our results motivate the adoption of error-mitigated quantum computation as a complementary tool for addressing concrete questions in many-body physics.

Looking ahead, anticipated improvements in quantum hardware fidelity, connectivity and execution rates \cite{ibmQuantumRoadmap}, as well as software advancements in noise characterization, noise-aware transpilation, error suppression and error mitigation, should extend the reach of error-mitigated quantum processors to larger and more flexible physical systems, longer evolution times, and richer observables. Even larger problem instances will require error-corrected quantum processors that can operate below break-even while still admitting at least tens of logical qubits. The limited number of physical qubits anticipated in the coming years, along with additional hardware constraints, will nevertheless constrain the logical fidelity that can be achieved, and we expect logical error mitigation methods to play a key role in quantum computing in this `early fault-tolerance' era \cite{Aharonov2025SALEM,Zhou2025SoftInformation,Dinca2025DecoderConfidence,Tsubouchi2026SyndromeAware,Umbrarescu2026InfiniteDistance,Kumar2026CoDesigning,Yuan2026Zeno}.

\section*{Acknowledgments}
We thank Jay Gambetta for his vision and for facilitating the development of this work;  Michael Foss-Feig for fruitful discussions and feedback on the manuscript; and Joseph Tindall for providing valuable feedback on classical simulations posted to the Quantum Advantage Tracker \cite{quantumAdvantageTrackerObservableEstimations} as well as for pointing us to \texttt{TensorNetworkQuantumSimulator.jl}. We acknowledge the use of IBM Quantum Credits via the IBM Quantum Startups Program for this work, and the use of Quantinuum System Model H2 and Helios via Quantinuum’s Startup Partner Program. Part of the numerical simulations was performed using the Supercomputer Fugaku (Project IDs hp260079, ra000011, and ra010014) and the HOKUSAI supercomputer at RIKEN (Project ID RB230105). 
A portion of this work at RIKEN is based on results obtained from Project No.~JPNP20017, supported by the New Energy and Industrial Technology Development Organization (NEDO). 
This work at RIKEN was also supported by JSPS KAKENHI Grants No.~JP21H04446 and No.~JP26K06972, JST COI-NEXT (Grant No.~JPMJPF2221), and the Program for Promoting Research on the Supercomputer Fugaku (Grant No.~MXP1020230411) by MEXT, Japan. 
L.B., T.S., and S.Y. also acknowledge support from the UTokyo Quantum Initiative, the RIKEN TRIP initiative (RIKEN Quantum), and the COE research grant in computational science from Hyogo Prefecture and Kobe City through the Foundation for Computational Science.

\clearpage

\bibliography{bibliography_revised.bib}

\clearpage

\appendix

\counterwithin{figure}{section}
\renewcommand{\thefigure}{\thesection.\arabic{figure}}

\startappendixtoc

\section*{Appendix}

\section*{Contents}
\setcounter{tocdepth}{2}
\renewcommand{\theequation}{\thesection\arabic{equation}}
\renewcommand{\thefigure}{\thesection\arabic{figure}}
\setcounter{equation}{0}
\setcounter{figure}{0}
\printappendixtoc

\section{IBM experimental details}
\label{app.ibm_exp_detail}

This appendix describes the IBM experiments used in the main text, including the circuit geometries, compilation procedure, device properties, and data aggregation across experimental runs. We also tabulate the QPU time required to achieve the target precision for each mitigated Floquet cycle.

\subsection{Quantum circuits and experimental geometries}

\begin{figure}[tb]
    \centering
    \includegraphics[width=\columnwidth]{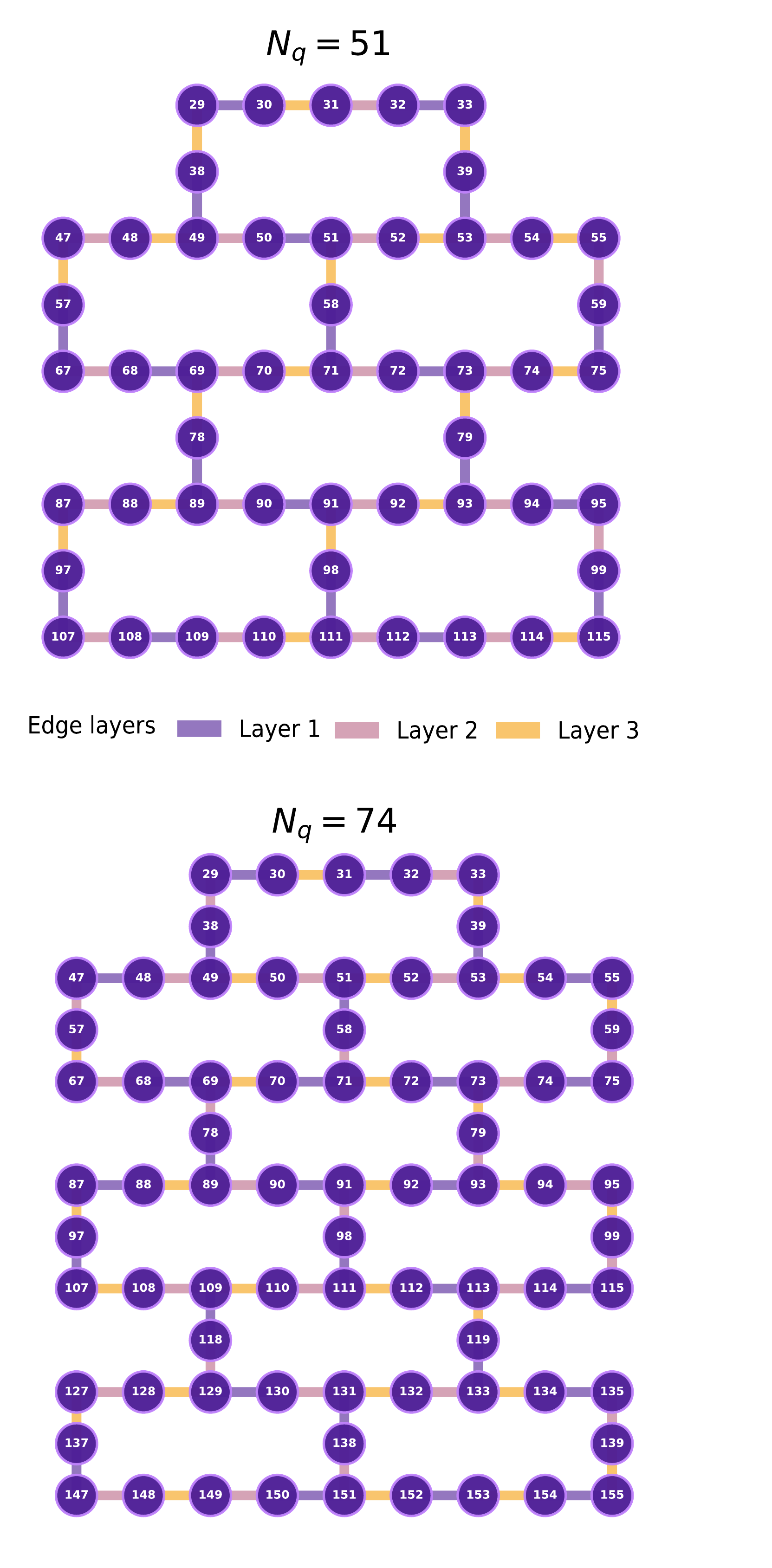}
    \caption{Heavy-hex subgraphs used for the IBM experiments on \texttt{ibm\_boston}. The upper and lower panels show the $51$- and $74$-qubit geometries, respectively. Edge colours indicate the three disjoint two-qubit gate layers, corresponding to the bond sets $E_1$, $E_2$, and $E_3$ used to implement the nearest-neighbour $R_{ZZ}$ interactions in one Floquet cycle.}
    \label{fig:ibm-boston}
\end{figure}

The quantum circuits executed in the IBM experiments implement the Floquet dynamics of a mixed-field Ising model on heavy-hex lattices. The main experiments use two device subgraphs containing $51$ and $74$ qubits, formed from six and nine closed heavy-hex loops, respectively (Fig.~\ref{fig:ibm-boston}). In each Floquet cycle, the nearest-neighbour $ZZ$ interactions are applied through three layers of parallel fractional-angle $R_{ZZ}$ gates. The corresponding edge partition is colour-coded in Fig.~\ref{fig:ibm-boston}, with the three colours denoting the bond sets $E_1$, $E_2$, and $E_3$ appearing in Eq.~\eqref{eq:unitary}.

The Floquet circuit is hardware-native by construction. Following the QESEM implementation of Ref.~\cite{QESEM}, we nevertheless apply the compilation and twirling steps required for mitigation: the fractional-angle $R_{ZZ}$ layers are partially Pauli-twirled and square-root Pauli-twirled, the intervening single-qubit rotations on each qubit are compressed into a fixed $U_3$ form between consecutive $R_{ZZ}$ layers, and the final measurements are twirled using complementary measurement pairs.

\subsection{Execution on IBM devices}

The IBM experiments described in the main text were executed on \texttt{ibm\_boston}, an IBM Heron r3 superconducting quantum processing unit (QPU). The device contains $156$ qubits arranged in a heavy-hex connectivity graph, whose edges support the native two-qubit operations used to implement the $R_{ZZ}$ layers. During the experimental runs, the average effective two-qubit infidelity over the $51$-qubit geometry ranged from $0.23\%$ to $0.31\%$, while over the $74$-qubit geometry it ranged from $0.22\%$ to $0.24\%$. These effective infidelities include the partially Pauli-twirled two-qubit
gate infidelity together with the infidelity of the four $SX$ gates entering the two $U_3$ compressions, two on each active qubit.

The IBM data used in the main text are organized into several data sets, grouped by system size and mitigation protocol. These comprise QESEM-Unbiased experiments on the $51$-qubit geometry at shallow and intermediate Floquet cycles, QESEM-Extrapolated experiments on the $51$-qubit geometry spanning the full range of probed Floquet cycles, and QESEM-Extrapolated experiments on the $74$-qubit geometry at late Floquet cycles. For each data set, measurements from multiple executions at different times are aggregated before forming the final mitigated observables. Each execution includes circuits for several Floquet cycles, allowing the same characterization data to be shared across multiple time steps and thereby reducing the characterization overhead. Table~\ref{tab:exp_series} summarizes the Floquet cycles, shot counts, mitigation method, and average precision for each data set.

\begin{table}[tb]
\centering
\caption{Summary of shot usage for the IBM experimental data sets. The final column reports the average absolute mitigated magnetization precision across the listed Floquet cycles.
}
\label{tab:exp_series}
{
\begin{tabular}{ |c||c|c|c|c|}
\hline
System & Floquet & Mitigation & Total  & Average\\
size & cycles & method & shots  & mitigation \\
$N_q$ & & & ($10^6$) & precision ($M$) \\
\hline
$51$ & $1$--$16$ & QESEM-Unbiased & $38$ & $1.5\times 10^{-2}$ \\
\hline
$51$ & $1$--$30$ & QESEM-Extrapolated & $83$ & $5\times 10^{-3}$ \\
\hline
$74 $& $15$--$27$ & QESEM-Extrapolated & $47$ & $5\times 10^{-3}$ \\
\hline
\end{tabular}
}
\end{table}

\subsection{QPU time and mitigation overhead}
The quasiprobabilistic mitigation and noise-amplification procedures used in QESEM-Unbiased and QESEM-Extrapolated, respectively, require executing many circuit variants for each target input circuit, corresponding to a specified number of Floquet cycles; see Ref.~\cite{QESEM} for details. As a result, the total number of shots alone does not fully capture the QPU time required for mitigation. Here we report the QPU time associated with each mitigated Floquet cycle and each mitigation method used in the IBM experiments.

Each experimental run contains circuits corresponding to several Floquet cycles. Each Floquet cycle defines a separate QESEM input circuit and is mitigated independently, but the circuits in a given run are executed together and share the same device characterization. This batching reduces the characterization overhead relative to treating each Floquet cycle as a separate experiment. The QPU times reported in Tables~\ref{tab:qesem_time} and~\ref{tab:qesem_zne_time} are therefore amortized over the set of Floquet cycles included in the same experimental run; estimating the time required to mitigate a single Floquet-cycle input circuit in isolation requires multiplying these values by a modest characterization-overhead factor.

The rapid growth of the QESEM-Unbiased cost with Floquet cycle motivates the use of the lower-overhead QESEM-Extrapolated protocol for the later-time $51$-qubit data and for the $74$-qubit experiments.

\begin{table}[tb]
\centering
\caption{QPU time per Floquet cycle for QESEM-Unbiased mitigation of the $51$-qubit dynamics. The precision column gives the target absolute precision of the mitigated magnetization.}
\label{tab:qesem_time}
{
\begin{tabular}{ |c||c|c|}
\hline
Floquet  & Mitigation  & QPU time \\
 cycle &  precision ($M$) & (minutes) \\
\hline
$1$ & $1.6\times 10^{-3}$ & $2.0$ \\
\hline
$2$ & $1.7\times 10^{-3}$ & $2.0$ \\
\hline
$3$ & $3.3\times 10^{-3}$ & $2.0$ \\
\hline
$4$ & $7.3\times 10^{-3}$ & $2.0$ \\
\hline
$5$ & $1.0\times 10^{-2}$ & $2.5$ \\
\hline
$6$ & $9.7\times 10^{-3}$ & $3.3$ \\
\hline
$7$ & $9.2\times 10^{-3}$ & $4.3$ \\
\hline
$8$ & $4.8\times 10^{-3}$ & $12.9$ \\
\hline
$9$ & $1.4\times 10^{-2}$ & $7.9$ \\
\hline
$10$ & $1.4\times 10^{-2}$ & $11.2$ \\
\hline
$11$ & $1.6\times 10^{-2}$ & $14.1$ \\ 
\hline
$12$ & $1.6\times 10^{-2}$ & $22.3$ \\
\hline
$13$ & $1.5\times 10^{-2}$ & $40.6$ \\
\hline
$14$ & $1.4\times 10^{-2}$ & $67.1$ \\
\hline
$15$ & $1.5\times 10^{-2}$ & $127.7$ \\
\hline
$16$ & $2.1\times 10^{-2}$ & $88.5$ \\
\hline
\end{tabular}
}
\end{table}

\begin{table}[tb]
\centering
\caption{QPU time per Floquet cycle for QESEM-Extrapolated mitigation of the $51$-qubit dynamics. The precision column gives the target absolute precision of the mitigated magnetization.}
\label{tab:qesem_zne_time}
{
\begin{tabular}{ |c||c|c|}
\hline
Floquet  & Mitigation  & QPU time  \\
cycle & precision ($M$) & (minutes) \\
\hline
1 & $3.1\times 10^{-3}$ & 2.6
\\
\hline
2 & $5.3\times 10^{-3}$ & 2.6
\\
\hline
3 & $4.6\times 10^{-3}$ & 2.6
\\
\hline
4 & $4.0\times 10^{-3}$ & 3.8
\\
\hline
5 & $3.8\times 10^{-3}$ & 4.4
\\
\hline
6 & $3.5\times 10^{-3}$ & 4.4
\\
\hline
7 & $4.0\times 10^{-3}$ & 4.8
\\
\hline
8 & $4.0\times 10^{-3}$ & 4.2
\\
\hline
9 & $4.3\times 10^{-3}$ & 5.0
\\
\hline
10 & $4.0\times 10^{-3}$ & 7.9
\\
\hline
11 & $4.0\times 10^{-3}$ & 8.5
\\
\hline
12 & $4.1\times 10^{-3}$ & 8.6
\\
\hline
13 & $5.0\times 10^{-3}$ & 8.2
\\
\hline
14 & $4.5\times 10^{-3}$ & 8.4
\\
\hline
15 & $5.0\times 10^{-3}$ & 8.2
\\
\hline
16 & $5.6\times 10^{-3}$ & 8.7
\\
\hline
17 &$3.2\times 10^{-3}$ & 36.3
\\
\hline
18 & $3.4\times 10^{-3}$ & 35.0
\\
\hline
19 & $3.4\times 10^{-3}$ & 39.9
\\
\hline
20 & $3.1\times 10^{-3}$ & 39.6
\\
\hline
21 & $4.6\times 10^{-3}$ & 14.1
\\
\hline
22 & $4.8\times 10^{-3}$ & 14.6
\\
\hline
23 & $4.0\times 10^{-3}$ & 36.7
\\
\hline
24 & $3.2\times 10^{-3}$ & 56.5
\\
\hline
25 & $3.5\times 10^{-3}$ & 44.0
\\
\hline
26 & $4.5\times 10^{-3}$ & 20.4
\\
\hline
27 & $3.4\times 10^{-3}$ & 39.1
\\
\hline
28 & $5.6\times 10^{-3}$ & 20.3
\\
\hline
29 & $3.6\times 10^{-3}$ & 57.1
\\
\hline
30 & $6.5\times 10^{-3}$ & 27.7
\\
\hline
\end{tabular}
}
\end{table}

\subsection{Active volumes and dilution factors}
The resources (time overhead) for noise mitigation of any Pauli-observable grow exponentially with the gate infidelity and the active volume $V_{\text{A}}$, which is the volume (number of gates) of the part of the circuit which affects the value of the observable (within its light-cone).
The noise dependence of the observable expectation value, however, decreases roughly exponentially with the infidelity and the \textit{effective} volume, which is always a fraction of the active volume. That is, the different noise terms in the circuit do not all contribute to the decay of observables even when inside the active volume. The scale difference between the two types of volumes is denoted by $\rho$:
\begin{equation}
    \rho \equiv V_{\text{eff}} \,/\, V_{\text{A}}
\end{equation}
which is the ratio of the two volumes and is termed the dilution factor \cite{QuantinuumDilution2025}.

The typical noise decay rate for an observable is approximately $\gamma = I_{F} \, V_{\text{eff}}$, where $I_{F}$ is the average infidelity per gate that observable experiences. For our experiments on the IBM Heron device, $I_{F}$ is of the order of $2 \times 10^{-3}$. The circuits relatively small dilution factors, $\sim 0.1$, allow estimating noisy and amplified observables to satisfactory precision with modest shot overhead. The estimates we have obtained allowed us to extrapolate the mitigated results to high accuracy. To demonstrate this quantitatively, consider the single-qubit Pauli-$Z$ observable $\langle Z_{71} \rangle$, which corresponds to the central qubit in the 51-qubit graph (see Fig.~\ref{fig:ibm-boston}).

\begin{figure}[tb]
    \centering
    \includegraphics[width=\columnwidth]{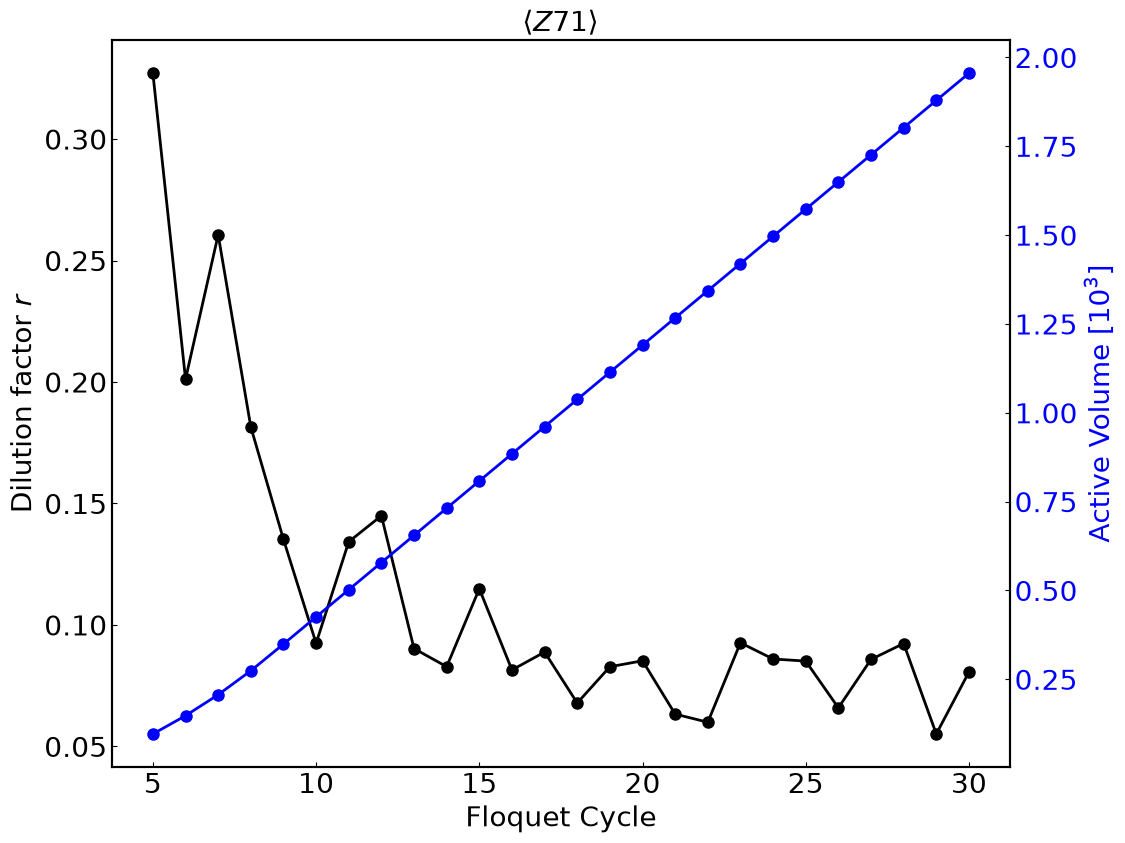}
    \caption{Blue axes: Active volume as a function of the Floquet cycle for the $\langle Z_{71} \rangle$ Pauli observable in the 51-qubit system. Black axes: dilution factor, $\rho$, for the same observable. The average gate infidelity is approximately $2.4 \times 10^{-3}$. }
    \label{fig:active_volume}
\end{figure}

In Fig.~\ref{fig:active_volume} we plot both the active volume $V_{\text{A}}$ and the dilution factor, $\rho$, as a function of the Floquet cycle for that observable. The light-cone of the center-qubit observable covers the entire computational graph starting from cycle 9, indicated by the linear growth of the active volume, at a rate of approximately 76 gates per cycle. The fluctuations in $\rho$ suggest that late-cycle bare expectation values cannot be reliably corrected by a simple rescaling calibrated at early cycles, where the tensor-network values are assumed to be accurate.

The distinction between active and effective volume also provides a useful way to quantify the circuit-volume reach enabled by mitigation. In the main text, we summarize this reach by a circuit-volume boost $B_V$, defined as the approximate increase in circuit volume that can be accessed at fixed target accuracy relative to unmitigated execution. This boost can be estimated in two complementary ways. First, at fixed precision, one can compare the active volumes reached with QESEM-Extrapolated and with unmitigated execution. For the $51$-qubit system, the QESEM-Extrapolated data reach $N_c=30$ with an average absolute precision of $5\times10^{-3}$, whereas the unmitigated magnetization deviates from the converged reference by more than this amount already at $N_c=2$. Using the active-volume map employed in Fig.~\ref{fig:fig1b}, this gives
\begin{equation}
    B_V^{\mathrm{reach}}
    =
    \frac{V_A(30)}{V_A(2)}
    =
    \frac{2000}{22.5}
    \simeq 90.
\end{equation}
This threshold-based estimate illustrates the increase in accessible circuit volume, but it is sensitive to the small active volumes at early cycles and to state-preparation and measurement errors, which require separate treatment.

A more robust estimate follows from the exponential
attenuation of the observable, $M_{\mathrm{unmit}}=M_{\mathrm{ideal}}e^{-I_FV_{\mathrm{eff}}}$.
Taking the mitigated estimate $M_{\mathrm{mit}}$ as an estimate of
$M_{\mathrm{ideal}}$, with uncertainty $\sigma_{\mathrm{mit}}$, the
circuit-volume boost corresponding to the achieved precision is
\begin{align}
    B_V(N_c)
    &=
    \frac{
        -\ln\left|
        M_{\mathrm{unmit}}(N_c)/M_{\mathrm{mit}}(N_c)
        \right|
    }{
        -\ln\left[
        1-\sigma_{\mathrm{mit}}(N_c)/|M_{\mathrm{mit}}(N_c)|
        \right]
    }
    \nonumber\\
    &\simeq
    \frac{
        \left|M_{\mathrm{mit}}(N_c)-M_{\mathrm{unmit}}(N_c)\right|
    }{
        \sigma_{\mathrm{mit}}(N_c)
    },
\end{align}
where the second line is the small-bias approximation. Averaging the full logarithmic expression over $N_c=17$--$30$ gives
$B_V\simeq29$, motivating the value $B_V\sim30$ quoted in the main text.
\section{Correlation length}
\label{app.correlation}
In this appendix, we estimate the correlation length governing the physics in the prethermal regime. We do so by calculating the correlation function
\begin{align}
    C(r) &= \frac{1}{N_r} \sum_{\substack{q<q':\\ d(q,q') = r}} \left\langle Z_q Z_{q'}\right\rangle,
    \label{eq:corr_def}
\end{align}
at various Floquet cycles. Here $d(\cdot,\cdot)$ is a distance function on the qubit graph and $N_r$ is the number of pairs of qubits with graph distance $r$ between them. In particular, we consider $r=1,2,3$ for which the 51-qubit system has $N_{r=1}=56$, $N_{r=2}=71$, and $N_{r=3}=86$.

\begin{figure}[tb]
\centering
\subfloat[]{
    \includegraphics[width=0.95\columnwidth]{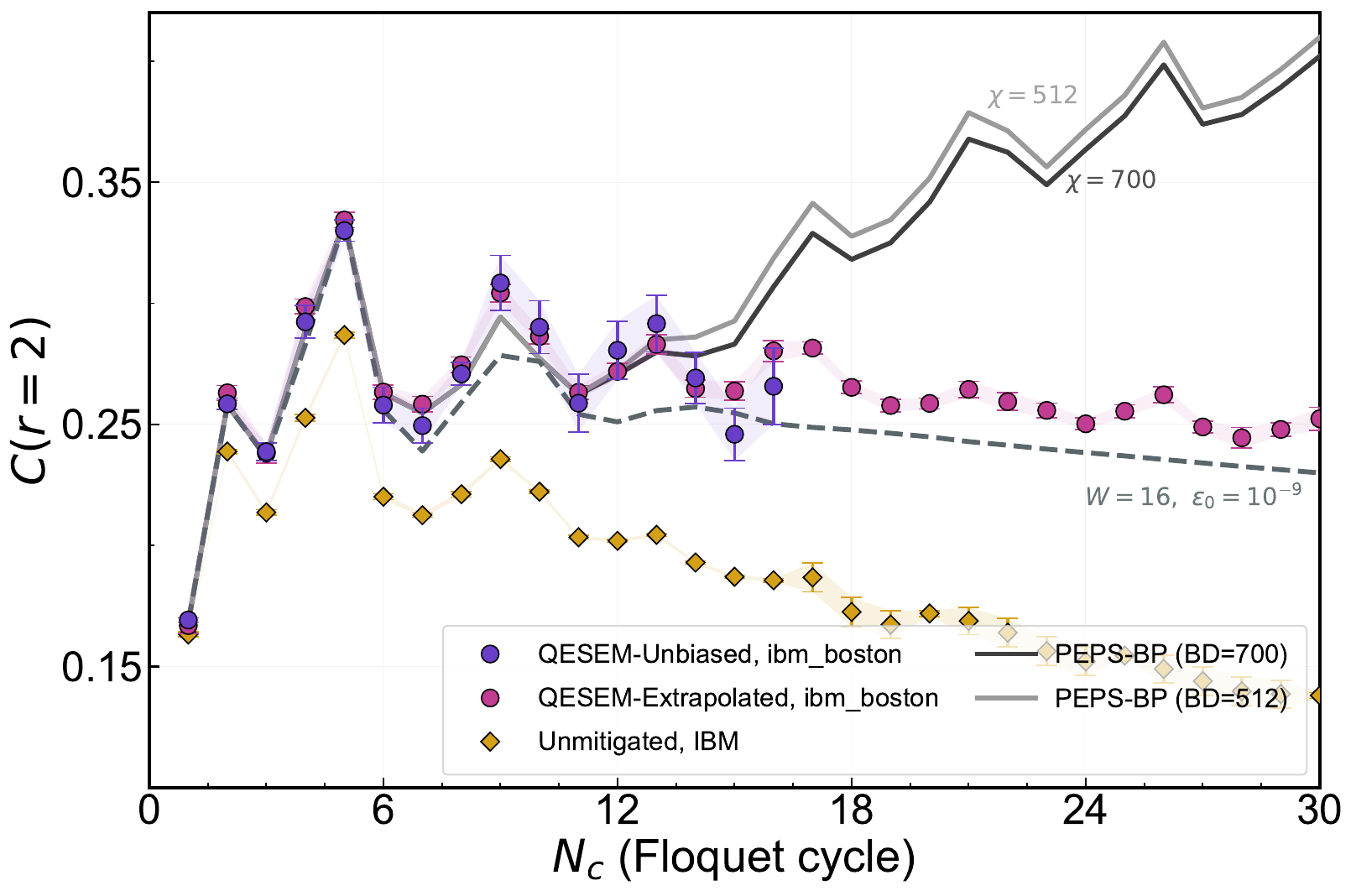}
    \label{fig:ZIZ}
}\\
\subfloat[]{
    \includegraphics[width=0.95\columnwidth]{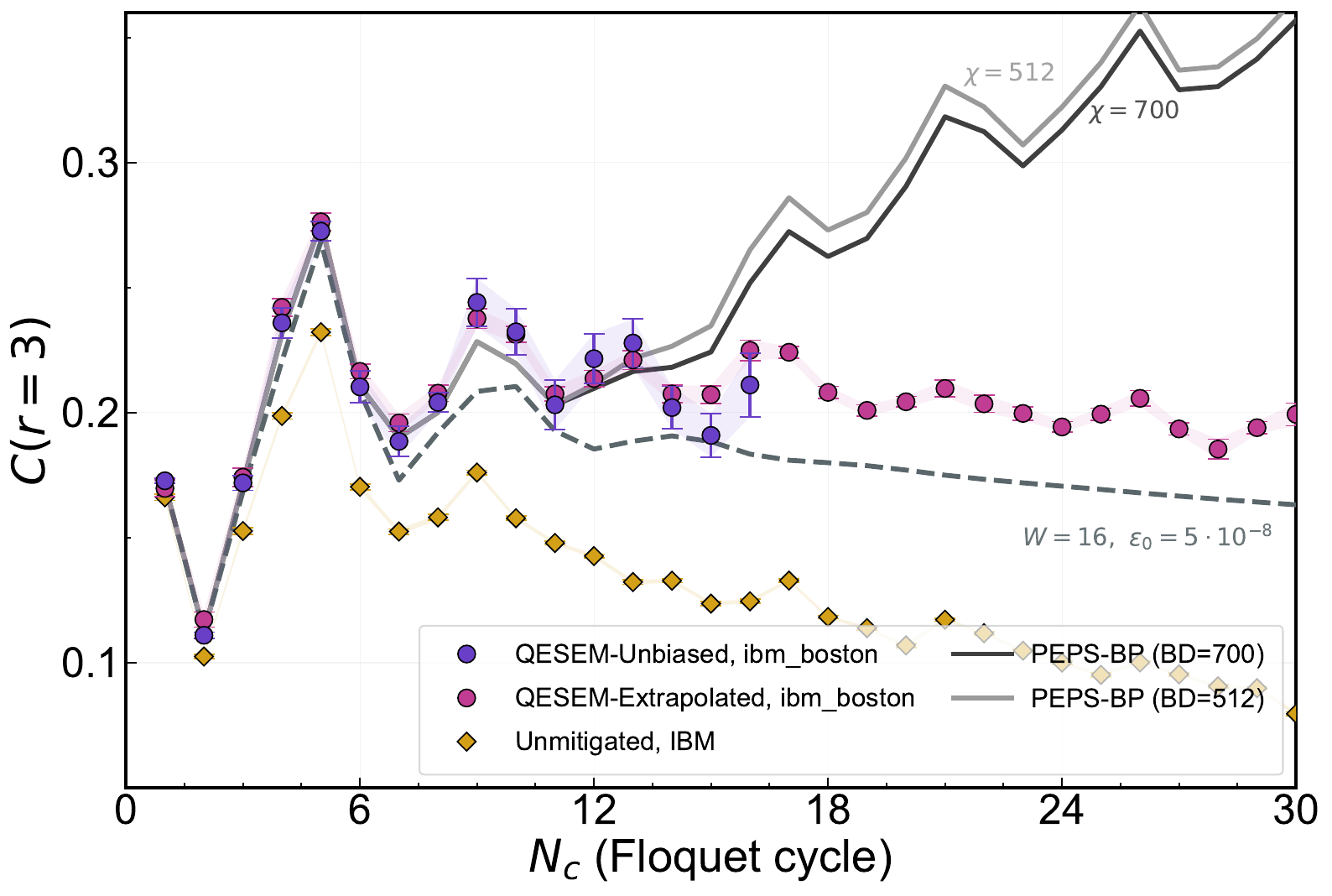}
    \label{fig:ZIIZ}
}
\caption{\textbf{\protect\subref{fig:ZIZ}} Error-mitigated quantum dynamics of the next-nearest-neighbour correlation function, $C(r=2)$, for the $51$-qubit system, compared with the corresponding PEPS-BP and SPP simulations. The PEPS-BP results agree closely with both the QESEM-Unbiased and QESEM-Extrapolated data for approximately the first twelve Floquet cycles, after which they diverge where the PEPS-BP develops the same nonphysical upward drift observed for the magnetization in Fig.~\ref{fig:magnetization} and the SPP fails to reproduce the late-cycle oscillatory behaviour. In contrast, the quantum data exhibit persistent late-cycle oscillations, while the correlations decay only slowly and remain appreciable at late times.
\textbf{\protect\subref{fig:ZIIZ}} Same as \textbf{\protect\subref{fig:ZIZ}}, but for the next-next-nearest-neighbour correlation function, $C(r=3)$. The qualitative behaviour is similar: the quantum dynamics display late-cycle oscillatory correlations, whereas the classical simulations fail to reproduce the late-cycle dynamics.}
\label{fig:ZIZ_ZIIZ}
\end{figure}

Figures~\ref{fig:ZIZ} and~\ref{fig:ZIIZ} show data for $C\left({r=2}\right)$ and $C\left({r=3}\right)$, respectively, including unmitigated results, QESEM-Unbiased results, QESEM-Extrapolated results, and results from classical simulation methods. Similar to the magnetization observables, the QESEM-Unbiased and QESEM-Extrapolated results remain in good agreement over the time interval for which QESEM-Unbiased is available, providing further evidence that the mitigation procedure remains effective for these higher-weight observables. In contrast, the classical simulations exhibit the same qualitative behaviour observed for the magnetization: PEPS-BP reproduces the early-time dynamics but loses convergence at later Floquet cycles, while SPP fails to reproduce the late-cycle dynamics. The slight discrepancies between the mitigated quantum data and the PEPS-BP results during the early Floquet cycles arise from the approximate belief-propagation contraction used to compute the expectation values on the loopy heavy-hex lattice. These approximation errors are expected to affect higher-weight observables more strongly than single-site quantities (see Fig.~\ref{fig:35q_peps_app}).

\par
These results extend the conclusions drawn from the magnetization dynamics; both correlation functions exhibit persistent late-cycle oscillations in the quantum data, demonstrating that the long-lived prethermal response is not limited to local magnetization observables but is also reflected in higher-weight correlations.
\par
In addition to the dynamics of the single-site observables discussed in the main text, namely the total magnetization, $M$, and the coordination-two magnetization, $M_{z=2}$, we also examine two-point correlation functions at fixed graph distance:

\begin{align}
    C_{z=2}^{\mathrm{conn.}}(r) &= \frac{1}{N_r^{(z=2)}} \sum_{\substack{q<q',\;q,q'\in V_{z=2}:\\ d(q,q') = r}}\left( \left\langle Z_q Z_{q'}\right\rangle - \left\langle Z_q\right\rangle \left\langle Z_{q'}\right\rangle\right),
\label{eq:corr_conn}
\end{align}
where $V_{z=2}$ is the set of coordination-two qubits and $N_r^{(z=2)}$ is the number of pairs in this set separated by distance $r$.
Because we restrict the finite-size-scaling analysis to coordination-two qubits ($z=2$), it is natural to define a system length scale from the averaged correlation values. We identify the characteristic correlation length as the scale over which $C_{z=2}^{\mathrm{conn.}}(r)$ decays with distance:
\begin{align}
   C_{z=2}^{\mathrm{conn.}}(r) \sim C_0^{\mathrm{conn.}} e^{-r/\xi'},
\end{align}
where $C_0^{\mathrm{conn.}}$ is a heuristic correlation amplitude and $\xi'$ is the characteristic correlation decay length.

\begin{figure}
    \centering
    \includegraphics[width=\linewidth]{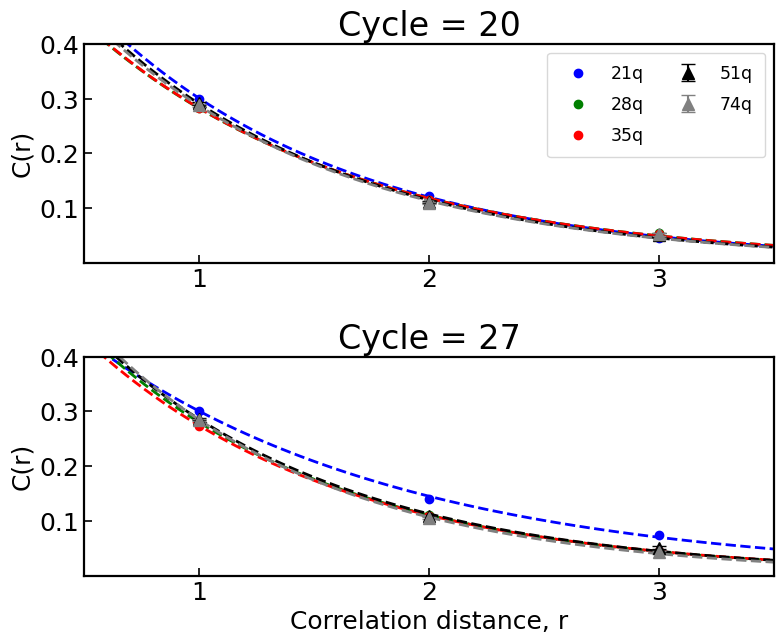}
    \caption{Two-point connected correlation $C_{z=2}^{\mathrm{conn.}}(r)$ [Eq.~\eqref{eq:corr_conn}] for different system sizes. Blue, green, and red circles: values from exact statevector simulations for system sizes 21, 28, and 35 qubits, respectively. Black and gray triangles: values (and their error bars) from IBM Heron experiments for system sizes 51 and 74 qubits, respectively. Each coloured dashed line is the best exponential fit derived from its corresponding data set (${\{r=1,2,3\}}$). Upper panel: Floquet cycle 20. Lower panel: Floquet cycle 27.}
    \label{fig:correlation_fit}
\end{figure}

Figure~\ref{fig:correlation_fit} shows the correlation function at distances $r\in\{1,2,3\}$ for different system sizes. For each system and Floquet cycle, we fit an exponential curve through each triplet of points and extract the estimated correlation length $\xi'$. Across different systems and cycles, $\xi'$ remains near an average value of $\sim 1.1$, with variations of a few percent between Floquet cycles. Thus, the correlation length remains approximately constant. The estimate of $\xi'$ is consistent with the finite-size-scaling length obtained in Sec.~\ref{sec.results} of the main text.

\section{QESEM-Extrapolated methodology}
\label{app.methods}

\subsection{Overview}

As the depth of the Floquet circuits increases, the active volume and accumulated infidelity grow with the number of gates that can affect the target observables. Consequently, the QPU time required to reach a fixed statistical uncertainty with \textit{QESEM-Unbiased} grows rapidly. For circuits containing more than 16 Floquet cycles, this cost became prohibitive (see Fig.~\ref{fig:fig1b}). We therefore adopted the more resource-efficient ZNE-based protocol described below.

Zero-noise extrapolation (ZNE) estimates a zero-noise expectation value from measurements obtained at systematically amplified noise strengths \cite{temme2017qem,li2017zne,Cai2023}. For each target observable, the expectation values measured at the native and amplified noise levels are fitted to a chosen response model, which is then evaluated at zero noise. The noise-scaling procedure and the extrapolation ansatz jointly determine the bias and sampling cost of the estimate \cite{giurgicatiron2020digitalzne,mari2021noisescaling}.

\paragraph{Exponential ansatz.}
For a Pauli observable $O=P$, we assume that its expectation value decays exponentially with the noise-amplification factor $f$,
\begin{equation}
\label{ZNE_ANSATZ}
    \langle O\rangle_f
    =
    \langle O\rangle_{\text{ideal}}\,e^{-\gamma_O f},
\end{equation}
where $\gamma_O$ is an observable-dependent effective decay parameter (also known as effective infidelity volume). This ansatz is motivated by the exponential decay of Pauli expectation values under Pauli noise and by established exponential and multi-exponential ZNE models \cite{Endo2018,Cai2021}. Because Eq.~\eqref{ZNE_ANSATZ} contains only two independent parameters, two distinct nonzero noise levels suffice to determine the extrapolated value.

For the results presented in the main text, we used the native noise level, $f=1$, and the amplified level, $f=2$. Their corresponding expectation values are denoted by $\langle O\rangle_{\mathrm{noisy}}$ and $\langle O\rangle_{\mathrm{amp}}$, respectively. The derivations that follow are valid for any noise amplification level $f>1$.

\subsection{Compilation}
\label{app.compilation}
To minimize the native circuit noise, we compiled the Floquet dynamics into native $R_{ZZ}(\alpha)$ gates and single-qubit gates. For the results presented in the main text, we used $\alpha=1.047$.

We tailored the noise of the $R_{ZZ}(\alpha)$ gates using randomized compiling \cite{wallman2016randomizedcompiling,QESEM}. Specifically, each $R_{ZZ}(\alpha)$ gate was conjugated by a Pauli operator sampled uniformly from the subgroup that commutes with $R_{ZZ}(\alpha)$,
\begin{equation}
    R_{ZZ}(\alpha)
    \longmapsto
    P\,R_{ZZ}(\alpha)\,P^{\dagger}
    =
    R_{ZZ}(\alpha),
\end{equation}
where
\begin{equation}
    P\in\mathcal{T}_{ZZ}
    \equiv
    \{II,IZ,ZI,ZZ,XX,XY,YX,YY\}.
\end{equation}
We refer to this procedure as \emph{partial Pauli twirling}. Because an $R_{ZZ}$ over-rotation commutes with the ideal gate, partial twirling does not remove this error component. We therefore calibrated the over-rotation using dedicated circuits before each run.

We also twirled the measurement operation. In 50\% of the circuits, we applied an $X$ gate to every measured qubit immediately before measurement and flipped the recorded outcomes, $0\leftrightarrow1$, in post-processing. This transformation leaves the ideal measurement statistics unchanged while symmetrizing the single-qubit assignment errors. In particular, for each qubit it enforces
\begin{equation}
    P_{01}=P_{10}\equiv P_M,
    \qquad
    P_{00}=P_{11}=1-P_M,
\end{equation}
where $P_{ij}$ is the conditional probability of reporting outcome $j$ when state $i$ was prepared, with $i,j\in\{0,1\}$. Thus, neglecting interqubit correlations, the measurement noise on each qubit is described by a single bit-flip probability. We denote this qubit-dependent probability by $P_M^i$ below.

Finally, we compressed each layer of algorithmic and compilation-induced single-qubit gates into the fixed-depth form
\begin{equation}
    G
    =
    R_Z(\phi_3)\,\mathrm{SX}\,R_Z(\phi_2)\,\mathrm{SX}\,R_Z(\phi_1),
\end{equation}
up to a global phase, where $\mathrm{SX}\equiv\sqrt{X}$. Because the $R_Z$ rotations are virtual, this representation keeps the effective single-qubit-gate noise model consistent between the characterization and application circuits. Further details of the compilation, characterization, and calibration procedures are given in Ref.~\cite{QESEM}.

\subsection{Noise amplification via PEA}
\label{appendix:PEA}
A central requirement of ZNE is the ability to construct an ensemble of circuits whose averaged expectation values for an observable reproduce the target circuit at a prescribed noise-amplification factor $f$ \cite{giurgicatiron2020digitalzne,mari2021noisescaling}. We use a characterization-based approach: hardware gate errors are first represented by local Pauli noise models, and those models are then used to sample a distribution of amplified-noise circuits. The characterization protocol is described in detail in Ref.~\cite{QESEM}. This noise-scaling strategy is commonly called probabilistic error amplification (PEA) \cite{mari2021noisescaling,kim2023utility,Ferracin2024}.

We model a noisy gate layer by a during-gate Pauli--Lindblad channel,
\begin{equation}
\label{during_noise_model}
    \widetilde{\mathcal{G}}
    =
    \exp\!\left[-i\,\operatorname{ad}_{h}+\mathcal{L}\right],
    \qquad
    \operatorname{ad}_{h}(\rho)=[h,\rho],
\end{equation}
with
\begin{equation}
\label{pauli_lindblad_generator}
    \mathcal{L}(\rho)
    =
    \sum_{P\in\mathcal{P}_n\setminus\{I\}}
    \lambda_P\bigl(P\rho P^{\dagger}-\rho\bigr),
    \qquad
    \lambda_P\ge 0.
\end{equation}
Here $h$ generates the ideal gate, $\mathcal{G}_{\mathrm{ideal}}=\exp[-i\,\operatorname{ad}_{h}]$, and $\mathcal{P}_n$ denotes the $n$-qubit Pauli group modulo phases. We use $\lambda_P$ for the Pauli-generator rates to distinguish them from the ZNE decay parameter $\gamma_O$.

For a Clifford gate that can be twirled by the full Pauli group without changing its ideal action, the noise can be represented as a Pauli channel applied after the ideal gate,
\begin{equation}
    \widetilde{\mathcal{G}}
    =
    \Lambda\circ\mathcal{G}_{\mathrm{ideal}},
    \qquad
    \Lambda=e^{\mathcal{L}},
\end{equation}
where
\begin{equation}
    \Lambda(\rho)
    =
    \sum_{P\in\mathcal{P}_n}
    \epsilon_P\,P\rho P^{\dagger},
    \qquad
    \epsilon_P\ge 0,
    \qquad
    \sum_P\epsilon_P=1.
\end{equation}
Amplifying the noise by a factor $f$ amounts to composing the native channel with the additional channel
\begin{equation}
    \Lambda^{f-1}
    =
    e^{(f-1)\mathcal{L}},
    \qquad
    \Lambda_f=e^{f\mathcal{L}}=\Lambda^{f-1}\circ\Lambda.
\end{equation}
For the factorized Pauli--Lindblad model in Eq.~\eqref{pauli_lindblad_generator}, this additional channel can be sampled by independently inserting each Pauli generator $P$ with probability
\begin{equation}
\label{pea_injection_probability}
    q_P(f)
    =
    \frac{1-e^{-2(f-1)\lambda_P}}{2}
    =
    (f-1)\lambda_P+O(\lambda_P^2).
\end{equation}
Thus, the familiar linear injection rule is the weak-noise limit of the exact generator-scaling construction \cite{vandenberg2023pec,kim2023utility}.

For fractional entangling gates, full Pauli twirling would modify the intended gate. Partial twirling therefore leaves a during-gate model of the form in Eq.~\eqref{during_noise_model}, which cannot in general be factored into an ideal fractional gate followed by a Pauli channel. We instead construct a multi-type circuit distribution whose elements include Pauli layers before or after the entangling gate, as well as elements in which the entangling operation is replaced by an available Pauli-layer operation. The distribution coefficients are chosen so that the ensemble-averaged channel targets $\exp[-i\operatorname{ad}_{h}+f\mathcal{L}]$. The full construction is given in Ref.~\cite{QESEM}; related theory for non-Clifford gates is developed in Ref.~\cite{Layden2024ErrorMitigationNonClifford}.

\subsection{SPAM mitigation}
In addition to gate errors, state-preparation and measurement (SPAM) errors contribute to the observed circuit noise. These errors are characterized together with the gate errors as part of our characterization protocol \cite{QESEM}.

Neglecting correlations between qubits, we characterize the measurement error on qubit $i$ by the symmetrized bit-flip probability $P_M^i$ introduced in Sec.~\ref{app.compilation}. For a $Z$-type Pauli observable $O=\prod_{i\in\mathcal{I}}Z_i$, the expectation value measured at any noise-amplification factor $f$ is corrected in post-processing as \cite{VanDenBerg2022}
\begin{equation}
\label{measurement-mitigation}
    \widehat{\langle O\rangle}_{\text{measurement mitigated},f}
    =
    \left[
        \prod_{i\in\mathcal{I}}
        \frac{1}{1-2P_M^i}
    \right]
    \widehat{\langle O\rangle}_{f}.
\end{equation}
Here and below, measurement mitigation is applied before the ZNE analysis; for brevity, we omit the ``measurement mitigated'' qualifier from subsequent expressions.

We model state-preparation errors as independent bit flips with probabilities $P_{\text{SP}}^i$ on the active qubits. Unlike measurement errors, state-preparation errors cannot generally be removed in post-processing because they propagate through the circuit and can affect observables on other qubits. For each active qubit $i$, the corresponding $X_i$ channel is
\begin{equation}
\label{state-preparation-channel}
    \Lambda_{\text{SP}}^{(i)}(\rho)
    =
    (1-P_{\text{SP}}^i)\rho
    +P_{\text{SP}}^i X_i\rho X_i,
\end{equation}
and the full state-preparation channel is the tensor product of these independent single-qubit channels.

We amplify state-preparation noise together with the gate noise. For circuit ensembles at amplification factor $f$, we insert an additional $X_i$ at the beginning of the circuit using the leading-order probability 
\begin{align}
\label{state-preparation-amplification}
    q^i_{X}(f) &= \frac{1-\left(1-2P_{\text{SP}}^i\right)^{f-1}}{2}\\
    &=
    (f-1)P_{\text{SP}}^i
    +O\!\left(\left(P_{\text{SP}}^i\right)^2\right),    
\end{align}
which is the state-preparation analogue of Eq.~\eqref{pea_injection_probability}. State-preparation errors are then mitigated, like gate errors, by extrapolating to zero noise.

\subsection{Extrapolation}

Let
\begin{equation}
    \widehat{o}_1
    \equiv
    \widehat{\langle O\rangle}_{\mathrm{noisy}},
    \qquad
    \widehat{o}_f
    \equiv
    \widehat{\langle O\rangle}_{\mathrm{amp}}
\end{equation}
be the estimated expectation values at noise levels $1$ and $f$, with variances $\sigma_1^2$ and $\sigma_f^2$, respectively. The total variance of an observable is composed out of the total shot-to-shot variance and the circuit-to-circuit variance. The latter comes from the distribution of circuits that are sampled using the PEA approach. When $\widehat{o}_1$ and $\widehat{o}_f$ are nonzero and have the same sign, Eq.~\eqref{ZNE_ANSATZ} gives the sign-preserving estimator
\begin{equation}
\label{ZNE-OBS-EV}
    \widehat{\langle O\rangle}_{\mathrm{mitigated}}
    \equiv
    \widehat{o}_0
    =
    \operatorname{sgn}(\widehat{o}_1)
    \exp\!\left[
        \frac{f\ln|\widehat{o}_1|-\ln|\widehat{o}_f|}{f-1}
    \right].
\end{equation}
For positive expectation values, this is equivalent to
$\widehat{o}_0=(\widehat{o}_1^f/\widehat{o}_f)^{1/(f-1)}$; for $f=2$, it reduces to $\widehat{o}_0=\widehat{o}_1^2/\widehat{o}_2$. The inferred decay parameter is
\begin{equation}
\label{ZNE-GAMMA-EST}
    \widehat{\gamma}_O
    =
    \frac{1}{f-1}
    \ln\!\left|\frac{\widehat{o}_1}{\widehat{o}_f}\right|.
\end{equation}

At fixed calibration parameters, the native- and amplified-noise ensembles are sampled independently, so their sampling covariance vanishes. A first-order delta-method expansion therefore gives
\begin{equation}
\label{ZNE-OBS-VAR}
    \operatorname{Var}(\widehat{o}_0)
    \simeq
    \frac{\widehat{o}_0^2}{(f-1)^2}
    \left[
        \frac{f^2\sigma_1^2}{\widehat{o}_1^2}
        +\frac{\sigma_f^2}{\widehat{o}_f^2}
    \right].
\end{equation}
Substituting Eq.~\eqref{ZNE_ANSATZ} then yields the equivalent form
\begin{equation}
    \operatorname{Var}(\widehat{o}_0)
    \simeq
    \frac{1}{(f-1)^2}
    \left[
        f^2 e^{2\widehat{\gamma}_O}\sigma_1^2
        +e^{2f\widehat{\gamma}_O}\sigma_f^2
    \right].
\end{equation}

The exponential estimator becomes ill-conditioned when either input distribution has appreciable support near zero or when the native- and amplified-noise estimates have different signs. In those cases, the logarithm and the inferred decay rate are unstable, and we do not apply Eq.~\eqref{ZNE-OBS-EV}. The observables reported in the main text were restricted to the stable regime in which both estimates have the same sign and negligible statistical overlap with zero.

\subsection{Bootstrapping composite observables}

Consider a composite observable that can be written as a weighted sum of Pauli operators, for example the average magnetization,
\begin{equation}
    O=\sum_{P\in\mathcal{P}}\alpha_P P.
\end{equation}
Because the exponential ansatz is formulated for individual Pauli expectation values, we extrapolate each Pauli term separately and then combine the mitigated estimates,
\begin{equation}
\label{multi-pauli-ev}
    \widehat{\langle O\rangle}_{\mathrm{mitigated}}
    =
    \sum_{P\in\mathcal{P}}
    \alpha_P\,
    \widehat{\langle P\rangle}_{\mathrm{mitigated}}.
\end{equation}
The variance cannot, in general, be obtained by summing only the individual Pauli variances. Defining $\widehat{p}_{0}\equiv\widehat{\langle P\rangle}_{\mathrm{mitigated}}$ and $\widehat{q}_{0}\equiv\widehat{\langle Q\rangle}_{\mathrm{mitigated}}$, the full expression is
\begin{align}
\label{multi-pauli-var}
    \operatorname{Var}\!\left(
        \widehat{\langle O\rangle}_{\mathrm{mitigated}}
    \right)
    &={}
    \sum_{P\in\mathcal{P}}
    \alpha_P^2\operatorname{Var}\!\left(\widehat{p}_{0}\right)
    \nonumber\\[-2pt]
    &\quad+
    \sum_{\substack{P,Q\in\mathcal{P}\\P\neq Q}}
    \alpha_P\alpha_Q
    \operatorname{Cov}\!\left(
        \widehat{p}_{0},
        \widehat{q}_{0}
    \right).
\end{align}
The covariance terms are non-negligible because Pauli observables extracted from the same circuit records are statistically correlated.

We estimate the joint uncertainty using the nonparametric bootstrap \cite{Efron1979Bootstrap}. Let $\mathcal{D}_1$ and $\mathcal{D}_f$ contain $N_1$ native-noise and $N_f$ amplified-noise circuit records, respectively. For each bootstrap replicate, we sample $N_1$ records with replacement from $\mathcal{D}_1$ and $N_f$ records with replacement from $\mathcal{D}_f$. Resampling entire records preserves the empirical correlations among all Pauli observables extracted from a given record. We then apply drift mitigation, extrapolate every Pauli term using Eq.~\eqref{ZNE-OBS-EV}, and combine the terms using Eq.~\eqref{multi-pauli-ev}.

Repeating this procedure $B=2000$ times gives bootstrap replicates of the mitigated estimator, $\widehat{o}_{0}^{\star(b)}$, with $b=1,\ldots,B$. We estimate the mitigated expectation value as
\begin{equation}
\label{bootstrap-mean}
    \widehat{\langle O\rangle}_{\mathrm{mitigated}}
    \equiv
    \widehat{o}_{0}
    =
    \frac{1}{B}
    \sum_{b=1}^{B}
    \widehat{o}_{0}^{\star(b)},
\end{equation}
and its variance as
\begin{equation}
\label{bootstrap-var}
    \operatorname{Var}\!\left(\widehat{o}_{0}\right)
    =
    \frac{1}{B-1}
    \sum_{b=1}^{B}
    \left(
        \widehat{o}_{0}^{\star(b)}
        -\widehat{o}_{0}
    \right)^2.
\end{equation}
Algorithm~\ref{alg:zne_bootstrap} summarizes the procedure.

\begin{algorithm}[t]
\caption{Bootstrap uncertainty estimation for a composite observable}
\label{alg:zne_bootstrap}
\footnotesize
\setlength{\abovedisplayskip}{3pt}
\setlength{\belowdisplayskip}{3pt}
\begin{algorithmic}[1]

\Require Native-noise records $\mathcal{D}_1$ of size $N_1$ and amplified-noise records $\mathcal{D}_f$ of size $N_f$.
\Require Amplification factor $f>1$ and composite observable $O=\sum_{P\in\mathcal{P}}\alpha_P P$.
\Require Number of bootstrap replicates $B$ and drift-mitigation map $\mathcal{M}_{\mathrm{drift}}$.

\Ensure Bootstrap estimate and variance of $\langle O\rangle_{\mathrm{mitigated}}$.

\For{$b=1,\ldots,B$}
    \State Draw $\mathcal{D}_1^{\star(b)}$ by sampling $N_1$ records with replacement from $\mathcal{D}_1$.
    \State Draw $\mathcal{D}_f^{\star(b)}$ by sampling $N_f$ records with replacement from $\mathcal{D}_f$.

    \For{$P\in\mathcal{P}$}
        \State Estimate $\widehat{p}_1^{\star(b)}\equiv\widehat{\langle P\rangle}_1^{\star(b)}$ from $\mathcal{D}_1^{\star(b)}$ and $\widehat{p}_f^{\star(b)}\equiv\widehat{\langle P\rangle}_f^{\star(b)}$ from $\mathcal{D}_f^{\star(b)}$.
        \State Apply drift mitigation:
        \[
            \bigl(
                \widetilde{p}_1^{\star(b)},
                \widetilde{p}_f^{\star(b)}
            \bigr)
            \gets
            \mathcal{M}_{\mathrm{drift}}
            \bigl(
                \widehat{p}_1^{\star(b)},
                \widehat{p}_f^{\star(b)}
            \bigr).
        \]
        \State Extrapolate $\widehat{p}_0^{\star(b)}$ from $\widetilde{p}_1^{\star(b)}$ and $\widetilde{p}_f^{\star(b)}$ using Eq.~\eqref{ZNE-OBS-EV}.
    \EndFor

    \State Compute
    $\displaystyle
        \widehat{o}_{0}^{\star(b)}
        =
        \sum_{P\in\mathcal{P}}
        \alpha_P\widehat{p}_0^{\star(b)}$.
\EndFor

\State Compute the bootstrap mean and variance using Eqs.~\eqref{bootstrap-mean} and~\eqref{bootstrap-var}.
\State \Return $\widehat{o}_{0}$ and $\operatorname{Var}(\widehat{o}_{0})$.

\end{algorithmic}
\normalsize
\end{algorithm}

\subsection{Drift mitigation}
\label{App:drift_mitigation}
The drift-mitigation procedure follows the adaptive, drift-robust execution described in Ref.~\cite{QESEM,qedma_drift_robust_2025}. The QESEM execution is divided into batches, each containing mitigation and characterization circuits. The characterization data from batch $i$ are used to construct the noise-amplification distribution for batch $i+1$; batch $0$ contains only characterization and native-noise circuits. Mitigation is performed separately for each batch. Let $\widehat{o}_{0}^{(i)}$ be the mitigated estimate from batch $i$, with variance $\sigma_i^2$. Assuming that the batch estimates are statistically independent, the reported estimate is the inverse-variance-weighted mean
\begin{equation}
\label{batch-weighted-mean}
    \widehat{o}_{0}
    =
    \frac{\sum_i \widehat{o}_{0}^{(i)}/\sigma_i^2}
         {\sum_i 1/\sigma_i^2},
    \qquad
    \operatorname{Var}(\widehat{o}_{0})
    =
    \left(\sum_i\frac{1}{\sigma_i^2}\right)^{-1},
\end{equation}
where the sum is performed over all batches that executed the required noisy and amplification circuits.

Because amplified circuits are executed after their characterization circuits, the QPU noise may drift between model construction and circuit execution. The characterized model used to sample the amplification circuits may therefore differ from the native noise present when those circuits are run.

Following Eq.~\eqref{ZNE_ANSATZ}, let $\gamma_O^{(i)}$ and $\gamma_O^{(i+1)}$ denote the effective native-noise decay parameters for an arbitrary Pauli observable $O$ during batches $i$ and $i+1$, respectively. Amplification circuits targeting factor $f$ are sampled using the model from batch $i$ and executed under the native noise of batch $i+1$. Their expectation value is therefore
\begin{equation}
\label{drifted-amplified-expectation}
    \langle O\rangle_{f}^{(i+1)}
    =
    \langle O\rangle_{\text{ideal}}
    \exp\!\left[-(f-1)\gamma_O^{(i)}-\gamma_O^{(i+1)}\right].
\end{equation}

We define the backward drift factor as the ratio of the native-noise expectation values measured in the two batches,
\begin{equation}
\label{eq:drift_factor}
    \operatorname{DF}_O^{(i)}
    \equiv
    \frac{\langle O\rangle_{1}^{(i)}}
         {\langle O\rangle_{1}^{(i+1)}}
    =
    \exp\!\left[-\gamma_O^{(i)}+\gamma_O^{(i+1)}\right].
\end{equation}
Multiplying the amplified expectation value by this factor maps it back to the noise model characterized in batch $i$,
\begin{align}
\langle O\rangle_{f}^{\mathrm{drift\text{-}mitigated},(i)}
&=
\operatorname{DF}_O^{(i)}
\,\langle O\rangle_{f}^{(i+1)}
\nonumber\\
&=
\langle O\rangle_{\mathrm{ideal}}
\,e^{-f\gamma_O^{(i)}}.
\end{align}
The drift correction is applied only when the native-noise reference estimates from batches $i$ and $i+1$ have the same sign and negligible statistical overlap with zero, matching the stability conditions used for ZNE.

\section{QESEM validation}
\label{app.validation}

\subsection{Characterization validation}

We first tested the predictive accuracy of the characterization protocol using the benchmarking procedure introduced in Ref.~\cite{QESEM}.
We considered a smaller, 28-qubit pretzel-shaped graph embedded in \texttt{ibm\_boston}. On this subsystem, we executed the characterization protocol and inferred a hardware noise model.
Our characterization procedure is based on fitting a multilocal Pauli model for the common $R_{ZZ}$ layers that compose each Floquet cycle. A wide variety of Clifford circuits are executed on the QPU, each designed to estimate different Pauli noise generators by selectively amplifying one generator at a time.

\begin{figure}[tb]
\centering
\subfloat[]{
    \includegraphics[width=0.87\columnwidth]{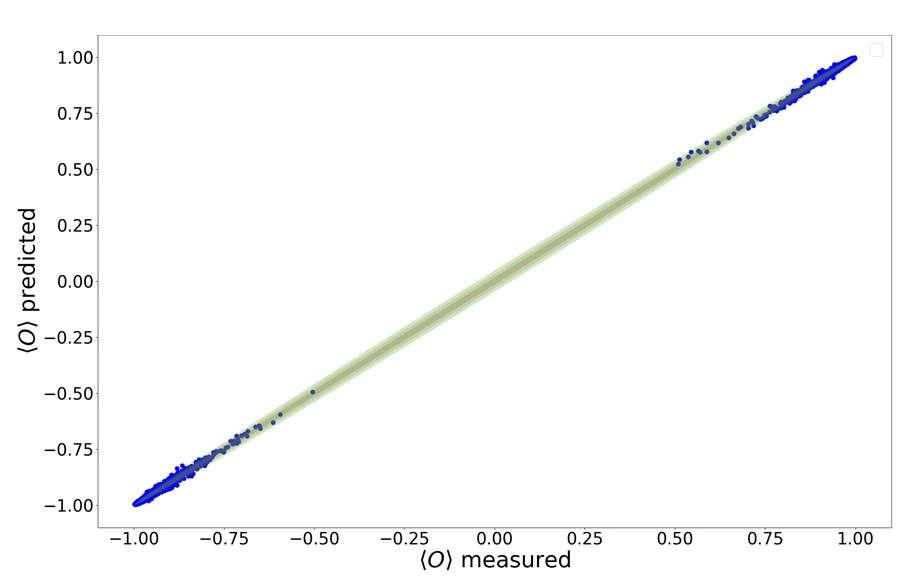}
    \label{fig:pacc_obs_28q}
}\\
\subfloat[]{
    \includegraphics[width=0.87\columnwidth]{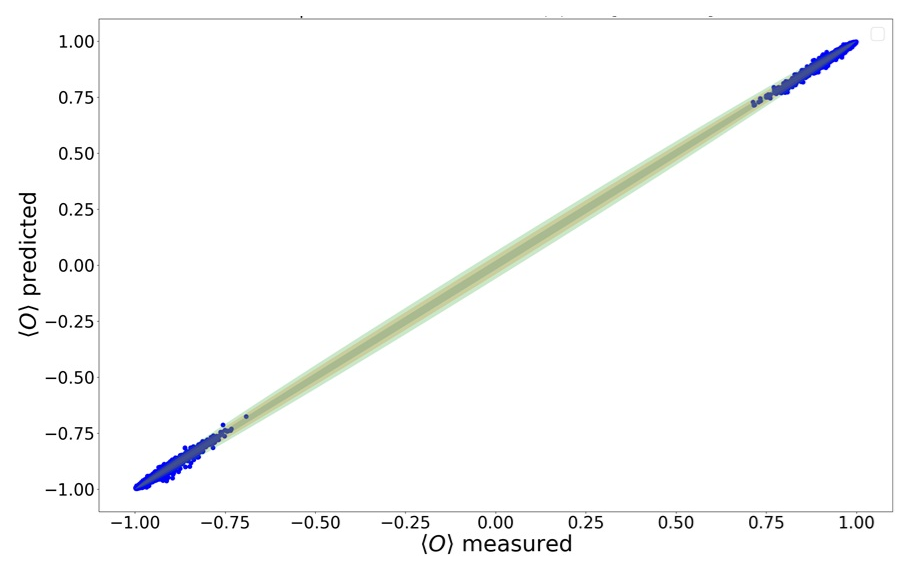}
    \label{fig:pacc_obs_51q}
}\\
\caption{Validation of the characterization model through measured-versus-predicted Pauli-observable expectation values. Each point corresponds to one Pauli observable, and together the points include all one- and two-local Pauli observables in the validation set. The measured expectation value is shown on the horizontal axis and the model-predicted expectation value on the vertical axis. The shaded bands indicate one-, two-, and three-standard-deviation regions around perfect agreement, computed from the shot-based uncertainties of the measured observables and excluding twirl variance; the outer, three-standard-deviation band is the most visible. Panel \textbf{(a)} shows the $28$-qubit characterization used for validation in this section, and panel \textbf{(b)} shows an example characterization from one of the $51$-qubit experiments used for the main results.}
\label{fig:obs-obs}
\end{figure}

In this manner, we gather a set of Pauli observable estimates. After fitting the QPU noise model, we obtain a mean value for each of the noise generators. By classically simulating the noisy QPU, we calculate predictions for these observables for each Clifford circuit that was executed on the QPU. We aggregate the experimental results and classical predictions in Fig.~\ref{fig:obs-obs}. The figure shows a scatter plot of the measured observable expectation values against the corresponding model predictions. Indeed, the set of observables aligns closely along the identity line $\langle O \rangle_{\text{pred}} =\langle O \rangle_{\text{measured}}$. We show the results from two selected experiments, one on a 28-qubit system and another on a 51-qubit system. The characterization performance is reproduced across different computational batches and extends well to a larger 74-qubit system. The statistical spread of the observables around the identity line agrees closely with the expected normal distribution: $68\%$, $92\%$, and $99\%$ of the observables lie within 1, 2, and 3 standard deviations of the line, respectively. 

\begin{figure}
    \centering
    \includegraphics[width=\linewidth]{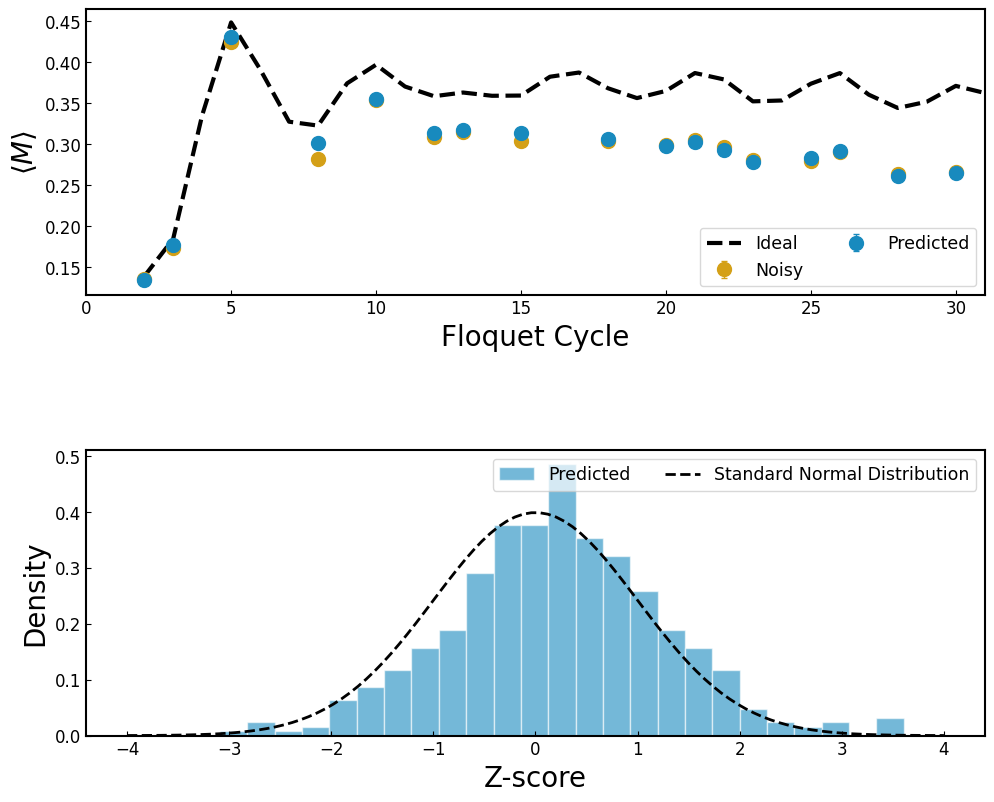}
    \caption{Characterization benchmark on the 28-qubit pretzel-shaped subsystem. Upper panel: predicted and experimentally measured noisy total magnetization as a function of the Floquet cycle. Lower panel: standardized residuals [Eq.~\eqref{eq:characterization-zscore}] for all single-qubit Pauli-$Z$ expectation values from Floquet cycles 1--30. The black curve denotes the standard normal density and serves as a visual calibration reference.}
    \label{fig:28q_pred}
\end{figure}

We later used the inferred 28-qubit model to simulate noisy Floquet circuits containing up to 30 cycles and compared the predicted expectation values with measurements from the same circuits executed on the QPU. The upper panel of Fig.~\ref{fig:28q_pred} compares the predicted and measured noisy total magnetization as a function of the Floquet cycle. The two curves agree closely over the full range of circuit depths.

As a more granular diagnostic, we computed a standardized residual (Z score) for each single-qubit Pauli-$Z$ observable at every collected Floquet cycle. For qubit $j$ and cycle $c$, we define
\begin{equation}
\label{eq:characterization-zscore}
    z_{j,c}^{\mathrm{pred}}
    =
    \frac{
        \widehat{\langle Z_j\rangle}_{\mathrm{pred},c}
        -
        \widehat{\langle Z_j\rangle}_{\mathrm{QPU},c}
    }{
        \sqrt{
            \operatorname{Var}\!\left(
                \widehat{\langle Z_j\rangle}_{\mathrm{pred},c}
            \right)
            +
            \operatorname{Var}\!\left(
                \widehat{\langle Z_j\rangle}_{\mathrm{QPU},c}
            \right)
        }
    }.
\end{equation}
The lower panel of Fig.~\ref{fig:28q_pred} shows that the resulting distribution is well-approximated by the standard normal distribution. Because residuals from different qubits and cycles can be correlated, we used this comparison as a calibration diagnostic rather than as an independent-sample goodness-of-fit test.

\subsection{QESEM-Extrapolated benchmark on a smaller system}

We next repeated the complete mitigation procedure described in Appendix~\ref{app.methods} on the same 28-qubit subsystem. At this size, the ideal expectation values can be computed exactly with a statevector simulator, providing a direct end-to-end validation of the mitigation protocol.

We considered two validation settings:
\begin{enumerate}
    \item \textit{Simulation-based mitigation.} We first characterized the physical QPU and inferred its noise model. We then simulated the circuit at the native noise level, $f=1$, and with all modeled noise-generator rates scaled to $f=2$. The two simulated data sets were combined using the same exponential extrapolation as in the experiment [Eq.~\eqref{ZNE-OBS-EV}].

    \item \textit{Full experimental mitigation.} We executed the native circuits on the QPU to obtain the $f=1$ expectation values. We also executed circuits sampled from the PEA distribution to obtain the $f=2$ expectation values. The mitigated estimates were then extrapolated using the same procedure as in the main experiment.
\end{enumerate}

\begin{figure}
    \centering
    \includegraphics[width=\linewidth]{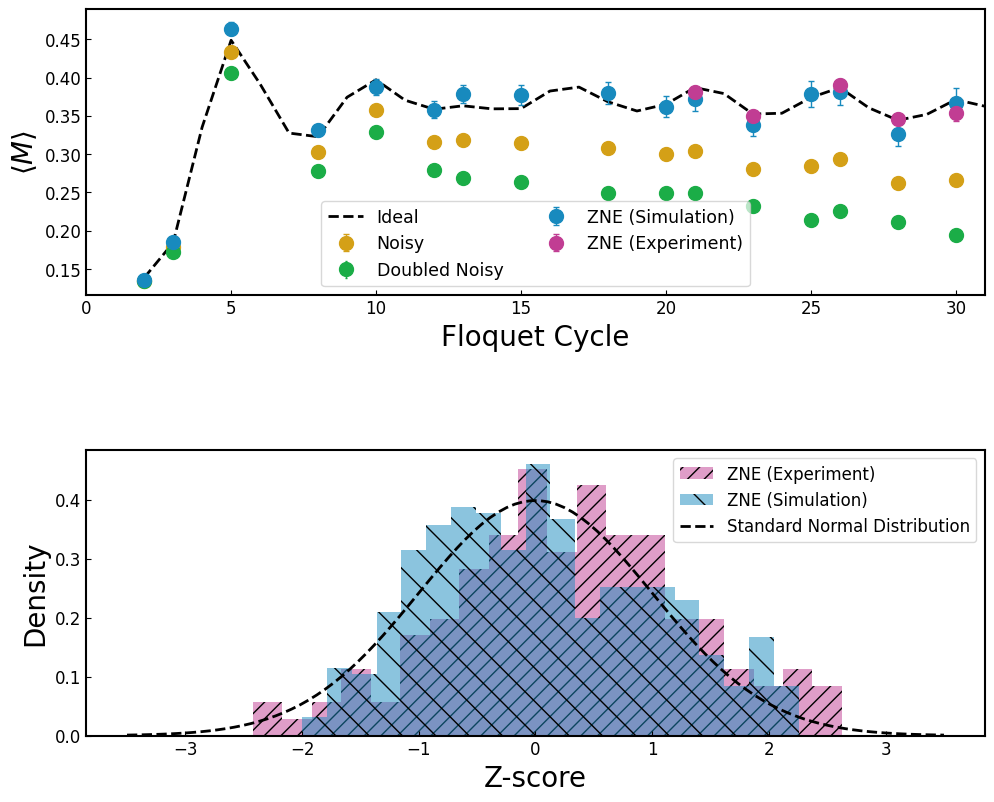}
    \caption{Mitigation benchmark on the 28-qubit pretzel-shaped subsystem. Upper panel: mitigated magnetization obtained from noisy simulation and from the full QPU experiment, compared with the exact statevector result. Lower panel: standardized mitigation residuals [Eq.~\eqref{eq:mitigation-zscore}] for all single-qubit Pauli-$Z$ expectation values and measured Floquet cycles. The black curve denotes the standard normal density and serves as a visual calibration reference.}
    \label{fig:28q_miti}
\end{figure}

The upper panel of Fig.~\ref{fig:28q_miti} presents the mitigated magnetization. In both validation settings, the mitigated estimates closely track the exact statevector values and substantially improve on the unmitigated results.

We also evaluated a standardized mitigation residual for each single-qubit Pauli-$Z$ expectation value at every measured Floquet cycle,
\begin{equation}
\label{eq:mitigation-zscore}
    z_{j,c}^{\mathrm{mit}}
    =
    \frac{
        \widehat{\langle Z_j\rangle}_{\mathrm{mitigated},c}
        -
        \langle Z_j\rangle_{\mathrm{ideal},c}
    }{
        \sqrt{
            \operatorname{Var}\!\left(
                \widehat{\langle Z_j\rangle}_{\mathrm{mitigated},c}
            \right)
        }
    }.
\end{equation}
The exact statevector values have no sampling uncertainty and therefore do not contribute to the denominator. The lower panel of Fig.~\ref{fig:28q_miti} shows distributions centered near zero with widths close to one for both the simulation-based and experimental validations. As above, correlations among observables reduce the effective number of independent samples, so the comparison with a standard normal distribution should be interpreted as a calibration diagnostic. Within this benchmark, the results are consistent with residual systematic errors smaller than the reported statistical uncertainty, whose typical scale is a few times $10^{-3}$.

\subsection{QESEM-Extrapolated benchmark for early cycles}

We further benchmarked the mitigation protocol on the full 51-qubit system at circuit depths that remain accessible to accurate classical simulation. Figure~\ref{fig:magnetization} compares the mitigated expectation values with results obtained using numerical simulation methods. Projected entangled-pair-state simulations with belief propagation (PEPS-BP) \cite{alkabetz2021tensor,tindall2023gauging} at bond dimension $D=700$ are converged through Floquet cycle 9 (Appendix~\ref{app.peps_bp}) and are considered accurate in this range. Indeed, we observe that the total magnetization values obtained with both QESEM-Unbiased and QESEM-Extrapolated agree closely with the PEPS-BP result.

\begin{figure}
    \centering
    \includegraphics[width=\linewidth]{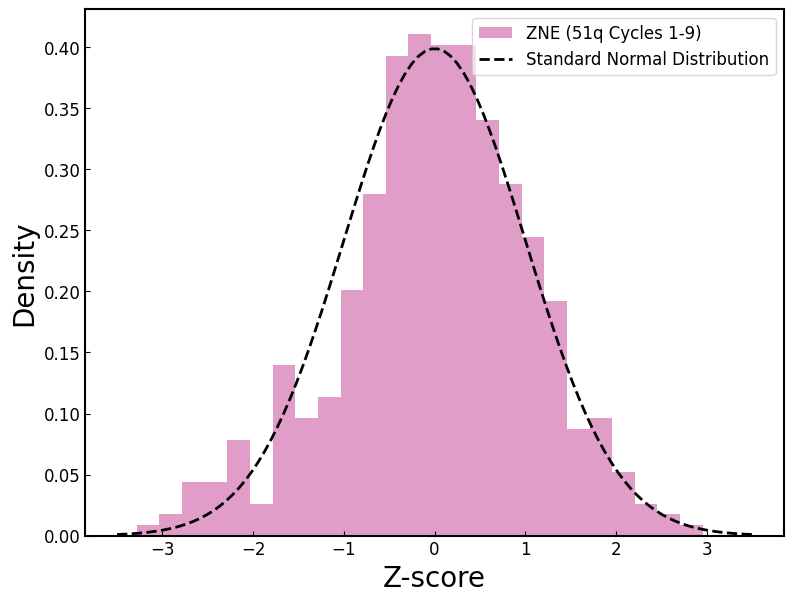}
    \caption{Z-score distribution for all single-qubit Pauli-$Z$ expectation values from Floquet cycles 1--9 in the 51-qubit experiment. The Z scores are computed relative to the converged PEPS-BP results obtained with bond dimension $D=700$ [Eq.~\eqref{eq:peps-bp-zscore}].}
    \label{fig:51q_z_score}
\end{figure}

At the level of individual qubits, we computed a Z score for every mitigated single-qubit Pauli-$Z$ observable from cycles 1--9. Neglecting the residual PEPS-BP numerical uncertainty relative to the experimental uncertainty, we define
\begin{equation}
\label{eq:peps-bp-zscore}
    z_{j,c}^{\mathrm{PEPS}}
    =
    \frac{
        \widehat{\langle Z_j\rangle}_{\mathrm{mitigated},c}
        -
        \langle Z_j\rangle_{\mathrm{PEPS\text{-}BP},c}
    }{
        \sqrt{
            \operatorname{Var}\!\left(
                \widehat{\langle Z_j\rangle}_{\mathrm{mitigated},c}
            \right)
        }
    }.
\end{equation}
Figure~\ref{fig:51q_z_score} shows the resulting Z-score distribution, which is close to the standard normal reference. Because observables extracted from the same execution records are correlated, the histogram is again used as a calibration diagnostic rather than as a collection of independent tests.

\subsection{Extrapolation ansatz validation}

We next tested the adequacy of the exponential ZNE ansatz for the Floquet circuits. For the data presented in the main text, the native-noise point, $f=1$, and the amplified point, $f=2$, were used to extrapolate the zero-noise expectation values. To validate the fitted response away from these calibration points, we additionally measured the circuits at amplification factors $f=1.5$ and $f=2.5$. We denote the calibration and validation sets by
\begin{equation}
    \mathcal{F}_{\mathrm{cal}}=\{1,2\},
    \qquad
    \mathcal{F}_{\mathrm{val}}=\{1.5,2.5\}.
\end{equation}

For each single-qubit Pauli-$Z$ observable $O$, we fitted the exponential model to the calibration points and evaluated the validation statistic
\begin{equation}
\label{eq:validation-statistic-T_o}
    T_O
    =
    \sum_{f\in\mathcal{F}_{\mathrm{val}}}
    \left(
        \frac{
            y_{\mathrm{model},f}-\widehat{o}_f
        }{
            \sigma_f
        }
    \right)^2,
\end{equation}
where
\begin{equation}
    y_{\mathrm{model},f}
    =
    \widehat{o}_0 e^{-\widehat{\gamma}_O f}
\end{equation}
is Eq.~\eqref{ZNE_ANSATZ} evaluated at $f$ using the parameters fitted to the calibration data, and $\sigma_f$ is the standard error of $\widehat{o}_f$.

We assessed the statistical consistency of the observed $T_O$ values using a parametric bootstrap \cite{Efron1979Bootstrap}, summarized in Algorithm~\ref{alg:bootstrap_exponential}. The bootstrap compares each experimental statistic with statistics generated under the fitted exponential model while propagating the uncertainty in the two calibration points.

In the validation experiment, we tested Floquet cycles 20 and 30 with three independent execution batches. For each Floquet cycle, the 51 single-qubit Pauli-$Z$ observables across the three batches yielded $51\times 3=153$ bootstrap $p$-values. Results for different qubits within the same batch are correlated.

\begin{algorithm}[t]
\caption{Parametric-bootstrap validation of a two-point exponential fit}
\label{alg:bootstrap_exponential}
\footnotesize
\setlength{\abovedisplayskip}{3pt}
\setlength{\belowdisplayskip}{3pt}
\begin{algorithmic}[1]
\Require Calibration and validation sets $\mathcal{F}_{\mathrm{cal}}$ and $\mathcal{F}_{\mathrm{val}}$.
\Require Data $\{(f,\widehat{o}_f,\sigma_f)\}_{f\in\mathcal{F}_{\mathrm{cal}}\cup\mathcal{F}_{\mathrm{val}}}$.
\Require Number of bootstrap replicates $B$.
\Ensure Bootstrap $p$-value for the exponential ZNE model.

\State Fit the calibration data by weighted least squares to the exponential model in Eq.~\eqref{ZNE_ANSATZ}.
\State Evaluate $y_{\mathrm{model},f}$ for all $f\in\mathcal{F}_{\mathrm{cal}}\cup\mathcal{F}_{\mathrm{val}}$.
\State Compute the observed statistic $T_O$ using Eq.~\eqref{eq:validation-statistic-T_o}.

\For{$k=1,\ldots,B$}
    \State Generate independent bootstrap calibration data for $f\in\mathcal{F}_{\mathrm{cal}}$:
    \[
        o_f^{\star(k)}
        \sim
        \mathcal{N}\!\left(y_{\mathrm{model},f},\sigma_f^2\right).
    \]

    \State Refit the exponential model to the bootstrap calibration data.
    \State Denote the fitted parameters by $\widehat{o}_0^{\star(k)}$ and $\widehat{\gamma}_O^{\star(k)}$.

    \State For $f\in\mathcal{F}_{\mathrm{val}}$, evaluate
    \[
        y_{\mathrm{model},f}^{\star(k)}
        =
        \widehat{o}_0^{\star(k)}
        e^{-\widehat{\gamma}_O^{\star(k)}f}.
    \]

    \State Generate independent bootstrap validation data for $f\in\mathcal{F}_{\mathrm{val}}$:
    \[
        o_f^{\star(k)}
        \sim
        \mathcal{N}\!\left(y_{\mathrm{model},f},\sigma_f^2\right).
    \]

    \State Compute
    \[
        T_O^{\star(k)}
        =
        \sum_{f\in\mathcal{F}_{\mathrm{val}}}
        \left(
            \frac{
                y_{\mathrm{model},f}^{\star(k)}-o_f^{\star(k)}
            }{
                \sigma_f
            }
        \right)^2.
    \]
\EndFor

\State Compute the Monte Carlo bootstrap $p$-value
\[
    p_{\mathrm{boot}}
    =
\frac{1+\sum_{k=1}^{B}\mathbf{1}\!\left(T_O^{\star(k)}\ge T_O\right)}{B+1}.
\]
\State \Return $p_{\mathrm{boot}}$.
\end{algorithmic}
\normalsize
\end{algorithm}

\begin{figure}[tb]
    \centering
    \includegraphics[width=\linewidth]{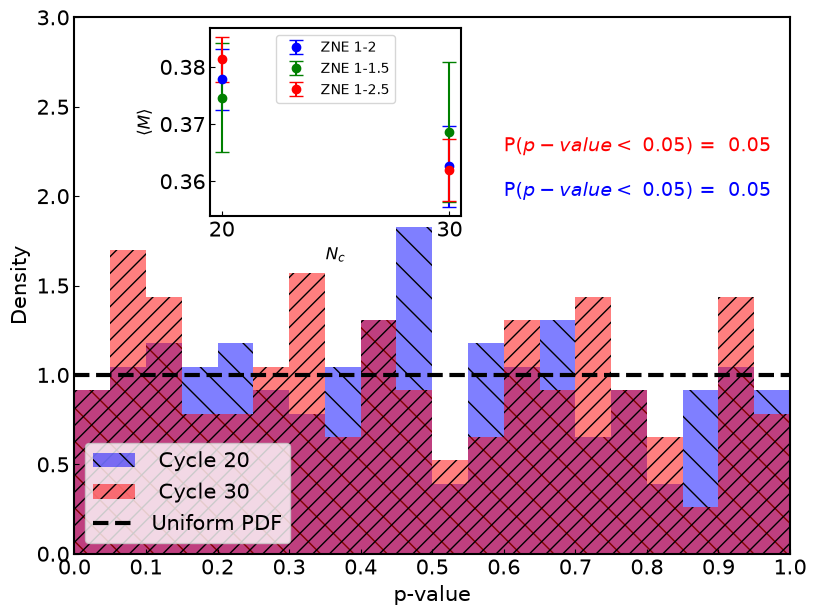}
    \caption{Validation of the exponential ZNE ansatz. Main panel: histograms of bootstrap $p$-values for the 51 single-qubit Pauli-$Z$ observables in each of three execution batches, shown separately for Floquet cycles 20 (blue) and 30 (red). Under a calibrated null model, the marginal $p$-value distribution is expected to be approximately uniform on $[0,1]$; the dashed horizontal line shows the corresponding density. Inset: mitigated total magnetization $\langle M\rangle$ for Floquet cycles 20 and 30, obtained from the amplification-factor pairs $\{1,2\}$ (blue), $\{1,1.5\}$ (green), and $\{1,2.5\}$ (red). All three estimates agree within statistical uncertainty.}
    \label{fig:zne_p_value_pfg}
\end{figure}

Figure~\ref{fig:zne_p_value_pfg} shows the $p$-value distributions across all observables and batches for Floquet cycles 20 and 30. The distributions are approximately uniform over $[0,1]$. At the nominal 5\% significance threshold, approximately 5\% of the observable--batch tests have $p<0.05$, and no observable yields $p<0.05$ in more than one batch. Because tests within a batch are correlated, these percentages should not be interpreted as outcomes of independent trials. Within the statistical resolution of this validation experiment, we find no evidence of a systematic departure from the exponential ansatz. The inset of Fig.~\ref{fig:zne_p_value_pfg} compares the total-magnetization estimates obtained from the amplification-factor pairs $\{1,2\}$, $\{1,1.5\}$, and $\{1,2.5\}$. For both Floquet cycles 20 and 30, all three extrapolations agree within statistical uncertainty.

\section{Floquet mixed-field Ising model parameter scan}
\label{app.scan}

\begin{figure}
    \centering
    \includegraphics[scale=0.25]{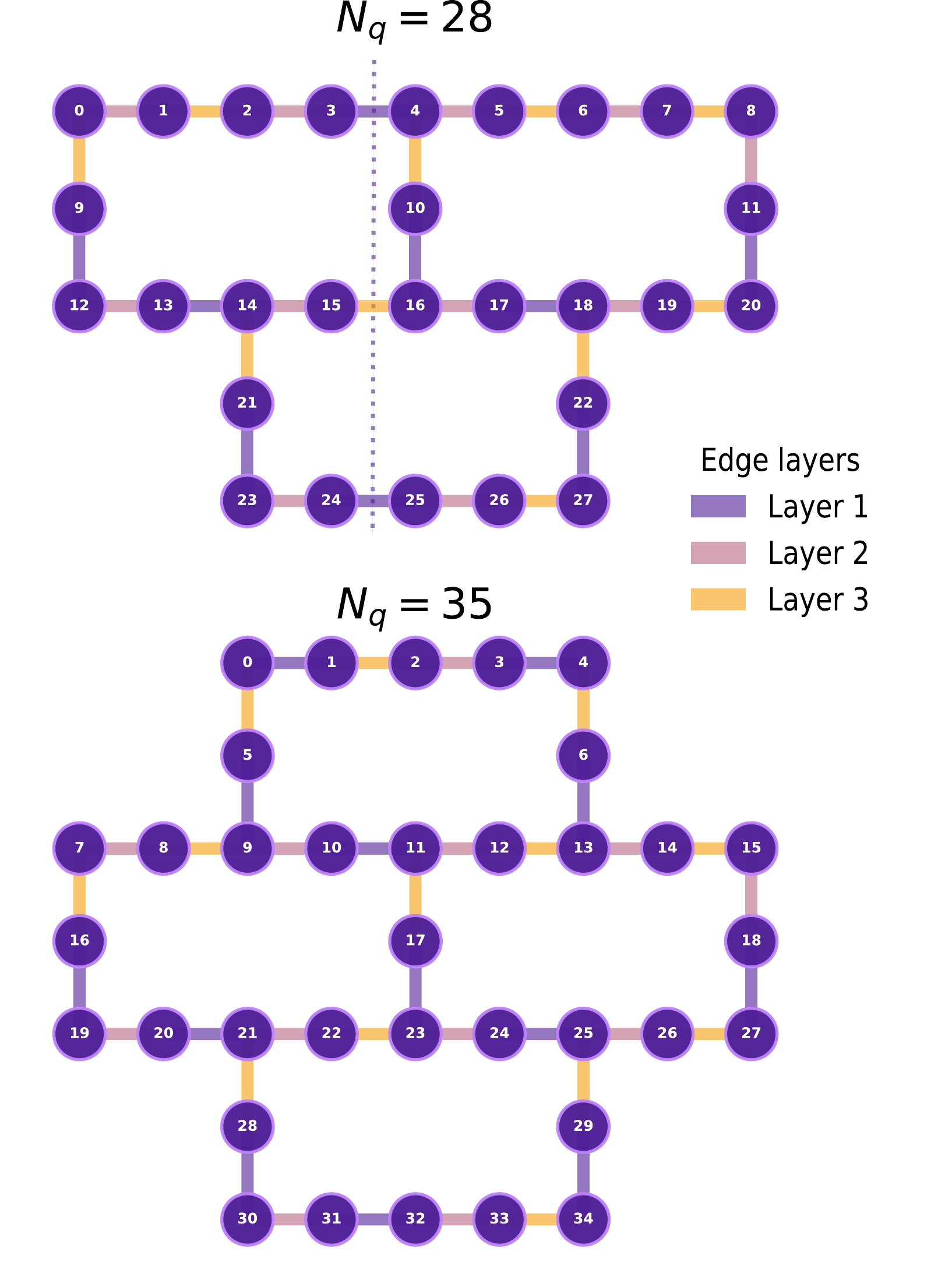}
    \caption{
    Small heavy-hex lattice systems used in the exact simulations supporting the Floquet parameter scan and finite-size comparison. The figure shows the 28- and 35-qubit systems, with different edge colours denoting the three edge layers $E_1$, $E_2$, and $E_3$ that define the Floquet circuit [Eq.~\ref{eq:unitary}]. The 28-qubit system consists of three heavy-hex loops: two upper loops side by side and a third loop below them. The two upper loops, corresponding to qubits \(0\)--\(20\), form the 21-qubit system used for the parameter-space exploration in Sec.~\ref{sec.setup}. The indicated cut defines the bipartition used to compute the state entanglement entropy for both the 21- and 28-qubit systems. The 35-qubit system is used as an additional exactly simulated size for comparison with the error-mitigated large-system dynamics.
    }
    \label{fig:small_systems}
\end{figure}

This appendix provides additional details on the small-system exact numerical studies underlying the parameter-space exploration discussed in Sec.~\ref{sec.setup}. The scans in Figs.~\ref{fig:param_scan_circ}d and \ref{fig:param_scan_circ}f use two complementary diagnostics of quantum complexity: the bipartite state entanglement entropy and the bipartite operator entanglement entropy.

For the state entanglement, we compute the von Neumann entropy directly from the evolved statevector,
\begin{equation}
    S_A=-\mathrm{Tr}\left(\rho_A\log_2\rho_A\right),
\end{equation}
where $\rho_A=\mathrm{Tr}_{\bar A}\left(|\psi\rangle\langle\psi|\right)$ is the reduced density matrix of subsystem $A$. The bipartitions used for the 21- and 28-qubit systems are shown in Fig.~\ref{fig:small_systems}. The 21-qubit patch is obtained from the 28-qubit patch by retaining qubits $0$--$20$, corresponding to the two
upper heavy-hex loops.

Operator complexity in the Heisenberg picture is quantified in Sec.~\ref{sec.setup} through the bipartite operator entanglement of an evolved single-qubit $Z$ operator. Specifically, we evolve the $Z$ operator
initially supported on the central qubit of the 21-qubit heavy-hex lattice, qubit 10 in Fig.~\ref{fig:small_systems},
for $N_c=7$ Floquet cycles using a PEPO representation; details of the PEPO evolution are given in App.~\ref{app.pepo_bp}.

\begin{figure}[htbp]
    \centering
    \includegraphics[
        width=0.8\linewidth,
        trim=3cm 1cm 13cm 0cm, 
        clip
    ]{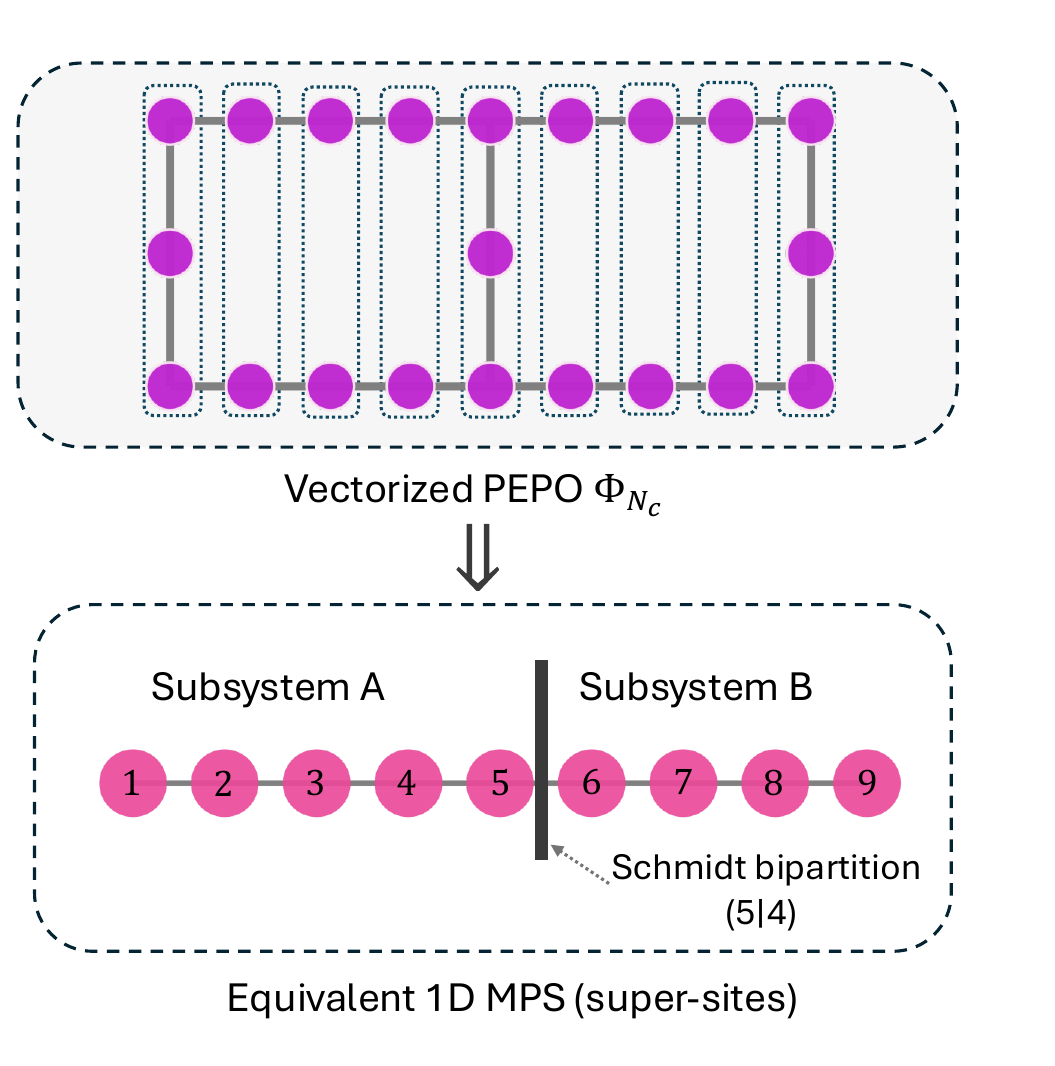}
    \caption{
    Mapping of the vectorized evolved PEPO to an equivalent one-dimensional MPS for evaluating the bipartite operator entanglement reported in Sec.~\ref{sec.setup}. The evolved operator $\Phi_{N_c}$, represented as a vectorized PEPO on the 21-qubit heavy-hex lattice, is mapped to a one-dimensional MPS by grouping the PEPO tensors into nine column-wise super-sites. The operator entanglement is obtained from the Schmidt spectrum across the indicated \(5|4\) super-site bipartition.
    }
    \label{fig:operator_ent}
\end{figure}

To evaluate the operator entanglement, the PEPO is represented in Liouville (vectorized) form and mapped to an equivalent one-dimensional MPS by grouping neighbouring PEPO tensors into nine super-sites following the lattice geometry. The resulting MPS is brought into mixed canonical form, and the Schmidt spectrum across the $5|4$ super-site bipartition shown in Fig.~\ref{fig:operator_ent}, is obtained from a singular value decomposition. The operator entanglement entropy is then  computed as
\begin{equation}
    S_{\mathrm{op}} = -\sum_\mu s_\mu^2 \log_2 s_\mu^2.
\end{equation}
where $\{s_\mu\}$ are the normalized Schmidt coefficients, 

\section{Finite-size scaling of the oscillation amplitude}
\label{app.scaling}

We extract the late-cycle oscillation amplitude from the coordination-two magnetization, $M_{z=2}$, shown in Fig.~\ref{fig:fit_to_decay_sin}. For each system size, we fit the late-cycle dynamics to the decaying-background-plus-oscillation form defined in Eq.~\eqref{eq:fit_sin}. The extracted amplitude $A$, period $T$, background parameters $C,\gamma$, and phase $\phi$ are summarized in Table~\ref{tab:finite_size_fits}.

The finite-size comparison uses five heavy-hex geometries. The $21$-, $28$-, and $35$-qubit amplitudes are extracted from exact statevector simulation, while the $51$- and $74$-qubit amplitudes are obtained from error-mitigated experiments. The small geometries are shown in Fig.~\ref{fig:small_systems}: the $21$-qubit system is formed by the two upper heavy-hex loops of the $28$-qubit systems, corresponding to qubits \(0\)--\(20\); the $28$-qubit system adds a third loop below them; and the $35$-qubit system is obtained by closing an additional loop above. The larger, $51$- and $74$-qubit, geometries are shown in Fig.~\ref{fig:ibm-boston}. Together, these systems preserve the local heavy-hex connectivity and form an approximately nested sequence, but their aspect ratios are not fixed. The scaling analysis below should therefore be interpreted as a phenomenological finite-size comparison rather than a controlled thermodynamic extrapolation.

\begin{table}[h!]
\centering
\caption{Fitted parameters of the late-cycle coordination-two magnetization,
$M_{z=2}$, for the system sizes studied in this work. The fits use
Eq.~\eqref{eq:fit_sin}. Uncertainties denote one standard error from
the fit. \label{tab:finite_size_fits}
}
{
\begin{tabular}{| c || c | c | c | c | c |}
\hline
$N_q$ & $C$ & $\gamma~[10^{-3}]$  & $A~[10^{-2}]$ & $T$ & $\phi$ \\
\hline
\hline
$21$ & $0.35 \pm 0.02$ & $0 \pm 2$ & $3.9\pm0.5$ & $4.9\pm0.1$ & $2.6\pm0.7$ \\
\hline
$28$ & $0.38 \pm 0.01$ & $2.8 \pm 0.5$ & $2.1\pm0.1$ & $4.55\pm0.04$ & $2.0\pm0.3$ \\
\hline
$35$ & $0.37 \pm 0.01$ & $2.4\pm0.6$ & $1.3\pm0.1$ & $4.85\pm0.08$ & $4.0\pm0.5$ \\
\hline
$51$ & $0.40 \pm 0.01$ & $5.1\pm 0.6$ & $1.3\pm0.1$ & $4.61 \pm 0.08$ & $2.4\pm0.6$ \\
\hline
$74$ & $0.40 \pm 0.01$ & $4.8\pm0.7$ & $0.8\pm0.1$ & $4.8 \pm 0.1$& $4.2\pm0.7$ \\
\hline
\end{tabular}
}
\end{table}

\begin{figure}[tb]
\centering
\subfloat[]{
    \includegraphics[width=0.87\columnwidth]{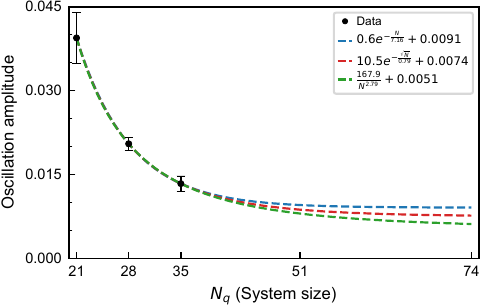}
    \label{fig:small_offset}
}\\
\subfloat[]{
    \includegraphics[width=0.87\columnwidth]{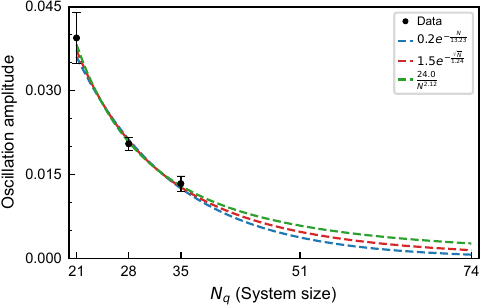}
    \label{fig:small_no_offset}
}\\
\subfloat[]{
    \includegraphics[width=0.87\columnwidth]{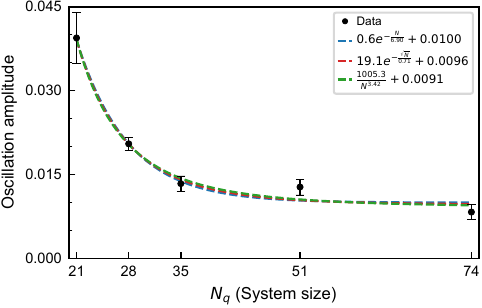}
    \label{fig:all_offset}
}
\caption{
    \textbf{\protect\subref{fig:small_offset}} Fits using only the exactly simulated 21-, 28-, and 35-qubit systems. All three phenomenological ansätze, Eq.\eqref{eq:fitting_funcs}, fit the small-system data but give different extrapolations to larger sizes.
    \textbf{\protect\subref{fig:small_no_offset}} Fits to the same small-system data with the offset fixed to $c=0$. The small-system data are compatible with both finite-offset and zero-offset forms, but the different ansätze extrapolate differently to larger sizes.
    \textbf{\protect\subref{fig:all_offset}} Fits including the error-mitigated 51- and 74-qubit data. The additional points constrain the large-size trend. The $\sqrt{N_q}$-exponential fit gives a finite offset $c=9.6\times10^{-3}$, while the algebraic fit requires a steep power $\alpha=3.42$, and the $N_q$-exponential fit gives a decay scale much larger than the independently extracted correlation length (App.~\ref{app.correlation}).
}
\end{figure}
\label{fig:finite_size_appendix}

We next fit the extracted amplitudes as a function of system size. Because the available heavy-hex patches have different aspect ratios, the effective finite-size scaling variable is not obvious from geometry alone. We therefore compare three phenomenological forms,
\begin{align}
\label{eq:fitting_funcs}
    A_1(N_q) &= a \exp\!\left(-\frac{N_q}{\xi}\right) + c, \\
    A_2(N_q) &= a \exp\!\left(-\frac{\sqrt{N_q}}{\xi}\right) + c, \\
    A_3(N_q) &= \frac{a}{N_q^\alpha} + c.
\end{align}
Here $c$ is a possible asymptotic offset. The form $A_2$ uses $\sqrt{N_q}$ as a proxy for linear system size, as would be natural for an approximately two-dimensional geometry. The form $A_1$ instead uses $N_q$ itself as the scaling variable, corresponding to a more rapid exponential dependence on system size and providing a useful comparison given the elongated finite geometries studied here. Finally, $A_3$ provides an algebraic alternative over the limited range of available sizes.

Figures \ref{fig:small_offset} and \ref{fig:small_no_offset} show fits obtained using only the exactly simulated 21-, 28-, and 35-qubit systems, with and without an offset. With only three amplitudes, the small-system data do not distinguish among the scaling ansätze, nor do they determine whether an offset is required. Both finite-offset and zero-offset fits describe the exact small-system amplitudes but give different extrapolations to 51 and 74 qubits. Thus, the exact small-system data alone do not determine the large-size trend.

Figure~\ref{fig:all_offset} repeats the finite-offset fits after including the error-mitigated 51- and 74-qubit amplitudes. The additional points strongly constrain the extrapolation. The $\sqrt{N_q}$-exponential fit gives
\begin{align}
    A(N_q) = 19.1\, e^{-\sqrt{N_q}/0.71} + 9.6\times10^{-3},
\end{align} 
with an offset resolved from zero at the $8.7\sigma$ level ($\Delta c \approx 10^{-3}$), as reported in the main text. 

The alternative fits clarify why we use the $\sqrt{N_q}$-exponential form as the primary finite-size fit. The algebraic fit with offset can be made over this finite range, but including the 51- and 74-qubit points drives the exponent to $\alpha=3.42$. Such a steep algebraic decay is difficult to regard as a natural finite-size trend for the approximately two-dimensional geometries studied here. The $N_q$-exponential fit also describes the amplitudes visually, but gives a decay scale $\xi_{N}\simeq 6.9$, an order of magnitude larger than the independently extracted correlation length from connected equal-time correlations (App.~\ref{app.correlation}). By contrast, the $\sqrt{N_q}$-exponential fit gives $\xi_{\sqrt{N}}\simeq0.71$, consistent with that independent correlation-length estimate.

\begin{figure}[tb]
\centering

\subfloat[]{
    \includegraphics[width=0.87\columnwidth]{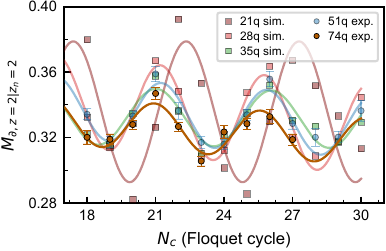}
    \label{fig:sin_fits_boundary_new}
}\\
\subfloat[]{
    \includegraphics[width=0.87\columnwidth]{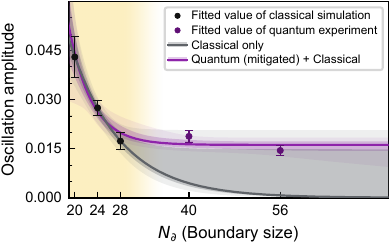}
    \label{fig:finite_size_boundary}
}
\caption{
    \textbf{\protect\subref{fig:sin_fits_boundary_new}} Boundary observable $M_{\partial,z=2|z_n=2}$, defined in
Eq.~\eqref{eq:boundary_z_2_zn_2_obs}, as a function of the Floquet cycle
$N_c \in [18,30]$, for exact $21$-, $28$-, and $35$-qubit simulations and
the error-mitigated $51$- and $74$-qubit experiments. Solid lines are fits to
Eq.~\eqref{eq:fit_sin}, a cosine oscillation on a linearly slowly decaying
background, using the same fitting model as in Fig.~\ref{fig:fit_to_decay_sin}.
    \textbf{\protect\subref{fig:finite_size_boundary}}  Oscillation amplitude of $M_{\partial,z=2|z_n=2}$ as a
function of the number of boundary qubits, $N_\partial$. Black markers
denote amplitudes extracted from classical simulations; purple markers
correspond to the error-mitigated quantum experiments. Solid curves are
exponential fits of the form
$A(N_\partial) = a\,e^{-N_\partial/\xi} + c$: the gray curve uses only
the classical data, the purple curve a combined fit to classical and
quantum data. The dark (light) gray shaded region indicates the $68\%$ ($95\%$) confidence band, obtained by parametric Monte Carlo
propagation, through resampling the fitted oscillation amplitudes within their
uncertainties and refitting. The classical-only extrapolation is
compatible with both vanishing and finite asymptotic offsets, whereas
the combined fit favors a finite and comparatively large offset.
}
\label{fig:finite_size_boundary_appendix}

\end{figure}

We next consider a boundary-restricted magnetization observable closely related to $M_{z=2}$. Specifically, we define
\begin{equation}
M_{\partial,z=2|z_n=2}=
\frac{1}{N_{\partial,z=2|z_n=2}}\sum_{\substack{i\in\partial\ z_i=2,; z_j=2\ \forall j\in n(i)}} Z_i,
\label{eq:boundary_z_2_zn_2_obs}
\end{equation}
where the sum runs over boundary qubits with coordination number $z_i=2$, restricted to those whose neighbouring qubits also have coordination number $z=2$. Here, $N_{\partial,z=2|z_n=2}$ denotes the number of qubits that satisfy these conditions.
We repeat the finite-size scaling analysis performed for $M_{z=2}$. Since this observable is defined on the boundary, the system size is naturally characterized by the number of boundary qubits, $N_{\partial}$, rather than the total number of qubits. Accordingly, we fit the oscillation amplitude to the exponential model

\begin{equation}
A(N_{\partial}) = a e^{-N_{\partial}/\xi}+c,
\label{eq:fitting_model_boundary}
\end{equation}

instead of the square-root scaling considered in the main analysis.
\par
Figure~\ref{fig:finite_size_boundary_appendix} shows the fitting results. The uncertainty of the finite-size extrapolation is estimated using parametric Monte Carlo uncertainty propagation. Specifically, for each realization, the oscillation amplitudes for each system size are independently resampled from Gaussian distributions whose means and standard deviations are given by the amplitudes and their corresponding uncertainties obtained from the oscillation fits. The resampled data set is then refitted to the exponential model, Eq.~(\ref{eq:fitting_model_boundary}). Repetition of this procedure many times results in an ensemble of extrapolated curves, from which we estimate the central $68\%$ interval ($16$th--$84$th percentiles), which defines the shaded dark gray uncertainty band in Fig.~\ref{fig:finite_size_boundary} and also in Fig.~\ref{fig:finite_size_scaling} of the main text. As can be seen, using only the classical simulation data, the uncertainty grows rapidly beyond the small system sizes available, leaving the asymptotic offset $c$ poorly constrained and consistent with both $c=0$ and a finite positive value. Including the error-mitigated quantum data, as seen before, substantially reduces the uncertainty of the extrapolation and favors a finite offset within this exponential model.

\section{Comparison to other mitigation methods}
\label{app.fail_em}
In this appendix we review the results obtained by applying naive error mitigation methods in the circuits studied in the main text, namely the mixed-field Floquet Ising magnet of Eq.~\eqref{eq:unitary}, and compare to QESEM. 

\subsection{Naive unitary folding}
\label{app.fail_em.unitary_folding}
Here we show that zero-noise extrapolation based on naive unitary folding does not reproduce the expected mitigated values obtained using QESEM.

In unitary folding, the standard approach exploits the ideal identities $(CZ)^2=I$ and $(CZ)^{2n+1}=CZ$. Under a sparse Pauli-Lindblad noise model, however, the noise associated with the $CZ$ gate accumulates as $n$ is increased. Thus, replacing each instance of a $CZ$ layer by $2n+1$ repetitions yields a circuit with the same ideal output but with amplified noise. The circuits executed in this work, however, use fractional-angle $R_{ZZ}$ gates. Each such gate can be transpiled into two $CZ$ gates, but doing so would incur a multiplicative infidelity cost for the circuits.

We therefore instead fold the fractional-angle $R_{ZZ}$ layers directly, using a construction that preserves the native $R_{ZZ}(\alpha)$ gate rather than replacing it by an inverse-angle gate $R_{ZZ}(-\alpha)$. This distinction is important because the noise associated with $R_{ZZ}(-\alpha)$ need not be the same as that associated with $R_{ZZ}(\alpha)$. By contrast, inserting single-qubit gates around additional $R_{ZZ}(\alpha)$ gates can flip the sign of the ideal interaction while retaining the same native two-qubit gate implementation. The added single-qubit gates are absorbed into the surrounding single-qubit $U_3$ compressions, as in the original circuit, so the folded circuit preserves the relevant native-layer structure while amplifying the $R_{ZZ}$ noise.

We implement the $R_{ZZ}$ gate folding, as
\begin{align}
\label{eq:rzz_folding}
    R_{ZZ}(\alpha) \rightarrow \left[R_{ZZ}(\alpha) (XI) R_{ZZ}(\alpha) (XI)\right]^n R_{ZZ}(\alpha),
\end{align}
where $XI$ denotes a Pauli-$X$ gate acting on the first qubit and the $R_{ZZ}$ angle $\alpha$ is left implicit. Conjugating an $R_{ZZ}$ gate by $XI$ flips the sign of $\alpha$, so the two $R_{ZZ}$ gates inside the bracket cancel ideally. At the same time, the Pauli-generator rates of the during-layer sparse Pauli-Lindblad noise associated with the $R_{ZZ}$ gates approximately combine additively, producing the desired noise amplification.

Together with readout-error mitigation, this procedure allows one to execute circuits at different amplification factors and extrapolate the measured observables to zero noise. Even amplification factors, and more generally fractional amplification factors, can be implemented by randomizing the folding procedure, but this increases the variance and the sampling cost required to obtain high-precision results. We therefore restrict the analysis here to amplification factors $1$ ($n=0$) and $3$ ($n=1$).

\begin{figure}[tb]
\centering
    \includegraphics[width=0.95\linewidth]{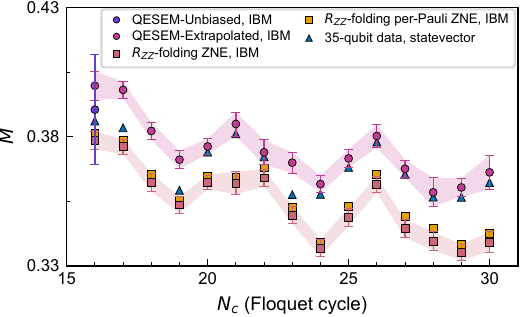}
    \caption{Comparison of QESEM with naive unitary-folding ZNE for the $51$-qubit Floquet magnet. The plot shows the magnetization for Floquet cycles $16$--$30$, comparing QESEM-Unbiased and QESEM-Extrapolated estimates with two ZNE procedures based on direct folding of the fractional-angle $R_{ZZ}$ layers. The exact 35-qubit statevector result is shown for reference. Both folding-based extrapolations remain systematically below the QESEM-Extrapolated results and further from the smaller-system exact dynamics, indicating that naive unitary folding does not provide a reliable noise-scaling procedure for these IBM Heron circuits.}
\label{fig:unitary_folding}
\end{figure}

Figure~\ref{fig:unitary_folding} compares the magnetization obtained from QESEM (-Unbiased and -Extrapolated) with two ZNE estimates based on the $R_{ZZ}$-layer folding procedure described above. In the first approach, the full magnetization is extrapolated directly. In the second, each single-qubit Pauli contribution to the magnetization is extrapolated separately and the results are then averaged. The latter procedure is closer in spirit to the QESEM-Extrapolated treatment of composite observables, but we do not assign reliable error bars to it because the correlations between the separately extrapolated Pauli estimates are not fully accounted for. Both folding-based estimates remain systematically below the QESEM-Extrapolated dynamics over Floquet cycles $16$--$30$ and lie further away from the exact 35-qubit statevector dynamics shown for reference. They therefore do not faithfully reproduce the expected mitigated dynamics.

\section{Quantinuum experimental details}
\label{app.quantinuum}

\begin{figure}
    \centering
    \includegraphics[width=\linewidth]{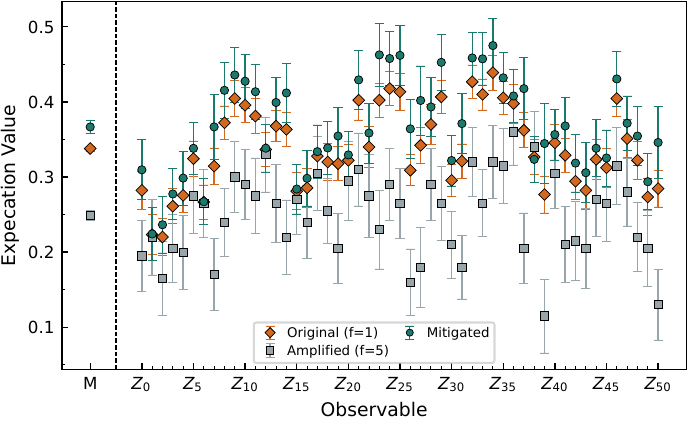}
    \caption{The per-Pauli extrapolation results for the 19-Floquet cycle circuit, illustrating the extrapolation procedure before the Pauli terms are combined into the final composite observables.}
    \label{fig:quantinuum_per_pauli_step_19}
\end{figure}

Floquet cycles 16, 17, and 19 of the 51-qubit system were also executed on Quantinuum's Helios and H2 devices. The number of shots used and obtained mitigation precision for each step are detailed in Tab.~\ref{tab:quantinuum_execution_details}.

\begin{table}[h]
\centering
\caption{Experimental details on Quantinuum devices.}
\label{tab:quantinuum_execution_details}
\begin{tabular}{c c c c}
\hline
Floquet cycles & QPU & Shots executed & Mitigation precision \\
\hline                              
16 & H2 & 1600 & $9\cdot 10^{-3}$ \\
17 & H2 & 1600 & $9\cdot 10^{-3}$ \\
19 & Helios & 2200 & $8\cdot 10^{-3}$ \\
\hline
\end{tabular}
\end{table}

In the mitigation experiments on Quantinuum's devices, several methods were employed to strengthen the reliability and accuracy of the results. First, coherent errors in the two-qubit gate-layers were suppressed by the application of pseudo twirling~\cite{Santos2024}. We suppressed the over-rotation error of single-qubit gates by introducing, with probability $p_{\text{shift}}=0.5$ a phase shift of $\pi$~\cite{haghshenas2026digitalmagnetism}. Additional coherent errors during idle periods were suppressed by Quantinuum's built-in dynamical decoupling framework.

To mitigate the remaining errors, we used a simplified variant of QESEM on Quantinuum devices, which will be described in full in a forthcoming paper~\cite{qedmaForthcoming}. The simplified version we employed in this work reduces to zero-noise extrapolation~\cite{temme2017qem,li2017zne} with unitary folding~\cite{giurgicatiron2020digitalzne}. Crucially, we folded the full Floquet cycle (Fig.~\ref{fig:floquet_cycle}), replacing each such cycle with a noise-amplified implementation that preserves the target unitary, $U_F$, while increasing the executed circuit depth. Doing so amplifies not only the gate-layer errors but also errors associated with qubit shuttling. Indeed, when $U_F$ is folded as a unit, the circuit structure and shuttling pattern between gates are preserved as closely as possible, ensuring that the amplified noise reflects the execution of the original circuit.

Shot allocations between the original circuit and its single amplified version were optimized to achieve the target statistical precision using minimal resources. The optimal ratio of shots between the amplified circuit and the original is given by
\begin{align}
    \frac{N_{s,\text{amp}}}{N_{s,\text{orig}}} = f^{-\frac{3}{2}}\exp{\left[\gamma(f-1)\right]}.
\end{align}
Here $\gamma$ is the effective infidelity afflicting the observable in the noisy circuit and $f$ is the noise amplification factor defined in Eq.~\eqref{ZNE_ANSATZ}. In this work, an amplification factor of $f=5$ was used throughout.

We calculate the noisy expectation values of all Paulis constituting the desired mitigated observable, $\hat{O}=\sum_i w_i P_i$, and extrapolate them independently using an exponential model
\begin{align}
    P_i(f) = P_{i,0} e^{-\Lambda_i f}.
\end{align}
Here, $P_{i,0}$ and $\Lambda_i$ are fitting parameters representing the ideal value of the extrapolated Pauli and the effective infidelity it accumulates in the circuit. We obtain the mitigated observable $\hat{O}$ as a weighted sum of the mitigated Paulis $O_0 = \sum_i w_i P_{i,0}$ (see App.~\ref{app.methods} for more details). Fig.~\ref{fig:quantinuum_per_pauli_step_19} shows an example of this per-Pauli extrapolation for the magnetization operator on 51 qubits after 19 Floquet cycles.

Estimating the error bar $\Delta O$ is not straightforward. Simply summing the variances of its various Paulis is insufficient due to correlations among them; doing so may provide merely a lower bound. Instead, we estimate $\Delta O$ by fitting and extrapolating $\hat{O}$ a second time, now as a whole. This auxiliary mitigation is aware of the correlations between Paulis, and thus produces a reliable error bar $\Delta O_{\text{aux}}$. Finally, we heuristically correct the auxiliary error bar as 
\begin{align}
    \Delta O = \Delta O_{\text{aux}} \frac{O_0}{O_{0,\text{aux}}},
\end{align}
since the mitigated value of the auxiliary mitigation $O_{0,\text{aux}}$ may differ from the per-Pauli value $O_0$.

We note that the choice of unitary gate folding, rather than PEA, as the noise amplification protocol for Quantinuum systems is motivated by two main considerations. First, some error sources on Quantinuum hardware, such as leakage, are either more difficult to characterize or more difficult to amplify using PEA, whereas unitary gate folding naturally captures their effect. Second, the dominant error sources on Quantinuum devices \cite{Moses_2023, ransford2025helios98qubittrappedionquantum} do not include the error generators that are incorrectly amplified by unitary gate folding \cite{Kim2023ScalableEm}. Consequently, unitary gate folding provides an effective noise amplification protocol for these systems.

\section{Effective Floquet Hamiltonian and Gibbs-state reference}
\label{app.effective_ham}

The periodically driven dynamics studied in this work are generated by the Floquet evolution operator $U_F$ of Eq.~(\ref{eq:unitary}). Each application of $U_F$ evolves the state through one driving period of duration $T$. Formally, such dynamics can be described by an effective Floquet Hamiltonian $H_F$, defined by
\begin{align}
    U_F \equiv e^{-iH_FT}.
\end{align}
For interacting many-body Floquet systems this logarithm is generally not unique and need not yield a simple local Hamiltonian. We therefore construct a local approximate effective Hamiltonian using a truncated Baker--Campbell--Hausdorff (BCH) expansion~\cite{reinsch1999simple}.

We begin by rewriting the Floquet evolution operator as
\begin{align}
    U_F = U_3 U_2 U_1,
\end{align}
where each $U_r$ consists of parallel single-qubit $R_X$ and $R_Z$ rotations on all qubits, followed by a layer of fractional-angle $R_{ZZ}$ rotations acting in parallel on a subset of disjoint lattice bonds $E_r$. It is useful to further rewrite each $U_r$ as a product of commuting local blocks,
\begin{align}
\label{eq:ur}
    U_r=
    \Big(\prod_{(q,q')\in E_r} M_{q,q'}\Big)
    \Big(\prod_{q \notin S(E_r)} Q_q\Big),
\end{align}
where $S(E_r)$ is the set of qubits participating in the active edges of layer $r$. The two-qubit blocks $M_{q,q'}$ and the one-qubit blocks $Q_q$ are
\begin{subequations}
\begin{align}
M_{q,q'} &= R_{ZZ}^{(q,q')}R_Z^{(q')} R_Z^{(q)} R_X^{(q')} R_X^{(q)},\\ 
Q_q &= R_Z^{(q)} R_X^{(q)},
\end{align}
\end{subequations}
where the superscripts denote the qubit support and the rotation angles are left implicit: $R_X\equiv R_X(\theta_x)$, $R_Z\equiv R_Z(\theta_z)$, and $R_{ZZ}\equiv R_{ZZ}(\theta_{zz})$.

For fixed $r$, all blocks $M_{q,q'}$ and $Q_q$ in Eq.~\eqref{eq:ur} commute with each other, because they have disjoint supports by construction. Each $U_r$ can therefore be expressed as $U_r=e^{D_r}$, with
\begin{align}
\label{eq:Dr}
D_r = \sum_{(q,q')\in E_r}\log(M_{q,q'}) + \sum_{q \notin S(E_r)}\log(Q_q).
\end{align}
The full Floquet operator may therefore be written as
\begin{align}
    U_F=e^{D_3}e^{D_2}e^{D_1}.
\end{align}

We combine these three exponentials using the BCH expansion in two stages,
\begin{subequations}
\begin{align}
    D_{12} &= \mathcal{Z}(D_2, D_1),\\
    D_{\mathrm{eff}} &= \mathcal{Z}(D_3,D_{12}),
\end{align}
\end{subequations}
where $\mathcal{Z}(A,B) \equiv \log(e^A e^B)$. We approximate $\mathcal{Z}(A,B)$ by truncating the BCH series at fourth order,
\begin{align}
\nonumber \mathcal{Z}(A,B) =& A + B + \frac{1}{2}[A,B] \\
& +\frac{1}{12} \left([A,[A,B]] + [B,[B,A]] \right) \\
\nonumber & - \frac{1}{24} [B,[A,[A,B]]].
\end{align}
The approximate effective Floquet Hamiltonian is then obtained as
\begin{align}
    \widetilde H_F = \frac{i}{T} D_{\mathrm{eff}}.
\end{align}
For subsequent PEPO calculations, we express $\widetilde H_F$ in the Pauli basis. The local block logarithms entering the generators $D_r$ [Eq.~\eqref{eq:Dr}] can be expanded as
\begin{subequations}
\begin{align}
\log(M_{q,q'})
&= \sum_{\alpha,\beta\in\{I,X,Y,Z\}} c_{\alpha\beta}^{(q,q')} \sigma_\alpha^{(q)}\sigma_\beta^{(q')}, \\
\log(Q_q) &= \sum_{\alpha\in\{I,X,Y,Z\}} c_{\alpha}^{(q)} \sigma_\alpha^{(q)},
\end{align}
\end{subequations}
After the BCH combination, this yields a sparse Pauli-string representation of $\widetilde H_F$.

We next construct the thermal reference state by imaginary-time evolution under the approximate effective Hamiltonian $\widetilde H_F$. Starting from the infinite-temperature state, $\rho(\beta=0)\propto I$, represented as a PEPO, we build the Gibbs operator by repeatedly applying small imaginary-time steps $e^{\Delta\beta\widetilde H_F}$.
The positive sign convention is used because the Floquet energy of the initial state lies above the infinite-temperature energy, so the evolution must move toward higher energy density. 

After each imaginary-time step, the PEPO is compressed using belief propagation with fixed bond dimension $D$. During the evolution, we periodically evaluate the energy density and magnetization. The imaginary-time evolution is terminated when the energy matches that of the initial state, $\rho_0=(|0\rangle\langle 0|)^{\otimes N_q}$,
\begin{align}
\frac{\mathrm{Tr}(\rho_{\mathrm{ITE}}\widetilde H_F)}{\mathrm{Tr}(\rho_{\mathrm{ITE}})} = \mathrm{Tr}(\rho_0\widetilde H_F).\end{align}
This energy-matching condition defines the effective inverse temperature $\beta_{\mathrm{eff}}$, yielding the Gibbs state of the approximate effective Floquet Hamiltonian $\widetilde H_F$,
\begin{align}
    \rho_{\mathrm{th}} = \frac{e^{\beta_{\mathrm{eff}}\widetilde H_F}}{\mathrm{Tr}\!\left[e^{\beta_{\mathrm{eff}}\widetilde H_F}\right]}.
\end{align}
The corresponding thermal magnetization,
\begin{align}
    M_{\rm th} = \mathrm{Tr}(\rho_{\rm th}M),
\end{align}
is used throughout the paper as a reference value for the prethermal plateau.

\begin{figure}[tb]
    \centering
    \includegraphics[
        width=\linewidth,
    ]{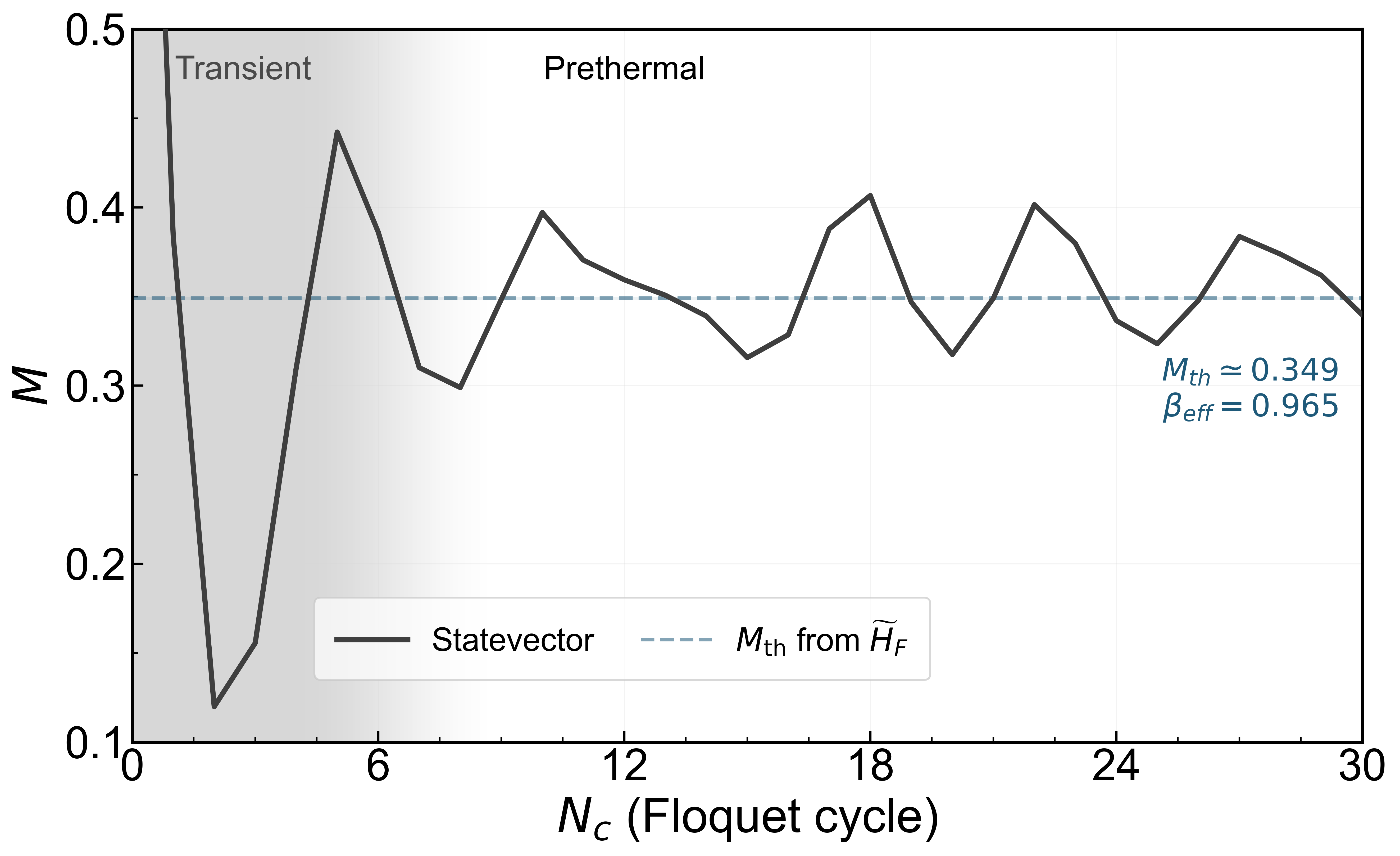}
    \caption{Floquet dynamics of the magnetization of the 21-qubit patch (Fig.~\ref{fig:small_systems}). After an initial transient, the magnetization approaches the thermal value predicted by the fourth-order effective Floquet Hamiltonian, $M_{\mathrm{th}}$, and exhibits long-lived oscillations around it. This behaviour is consistent with an intermediate-time prethermal regime.
}
\label{fig:prethermalization_21q}
\end{figure}

For the 51-qubit system, we use an imaginary-time step $\Delta\beta=0.05$ and PEPO bond dimension $D=64$. Matching the Floquet energy of the initial state gives
\begin{align}
    \beta_{\mathrm{eff}}&\approx1.016, \\
    \frac{E_{\mathrm{th}}}{N}&\approx0.7129, \\
    M_{\mathrm{th}}&\equiv\mathrm{Tr}(\rho_{\mathrm{th}}\,M)\approx0.410.
\end{align}
As a reference, Fig.~\ref{fig:prethermalization_21q} shows the dynamics of a 21-qubit system, where the Floquet evolution can be computed exactly. In this smaller system, the magnetization approaches the effective-Hamiltonian Gibbs-state reference value after an initial transient of roughly $N_c\approx9$
Floquet cycles and then exhibits long-lived oscillations around it. The same timescale is consistent with the approach to the prethermal reference observed in the 51-qubit data of Fig.~\ref{fig:magnetization}. This behaviour is consistent with prethermalization under the effective Floquet Hamiltonian before the onset of eventual Floquet heating.

\newcommand{\ev}{\mathrm{ev}}
\newcommand{\supp}{\mathrm{supp}}

\section{Sparse Pauli-Path Simulations}
\label{app.spp}

Sparse Pauli-path techniques have gained notable traction in the domain of numerically simulating quantum circuits and similar quantum many-body dynamics in the Heisenberg picture. We provide a brief reiteration of this class of methods~\cite{begusic2023fastconverged,begusic2025sparsepauli,rudolph2026paulipropagationcomputationalframework,Loizeau25paulistrings,gharibyan2025practicalguideusingpauli}.
The general structure of sparse Pauli-path simulation methods, as we employ it in this work, consists of obtaining an efficient representation of an arbitrary operator 
\begin{equation}
O=\sum_I O_I P_I,
\end{equation}
where $I$ sums over all $4^n$ possible Pauli strings that we denote as $P_I$ and their corresponding coefficients $O_I$. 
In order to reduce computational requirements, Pauli strings that are deemed negligible for obtaining an expectation value of interest, are heuristically truncated. In this work we discard Pauli strings if
\begin{equation}
|O_I| < \epsilon_0 \max_J |O_J| \frac{|O|}{N_\mathrm{ref}}.
\end{equation}
This heuristic is relative in the sense that the threshold for truncation is proportional to the largest coefficient in the entire representation of O.
It is also responsive in the sense that the threshold for truncation increases as the overall number of Pauli strings $|O|$ increases. $N_\mathrm{ref}$ acts as a parameter that scales the responsiveness which can be absorbed into $\epsilon_0$. 
In this work we fix $N_\mathrm{ref}=10^8$. 
$\epsilon_0$ is a parameter that we choose freely to tune the overall truncation strength.
The responsive nature of this heuristic has the desirable effect that the overall number of Pauli strings throughout a simulation is stabilized to some extent. 
As $|O|$ increases, truncation becomes stronger which reduces the rate of accumulating more Pauli strings.
As $|O|$ decreases, truncation becomes weaker which increases the rate of accumulating Pauli strings.
Furthermore, we utilize weight-based truncation that discards Pauli strings if their weight $W$, defined as the count of non-identity local Pauli operators, is larger than a threshold $W_0$ which we choose freely.

Simulating the dynamics of some operator O in the Heisenberg picture relies on the straight-forward action derived from the Baker-Campbell-Hausdorff relation

\begin{widetext}
\begin{equation}
e^{i \theta/2 P_J} P_I e^{-i \theta/2 P_J} = 
\begin{cases}
	P_I, &[P_I, P_J] = 0\\ 
	\cos(\theta) P_I - i P_I P_J \sin(\theta), &[P_I, P_J] \not = 0
\end{cases}
\end{equation}
\end{widetext}

Since any unitary dynamics can be expressed using transformations that are generated by individual Pauli strings, this update rule is sufficient to perform arbitrary unitary simulations in the Heisenberg picture. Several numerical frameworks exist that implement this or a similar algorithmic structure. 
In this work we utilize the high-performance ORQA framework~\cite{broers2025scalablesimulationquantummanybody,Broers_2026} executed on the supercomputer Fugaku, as well as BlueQubit's Pauli-path GPU simulator (``pauli-path.gpu``) on an H100 GPU.

\subsection{SPP Convergence Analysis}
\label{app.sppconvergence}
For the circuit that we study in the main-text, we find that SPP simulations do not converge within the truncation parameters accessible within our computational resources. Even beyond a trillion Pauli strings, these non-converged results show suppressed oscillatory behaviour that is out-of-phase with the corresponding quantum data. Furthermore, for finite $W_0$, we observe a trend of overshooting past the expected results. 
This trend becomes more prominent as $W_0$ decreases.
In general, with finite $W_0$, the results only converge to an approximate solution as $\epsilon_0\rightarrow 0$, which only becomes exact as $W_0\rightarrow N$. 

\begin{figure}
    \centering
    \includegraphics[width=\linewidth]{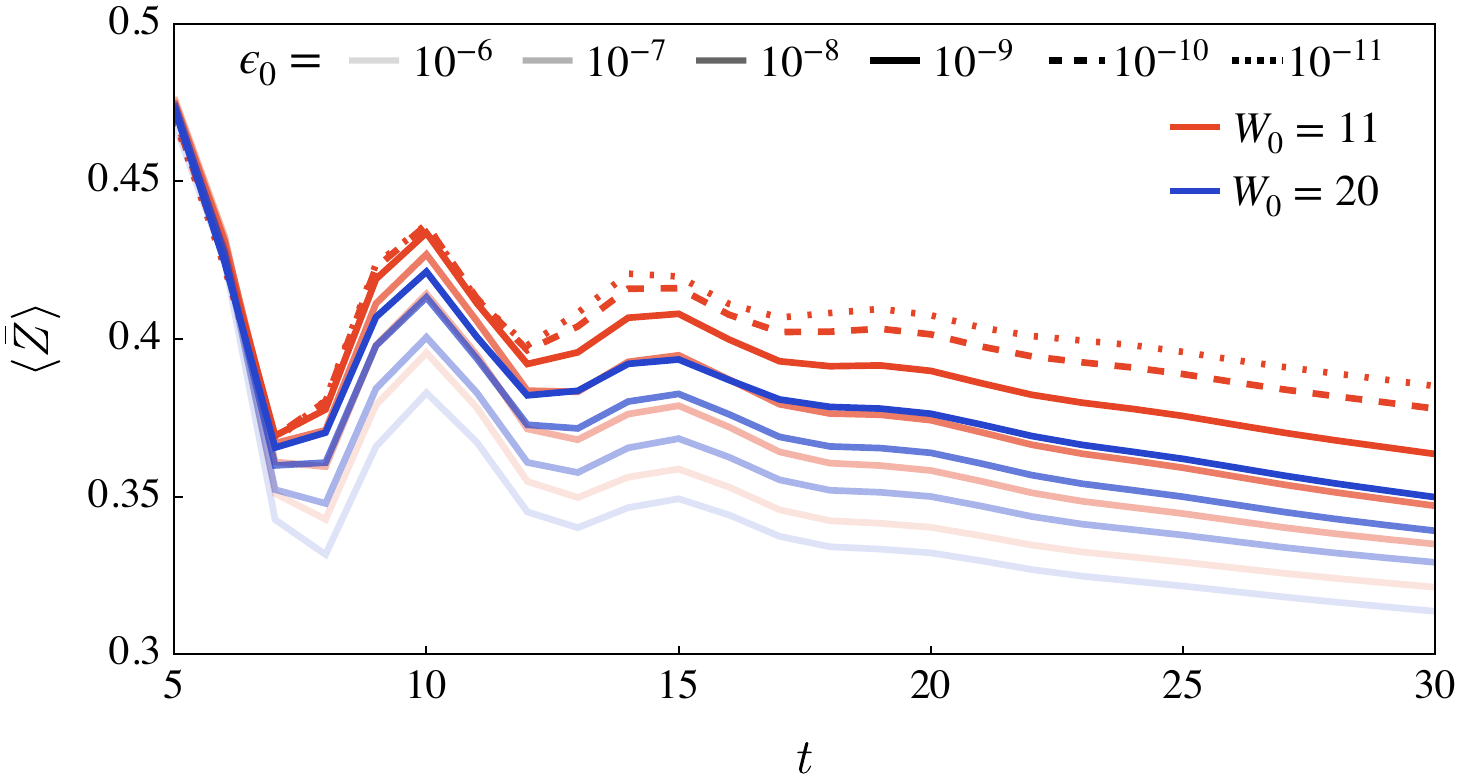}
    \caption{The average magnetization obtained from sparse Pauli path simulations with $W_0=11$ (red) and $W_0=20$ (blue) for various values of $\epsilon_0$.}
    \label{fig:n51_overshooting}
\end{figure}

In Fig.~\ref{fig:n51_overshooting}, we show the results for the cases of $W_0=11$ and $W_0=20$ for values of the truncation parameter $\epsilon_0=10^{-6}$ to $10^{-9}$. The dashed and dotted lines correspond to additional simulations of $W_0=11$ with $\epsilon_0=10^{-10}$ and $10^{-11}$, respectively.
We see a clear trend of the magnetization increasing as $\epsilon_0$ is reduced, overshooting the results that we expect from our quantum experiment. 
In the case of $W_0=20$, we find that the high-weight Pauli strings are not significantly occupied, such that further increasing $W_0$ does not provide any benefit. 
While in some SPP simulations, weight-based truncation can improve convergence, in our case the choice of $W_0=11$ leads to worse results. 

The trade-off between weight-based and coefficient-based truncation gives a set of choices for the values of $W_0$ and $\epsilon_0$ that are numerically accessible.
While it is always possible to \textit{ad hoc} pick the parameters that best fit the experimental data, it is more noteworthy that we do not find convergence within the numerical capabilities provided by our computational resources. 
The most reliable results are the ones with large enough $W_0$, and smallest possible $\epsilon_0$, which in our case is $W_0=20$ and $\epsilon_0=10^{-9}$.
Hence we find that the late-cycle oscillatory behaviour of the quantum data is not within reach of the SPP simulation strategy as we employ it here, retaining on the order of trillions of Pauli strings.

To better understand the lack of convergence, we analyze the same circuit as in the main-text but reduced to a $N=21$ qubit patch that consists of two heavy-hex loops. 
This system has a maximum number of $4^{21}$ (approximately $4.4$ trillion) Pauli strings which is within our computational limitations using the supercomputer Fugaku, such that we can solve this case exactly.
When comparing with the reference statevector simulation, we find persistent late time oscillations in this system which are qualitatively consistent with our quantum data in the case of $N=51$.

\begin{figure}
    \centering
    \includegraphics[width=\linewidth]{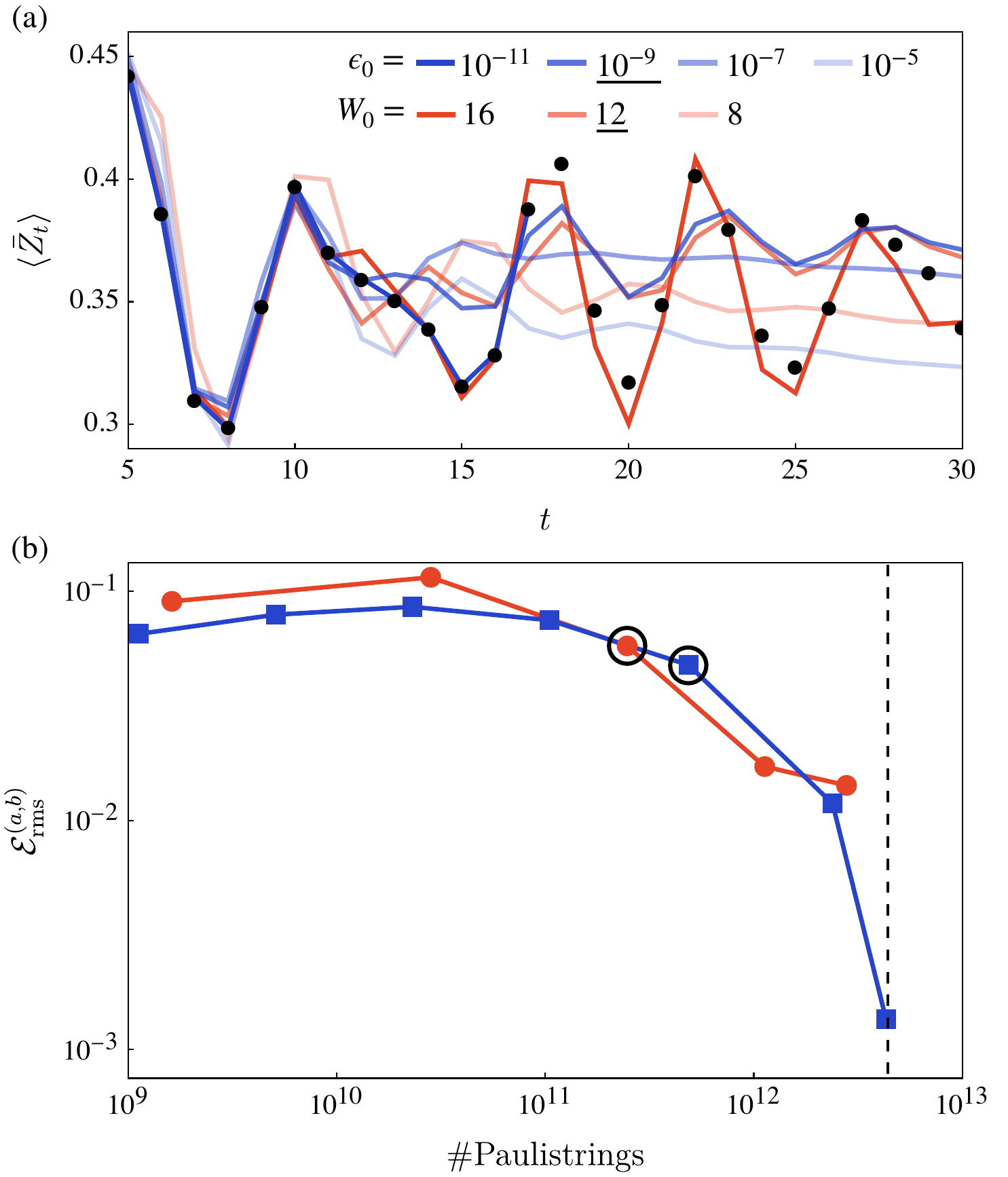}
    \caption{Convergence behaviour in the case of $n=21$ with different truncation. 
    Red curves are obtained with weight-based truncation alone, and blue curves are obtained with coefficient-based truncation alone.
    (a) The average magnetization as a function of time. 
    Black dots correspond to the exact solution.
    The parameter values that are underlined correspond to the circles in (b).
    Note that the simulation corresponding to the smallest $\epsilon_0$ terminates at time-step 17.
    (b) The root-mean-square error over the time-steps from $a=10$ to $b=17$ as a function of the number of Pauli strings retained in the corresponding simulation. The dashed vertical line indicates the maximum number of Pauli strings for 21 qubits, i.e., $4^{21}$. The data points that are circled correspond to the simulations in Fig.~\ref{fig:weight_resolved_n21}.}
    \label{fig:error_n21}
\end{figure}

In Fig.~\ref{fig:error_n21}, we show the convergence behaviour with respect to weight-based and coefficient-based truncation. First, we show the average magnetization $\langle \bar Z _t\rangle$ as a function of time for different truncation thresholds $W_0$ and $\epsilon_0$. 
Here we find results that are consistent with our observations in the case of $N=51$ in the main-text. 
The exact solution displays prominent late-cycle oscillations, whereas moderately truncated simulations show significantly damped and out-of-phase values that fail to reproduce the exact results.
Only with very weak truncation do the simulations begin to converge, reproducing the oscillatory behaviour correctly.
We find however, that this convergence requires a $\mathcal{O}(1)$ fraction of the total possible number of Pauli strings, i.e., on the order of trillions of Pauli strings for $N=21$.

To quantify the convergence behaviour, we further show the root-mean-square of the relative error
\begin{equation}
    \mathcal{E}^{(a,b)}_\mathrm{rms}=\left(\frac{1}{b-a}\sum_{t=a}^{b}  \left(\frac{\langle\bar Z_t\rangle_\mathrm{SPP}-\langle \bar Z_t\rangle_\mathrm{Exact}}{\langle \bar Z_t\rangle_\mathrm{Exact}}\right)^2\right)^{\frac{1}{2}}
\end{equation}
in the time-window from $a=10$ to $b=17$ as a function of the number of Pauli strings retained through-out the simulations.
Here we clearly see that for either truncation strategy, convergence happens very late and requires a significant portion of all possible Pauli strings.
Therefore, the consistency in the results, and the effects of truncation in the cases of $N=21$ and $N=51$,  suggest that in the latter a significant portion of the total number of $4^{51}\approx 5\times 10^{30}$ Pauli strings is necessary to reproduce the late time oscillations which we observe in our experimental error-mitigated quantum data. 

\begin{figure*}
    \centering
    \includegraphics[width=\linewidth]{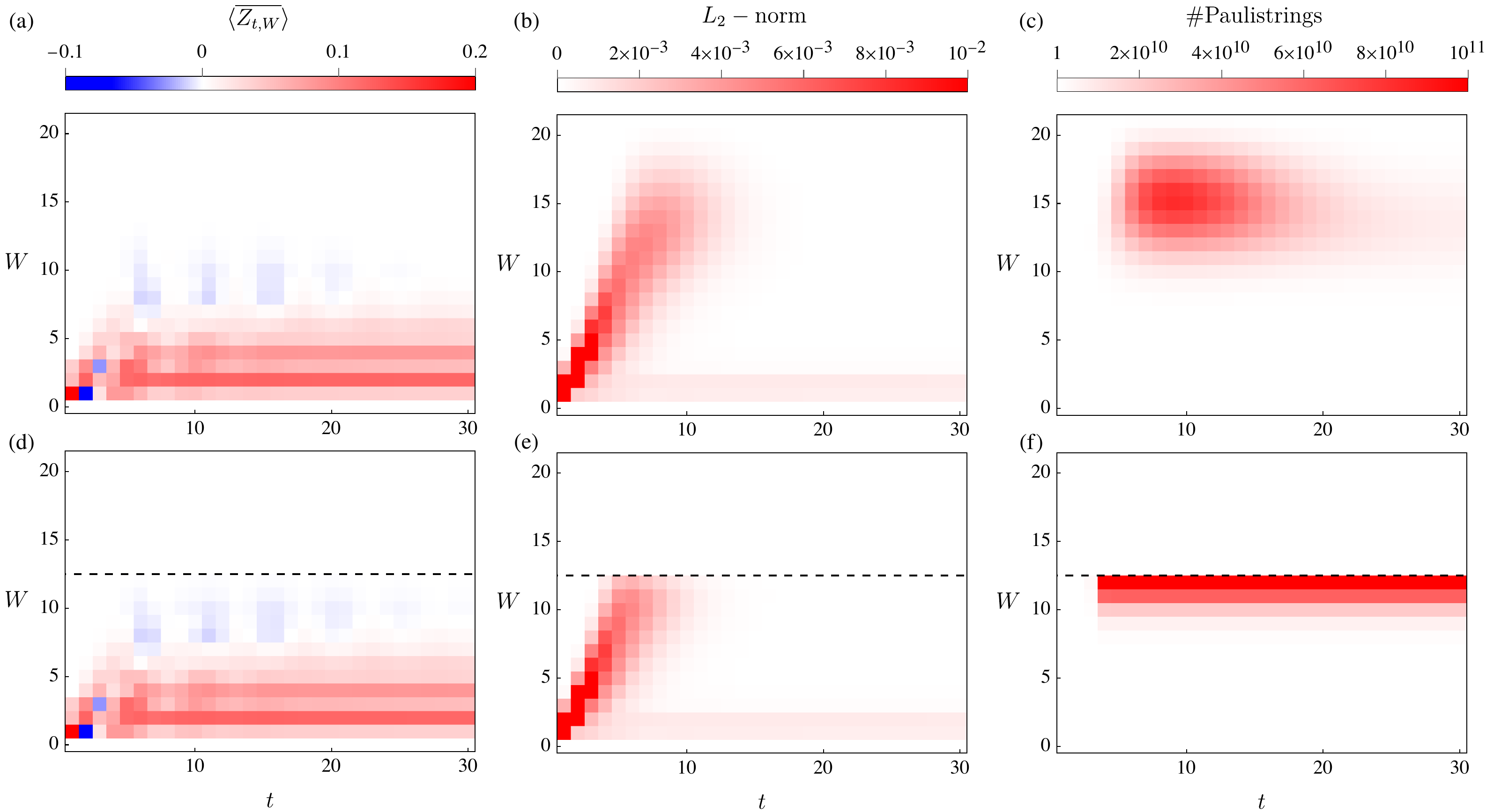}
    \caption{Weight-resolved results of SPP simulations for the $N=21$ qubit case. The top row (a, b, c) corresponds to $W_0=21$ (no weight-based truncation) and $\epsilon_0=10^{-9}$. The bottom row (d, e, f) corresponds to $W_0=12$ and $\epsilon_0=0$ (no coefficient-based truncation). 
    The left column (a, d) shows the contributions to the average magnetization resolved by the weight of Pauli strings.
    The center column (b, e) shows the $L_2$-norm resolved by the weight of Pauli strings.
    The right column (c, f) shows the number of Pauli strings resolved by their weight.}
    \label{fig:weight_resolved_n21}
\end{figure*}

Fig.~\ref{fig:weight_resolved_n21} shows the weight-resolved SPP simulation results for 21 qubits with weak truncation, either weight-based with $W_0=12$ or coefficient-based with $\epsilon_0=10^{-9}$. 
We show the contributions to the average magnetization resolved by weight $\langle \overline{Z_{t,W}}\rangle$, the $L_2$ norm of the simulated observable resolved by weight, and the number of Pauli strings retained throughout the simulations resolved by weight.

In the contributions to the average magnetization, we see two distinct features.
Low-weight Pauli strings with $W \leq 8$ show a steady distribution that barely changes during the long-time dynamics. These contributions lead to a constant offset in the total average magnetization.
Medium-weight Pauli strings with $8<W\leq12$ display negative oscillatory contributions, which leads to the late-cycle oscillations observed in the average magnetization in Fig.~\ref{fig:error_n21}~(a). 
As explained above, this feature is very susceptible to truncation and is only reproduced under either truncation strategy if a substantial portion of all possible Pauli strings is retained. The simulations we choose to show here are just on the verge of capturing these oscillations, as can be seen by comparing with Fig.~\ref{fig:error_n21}.

The $L_2$-norm distribution in Figures~\ref{fig:weight_resolved_n21}~(b) and (e) also displays two clearly distinct features. 
First, low-weight Pauli strings maintain a persistent fraction of the $L_2$-norm throughout the dynamics. 
This feature is again robust against stronger truncation.
Second, high-weight Pauli strings are generated very quickly during early times, and their $L_2$-norm contributions dissipate after $t=10$, as the magnitude in coefficients spreads over the significant majority of the high-weight Pauli strings which leads to them being truncated. 
In the case of weight-based truncation, the Pauli strings are truncated and cannot reach the regime of large weights.
In the exact dynamics, the total $L_2$-norm is constant such that the qualitative histogrammatic profile around $t=10$ persists for long-time dynamics.

The distributed count of Pauli strings in Figures~\ref{fig:weight_resolved_n21}~(c) and (f) is completely dominated by large weights, due to the fact that combinatorially those weights have the largest amount of possible Pauli strings. 
The total number of Pauli strings with a weight $W$ in a system of $N$ qubits is given by
\begin{equation}
    N_W=\binom{N}{W}3^W.
\end{equation}

While the $L_2$-norm does not reflect this, almost all Pauli strings have a weight above $W\approx 10$. 
In particular in the case of weight-based truncation, the distribution of Pauli strings is steady, while the $L_2$-norm of high-weight Pauli strings vanishes. 
This indicates that the high-weight Pauli strings take on a very flat distribution. 
In contrast, the amount of low-weight Pauli strings is necessarily very small while their $L_2$-norm is relatively large, which makes them robust to truncation.
However, it is the high-weight Pauli strings that sustain the oscillatory features in the average magnetization.
With stronger truncation, the flat distribution of high-weight Pauli strings is removed regardless of whether the heuristic is weight-based or coefficient-based. 

\begin{figure}
    \centering
    \includegraphics[width=\linewidth]{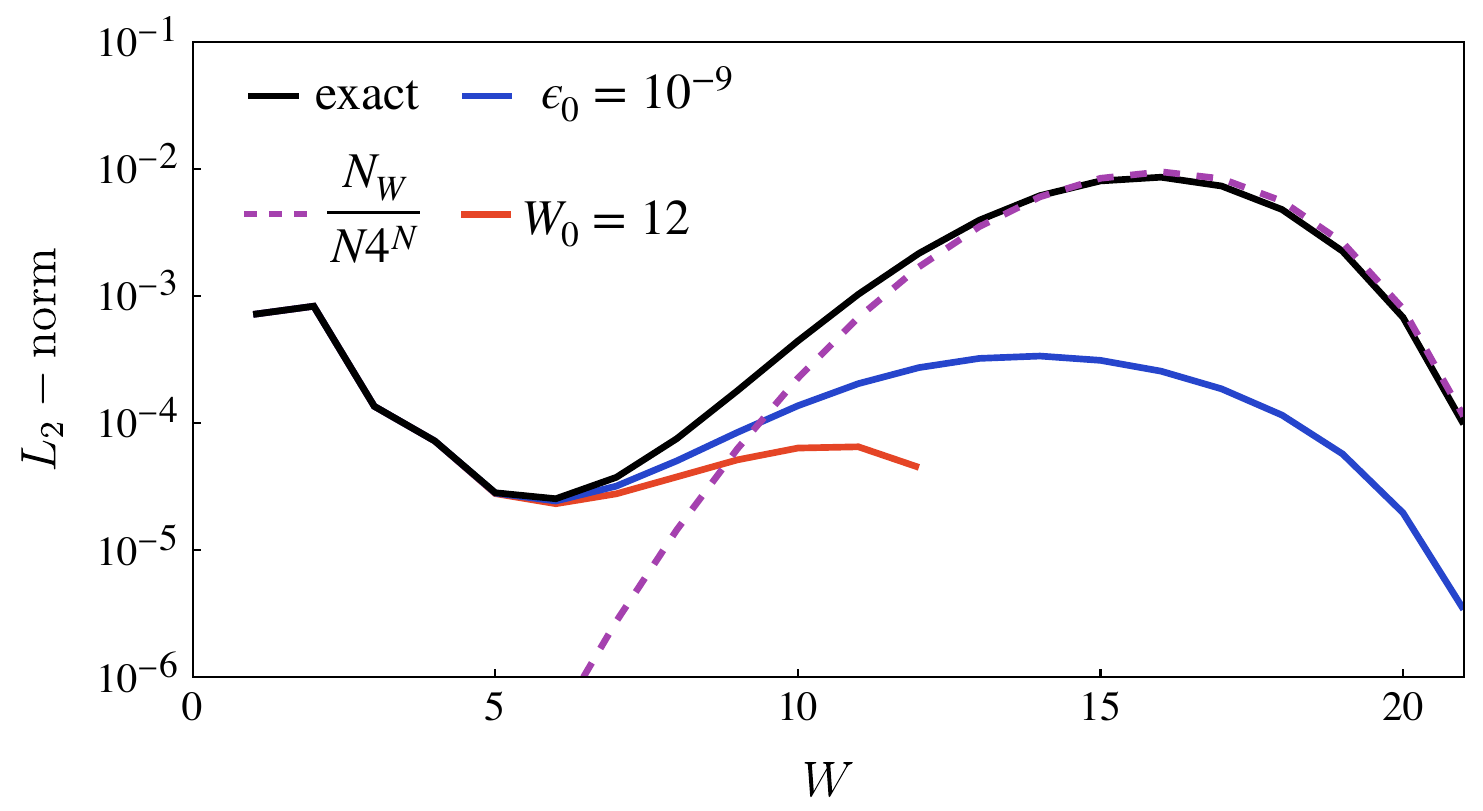}
    \caption{The weight-resolved $L_2$-norm at $t=15$ obtained from the simulations shown in Fig.~\ref{fig:weight_resolved_n21} (red and blue), as well as exactly with all $4^{21}$ Pauli strings (black), as well as a analytical uniform distribution (dashed purple). 
    The truncated results correctly capture the low-weight Pauli strings, but fail to resolve the high-weight Pauli strings. The exact solution for the $L_2$-norm agrees well with the fully uniform distribution.}
    \label{fig:n21_l2}
\end{figure}

In Fig.~\ref{fig:n21_l2}, we show the $L_2$-norm resolved by weight at the time-step $t=15$ for the same simulations as we show in Fig.~\ref{fig:weight_resolved_n21}. Additionally, we show the exact solution from a simulation that we perform without any truncation whatsoever, retaining all $4^{21}$ Pauli strings. Further, we show the results produced analytically by assuming a completely uniform distribution over all Pauli strings. 
As already discussed above, we see here that the truncated simulations reproduce the $L_2$-norm contributed by low-weight Pauli strings correctly, while they fail to do so for the high-weight Pauli strings. 
The uniform distribution over all possible Pauli strings produces a structure directly proportional to $N_W$, which we find coincides with the high-weight $L_2$-norm contributions obtained form our exact simulation. 
This further underlines our understanding of why these simulations are sensitive to truncation and computationally extremely demanding.

Consequently, heuristic truncation that reduces the number of retained Pauli strings immediately omits the high-weight Pauli strings, and therefore, the oscillatory features in the late-cycle dynamics, while only retaining the steady magnetization originating from low-weight Pauli strings. 
This is also the behaviour we observe in the case of $N=51$.
We therefore conclude that reducing the truncation to the point where we would reproduce the oscillating average magnetization that we see in our experimental data, requires computational resources many orders of magnitude beyond contemporary supercomputers.

We acknowledge that it is still reasonable to assume that the more intricate structures of Pauli strings that are strictly necessary to obtain the correct expectation values may very well be a small fraction of all generated Pauli strings. 
If so, significantly more involved truncation strategies would need to sift through this potential dead-wood of Pauli strings in a way that reduces the computational demand again, while negligibly affecting the accuracy of the results.
However, with the conventional truncation heuristics at hand there appears no possibility to converge our SPP simulations in the case of $N=51$ while retaining on the order of a few trillion Pauli strings.

\subsection{Noisy SPP Simulations}
\label{app.noisy_spp}

\begin{figure*}[!t]
    \centering
    \includegraphics[width = \linewidth]{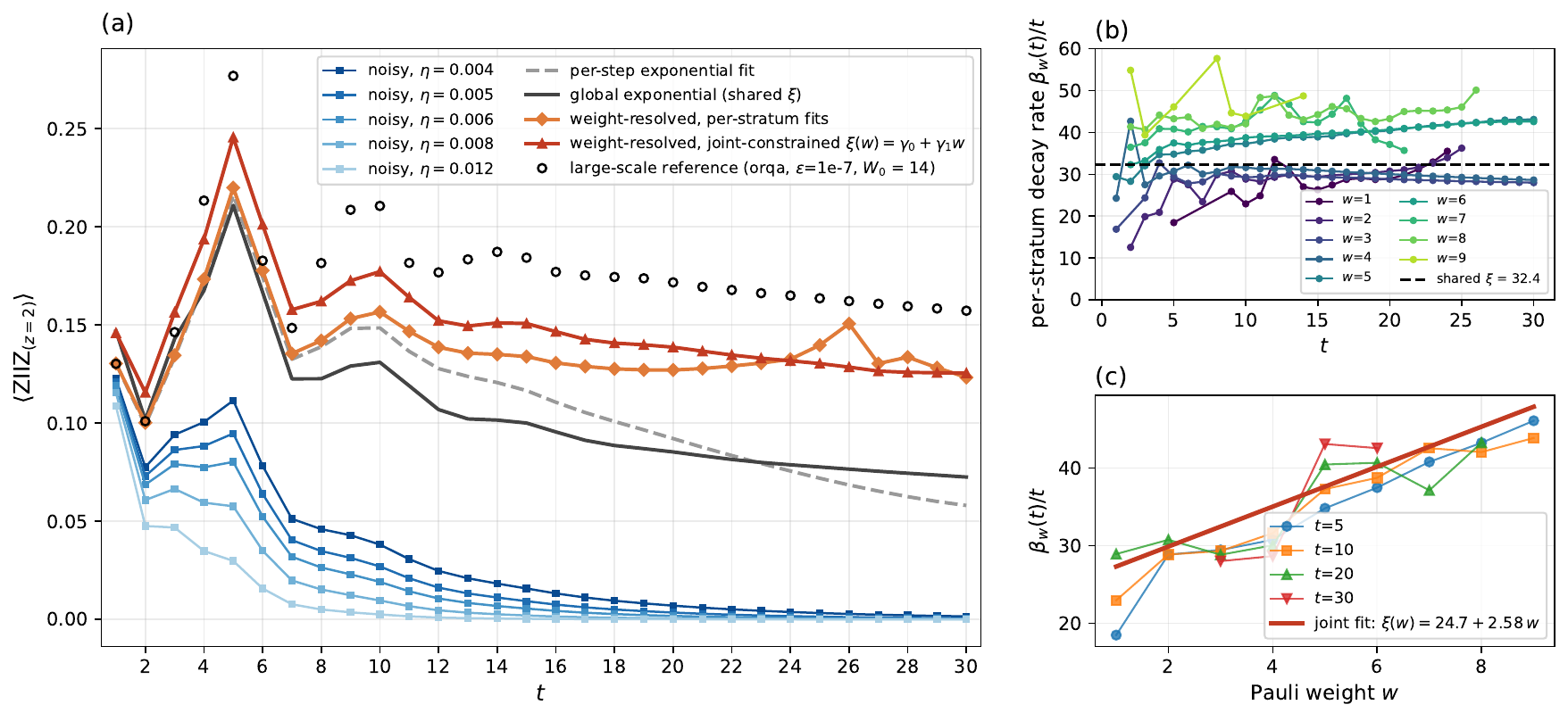}
    \caption{Pauli-path ZNE for $\langle ZIIZ_{(z=2)}\rangle$ on the 51-qubit patch ($T=30$ steps, weight cutoff $W_0=15$, relative
    coefficient threshold $\delta_{rel} = 8\times10^{-6}$).
    (a) Noisy SPP expectation values for symmetric depolarizing rates
    $\eta \in \{0.004, 0.005, 0.006, 0.008, 0.012\}$, together with four zero-noise extrapolations fit in log space over the five rates: independent per-step exponential fits, a global exponential with a single shared decay rate $\xi$, and the two weight-resolved methods, which extrapolate each final-weight stratum $s_w(\eta,t)$ separately. The simulations involved between $5.5 \times 10^8$ ($\eta = 0.012)$ and $9.8 \times 10^8$ ($\eta = 0.004$) Pauli strings at peak. Open circles show the large-scale noiseless reference (ORQA run on Fugaku, $\epsilon = 10^{-7}$, $W_0=14$, ${\sim 1.3 \times }10^{11}$ Pauli strings at peak).
    (b) Per-stratum decay rates $\beta_w(t)/t$ for $w \le 9$.
    (c) The same rates versus Pauli weight $w$ at four times, with the joint fit $\xi(w) = 24.7 + 2.58\,w$ (red line): the approximately linear weight dependence is predicted by the path-multiplicity picture, which motivates the joint-constrained method.
    }
    \label{fig:pps_zne_main}
\end{figure*}

In the SPP simulations of the quantum dynamics studied in this work, we use both coefficient and weight truncation to avert the exponential growth of Pauli terms. Truncation and the resulting trade-off in accuracy are unavoidable, as the available memory is limited. This trade-off varies in severity, depending on the complexity of the problem. In the previous section, we saw that truncation irreversibly destroys the coherent higher-weight Pauli structures responsible for late-cycle oscillations.

A potential line of inquiry, which at the outset seems complementary to truncation strategies, is to study the effect of noise in Pauli-path simulations. The motivation comes from studies showing that Pauli-path methods are effective in simulating noisy quantum circuits (in the average case) in certain noise regimes. For depolarizing noise for instance, a Pauli path that encounters $m$ noise factors is damped by $\lambda^m$, which concentrates the path sum on low-weight paths and yields a polynomial-time classical algorithm for noisy random circuit sampling as well as for expectation values of noisy quantum circuits ~\cite{Aharonov2023, Schuster2025apolynomialtime, Shao2024simulating, Angrisani2026simulating}. Conversely, truncation in noiseless Pauli-path simulations incurs an error that behaves, on average over circuits, like that of a noise channel ~\cite{Angrisani2025classically}.  

The foregoing discussion suggests injecting noise into the SPP simulations, which enters the expectation value through an explicit parameter $\lambda$. By varying this noise damping level $\lambda$, we can extrapolate to the zero-noise limit of the simulations. Zero-noise extrapolation is then the natural protocol for removing the injected damping -- mirroring similar protocols used in hardware error mitigation (\emph{cf.} Sec.~\ref{app.methods} and the references therein). Unlike in the hardware setting, here the noise channels and strengths are known exactly, and can even be chosen based on the quantum circuit. We carry out this technique in the setting of products of single-qubit Pauli noise channels.

We find that such an extension of the Pauli-path method can only extrapolate within the family of paths that survive truncation. As the noise level $\eta \to 0$, the noisy sum approaches the truncated noiseless sum. This provides a moderate memory advantage, but it does not lead to true noiseless answer due to the simple fact extrapolation cannot remove the effects of truncation. Unlike hardware ZNE where the noise can only be amplified, in SPP, well-chosen noise reduces the simulation cost: damping suppresses coefficients such that noisy runs survive truncation with fewer Pauli terms. ZNE therefore trades a controlled extrapolation bias for a reduction in memory and runtime: several cheap noisy runs can (in principle) replace a more expensive noiseless run. Our numerical experiments showed memory savings of 30\% to 75\%, depending on the truncation policy.          

The experiments we report here were carried out with BlueQubit's Pauli-path GPU simulator. We use both weight and coefficient truncation. For coefficient truncation, we use relative truncation with respect to Frobenius norm of the evolving Pauli sum. Writing $O^{(k)} = \sum_P c_P P$ for the branched-and-merged Pauli sum after applying the $k$-th gate, we prune Pauli strings $P$ which satisfy $|c_P| \leq \|O^{(k)}\|_F\,\delta_{rel}$, for a chosen threshold $\delta_{rel}$ (which is fixed throughout the circuit simulation), where $\|O^{(k)}\|_F = (\sum_{P} |c_P|^2)^{1/2}$ is the normalized Frobenius norm.

\emph{The noise model}: After each two-qubit gate in the circuit, single-qubit Pauli noise channels with probabilities $\eta = (p_x, p_y, p_z)$ are applied independently to each of the two qubits acted on by the gate. In other words, we apply the channel $\varepsilon \otimes \varepsilon$, where  $$\varepsilon (\rho) = (1-p_x - p_y - p_z) \rho + p_x X \rho X + p_y Y\rho Y + p_z Z \rho Z.$$ 

Single-qubit Pauli channels are diagonal in the Pauli basis; the Pauli-transfer-matrix eigenvalues are 
\begin{align*}
    & \lambda_X = 1 - 2(p_y + p_z), \quad  \lambda_Y = 1 - 2(p_x + p_z), \\ & \lambda_Z = 1 - 2(p_x + p_y), \quad \lambda_I = 1.
\end{align*}

For simplicity of discussion, we focus on the case of symmetric depolarizing noise, $p_x = p_y = p_z$, so that every non-identity Pauli operator on a noise site is damped by the common factor $$\lambda = \lambda_X = \lambda_Y = \lambda_Z = 1 - 4\eta.$$ 

Note that $\eta$ is the rate of each individual Pauli error, so the total error
probability per noise event is $3\eta$; in terms of the standard depolarizing
parameter $p$ (for which $\lambda = 1 - \tfrac{4}{3}p$) our $\eta$ equals $p/3$.

In the Heisenberg picture the effect of one noise event on a Pauli string $P$ is a pure rescaling of its coefficient: since the channel acts only on the two qubits $Q_g$ supporting gate $g$, the coefficient of $P$ is multiplied by
\begin{equation}
  \lambda^{\,|\supp(P) \cap Q_g|}\,,
  \qquad |\supp(P) \cap Q_g| \in \{0, 1, 2\}.
  \label{eq:pergate}
\end{equation}

Thus, applying noise has the following effect.
If $n_j$ denotes the number of two-qubit gates that act on a qubit $j$ during one Trotter step, a Pauli string $P$ of weight $w$ is damped by 

\begin{equation}
    \prod_{j \in \supp(P)} \lambda ^{n_j} \approx \lambda ^{\bar n w} =  (1-4\eta)^{\bar n w} \approx e^{-4 \bar n \eta w},
    \label{eq:damping_frozen}
\end{equation}  
where $\bar n$ is the mean gate incidence per qubit per step. ($\bar n = 2 \times \frac{56}{51} \approx 2.2$ for the 51-qubit patch.)

The above analysis supposes that $P$ is frozen throughout the Trotter step; to move past this caveat, and to obtain an analytically correct expression for the damping of $P$ after one Trotter step, we look at the path-multiplicity of $P$. Let $g_0, g_1, \ldots, g_M$ be a sequence of Pauli rotation gates comprising a Trotter step. Let $\gamma = (P^{(0)} \to \cdots  \to P^{(M)})$ be a Pauli path through this gate sequence, with noiseless amplitude $A_\gamma$ (a product of the visited $\cos \theta_g$, $\pm \sin \theta_g$ entries of the gates). This path accumulates a damping factor $\lambda^{m_\gamma}$, where 
\begin{equation}
    m_\gamma = \sum_{g} |\supp(P^{(g)}) \cap Q_g|
\end{equation}    
counts the noise factors collected along the sequence. Our simulator does not keep track of the paths; rather the contribution of paths arriving at the same string are collapsed into one aggregated coefficient (the merge). The multiplicity structure survives this aggregation in the sense we explain below.   

\begin{figure*}
    \centering
    \includegraphics[width=\linewidth]{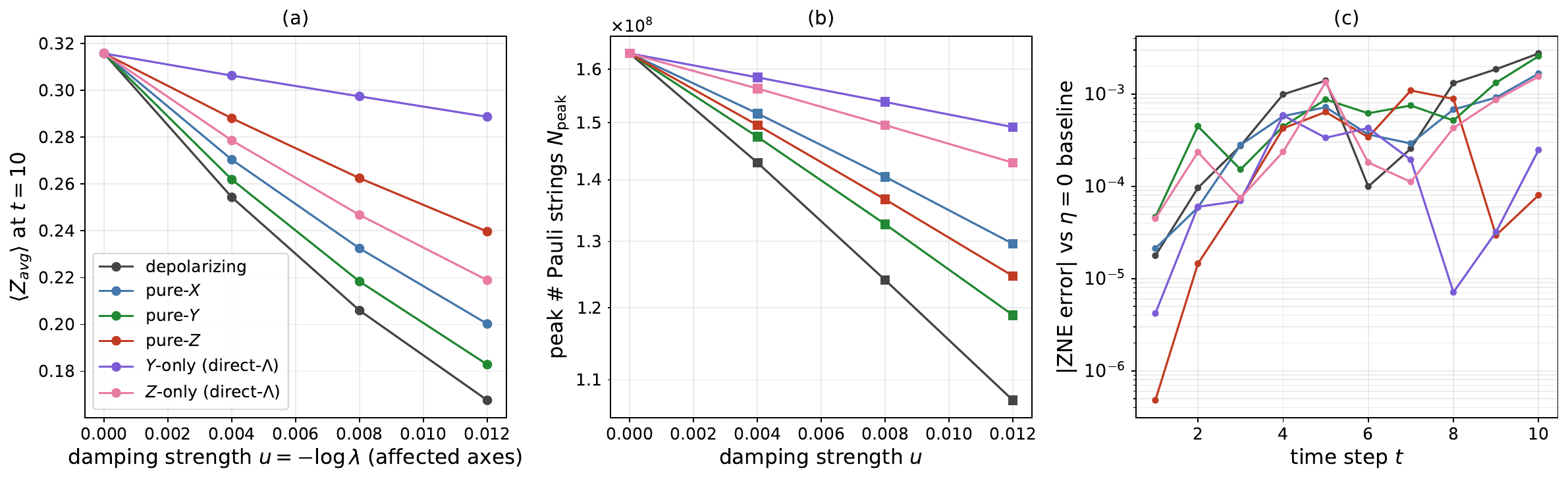}
    \caption{Axis-resolved SPP noise probe for $\langle Z_{\mathrm{avg}} \rangle$ on the 51-qubit patch ($T=10$ Trotter steps, relative truncation $\delta_{rel} = 2\times10^{-5}$, no weight cutoff). We choose the noise parameter $\eta = (p_x, p_y, p_z)$ to damp the three Pauli axes independently; we sweep four directions -- isotropic (depolarizing noise) and the three pure-rate channels -- each at matched damping $\lambda = e^{-u}$ on the affected axes. We additionally sweep two single-axis directions, $Y$-only and $Z$-only; these do not correspond to any physical CPTP channels. (b) The peak Pauli-string count $N_{\mathrm{peak}}$ per run; its slope with $u$ is the memory saved per unit damping in each direction. (c) shows the error of the three-point log-linear ZNE of each family against the measured $\eta=0$ baseline (at the same truncation policy).}
    \label{fig:pps_zne_probe}
\end{figure*}

Each coefficient $c_P$, after time step $t$, can be expressed as a sum over families of paths of different multiplicities that merge to $P$:
\begin{equation}
  c_P(\lambda, t) = \sum_{m \ge 0} \sigma_{P,t}(m)\, \lambda^m, \quad \sigma_{P,t}(m) = \sum_{\substack{\gamma \to P \\ m_\gamma = m}} A_\gamma,
  \label{eq:perstring}
\end{equation}

Let us introduce the equivalent log-damping variable
\begin{equation}
  u \;\coloneqq\; -\log \lambda \;=\; -\log(1 - 4\eta) \;=\; 4\eta + O(\eta^2).
  \label{eq:u}
\end{equation}

Then $c_P(u, t) = \sum_{m \ge 0} \sigma_{P,t}(m) \, e^{-mu}$; 
and summing over Pauli strings $P$ that has non-trivial overlap with the initial state $\rho_0$, we get a similar decomposition for the expectation value.
\begin{equation}
    \ev(u, t) = \sum_{m \ge 0} \sigma_t(m) \, e^{-mu}, \quad \sigma_t = \sum_{\substack{P \\ \mathrm{Tr}(\rho_0 P)  \neq 0}} \sigma_{P, t}.
    \label{eq:ev_laplace}
\end{equation}  

Equation \eqref{eq:ev_laplace} is saying that the expectation value $\ev$ (for the truncation-free dynamics) is a polynomial in $\lambda = e^{-u}$ of enormous degree, or equivalently as a Laplace transform of the signed multiplicity measures $\sigma_t$. Zero-noise extrapolation is the inverse-Laplace problem of estimating $\sigma_t$'s total mass $\ev_0(t) = \sum_m \sigma_t(m)$ from a few evaluations at $u > 0$. 

In order to extrapolate from a finite set of noisy runs with varying noise levels $\eta$ to the zero-noise limit $\eta=0$, one can proceed in a few  different ways: 1.) coarser, per-step extrapolation of expectation value, or 2.) refined, weight-resolved extrapolation. We explain both methods, as well as an intermediate method, called \emph{joint-constrained fit}, which models the per-stratum decay parameter as linear in $t$ and $w$. 

We write $\ev(\eta, t)$ for the SPP-simulated expectation value for noise rate $\eta$ at time $t$ (with the truncation parameters being fixed). 

In the first method, we fit at each step $t$, the log-linear model
\begin{equation}
    \log \ev(\eta, t) = \log \ev_0(t) - b(t)u(\eta),
    \
\end{equation}
by least squares over the noise rates in the scan, and report the intercept $\ev_0(t)$ as the zero-noise estimate. Empirically, the fitted slope is linear in $t$, $b(t) \approx \xi\,t$, with $\xi$ independent of $t$ when working with low noise-levels $\eta$; we therefore also perform a joint-fit in which the constraint $b(t) = \xi t$ is imposed globally, with one shared parameter $\xi$ and one intercept per step. 

As we saw in Eq.~\eqref{eq:damping_frozen}, the level of noise-induced dampening of Pauli coefficients in noisy circuit evolution depends on the weight of the Pauli strings, and therefore a more refined approach for extrapolating is to break up the expectation value into weight-resolved contributions from Pauli-strata of different weights $w$:

\begin{equation}
    \ev(\eta, t) = \sum_w s_w(\eta, t),
    \quad s_w(\eta, t) \coloneqq \sum_{\substack{P \ \text{$Z$-type},\\ |P| = w}}\!\! c_P(\eta, t),
\end{equation}
where $c_P(\eta, t)$ is the coefficient of the surviving string $P$ at time $t$ and $|P|$ its weight. (The simulator gives direct access to such a refinement that a hardware experiment cannot observe.) Note that in the description of $s_w(\eta, t)$, we have utilized the fact that we are starting with an initial state on which only $Z$-type strings have nonzero expectation ($|0\rangle^{\otimes n}$ in our case).

We then write 
\begin{equation}
    \log (s_w(\eta, t)) = a_w(t) - \beta_w(t) u(\eta),
    \label{eq:strata_fit}
\end{equation}
and find the best fit per stratum $w$ and per step $t$. This gives a recombination of the intercepts into the zero-noise estimate $\ev_{\mathrm{zne}}(t) = \sum_w e^{a_w(t)}$. Because of the sign issue, we fit \eqref{eq:strata_fit} only on strata that have a consistent sign. In our case, a bin $(t, w)$ enters the fit only if $s_w > 0$ at every noise rate in the scan. We moreover impose that the stratum contains at least $N_{\min}$ strings ($N_{\min} = 20$ in our experiments); the remaining (negligible) bins are carried at their value at the smallest simulated noise rate.

Recall from Eq.~\eqref{eq:damping_frozen} that a frozen Pauli string accumulates a damping factor  $\approx \lambda^{\bar n w}$ over a time step; over $t$ steps a family whose weight profile averages $\kappa w$ (relative to its final weight $w$) accumulates $\bar n \kappa w t$ noise factors. This gives a plausible explanation for the observed linearity:
\begin{equation}
    \mathbb{E}[m \mid w, t] \;\approx\; \bar n\, \kappa\, w\, t \quad \Longrightarrow \quad
    \beta_w(t) \approx \bar n \kappa w t.
    \label{eq:beta_linearity}
\end{equation}

Instead of fitting Eq.~\eqref{eq:strata_fit} independently in each admissible bin $(t, w)$ (with two parameters per bin), we can impose the form suggested by Eq.~\eqref{eq:beta_linearity} globally: 
\begin{equation}
    \log s_w(\eta, t) = \alpha_{t, w} \, - \, (\gamma_0 + \gamma_1 w)\,t \,u(\eta);
    \label{eq:joint_fit}
\end{equation}
a single linear least-squares problem over all admissible bins, with intercept $\alpha_{t, w}$ per bin and only two global noise parameters $(\gamma_0, \gamma_1)$, yielding the \emph{joint-constrained} weight-resolved fit. We remark that the aggregate global-$\xi$ fit mentioned earlier is the special case $\gamma_1 = 0$ of Eq.~\eqref{eq:joint_fit} (all strata damped identically, so the strata sum collapses to a single exponential), and the unconstrained per-stratum fit is the opposite limit in which every bin's slope is free. The joint-constrained fit is the middle-ground, and it seems to be numerically best method (among  the four).

Fig. \ref{fig:pps_zne_main} shows the Pauli-path ZNE method applied to the observable $(ZIIZ)_{z=2}$, i.e., the average of distance-3 $ZIIZ$-correlators for qubit pairs with coordination number $(2, 2)$ on the 51-qubit patch (there are 36 such pairs). The weight-resolved extrapolations track the reference markedly better than the aggregate fits at late times. (Running the $\eta=0$ simulation with $W_0 = 15$ and $\delta_{rel} = 8 \times 10^{-6}$ was not feasible, as the memory requirement is beyond the limit of an Nvidia H100 GPU.)

\begin{figure}[!t]
    \centering
    \includegraphics[width=1.0\linewidth]{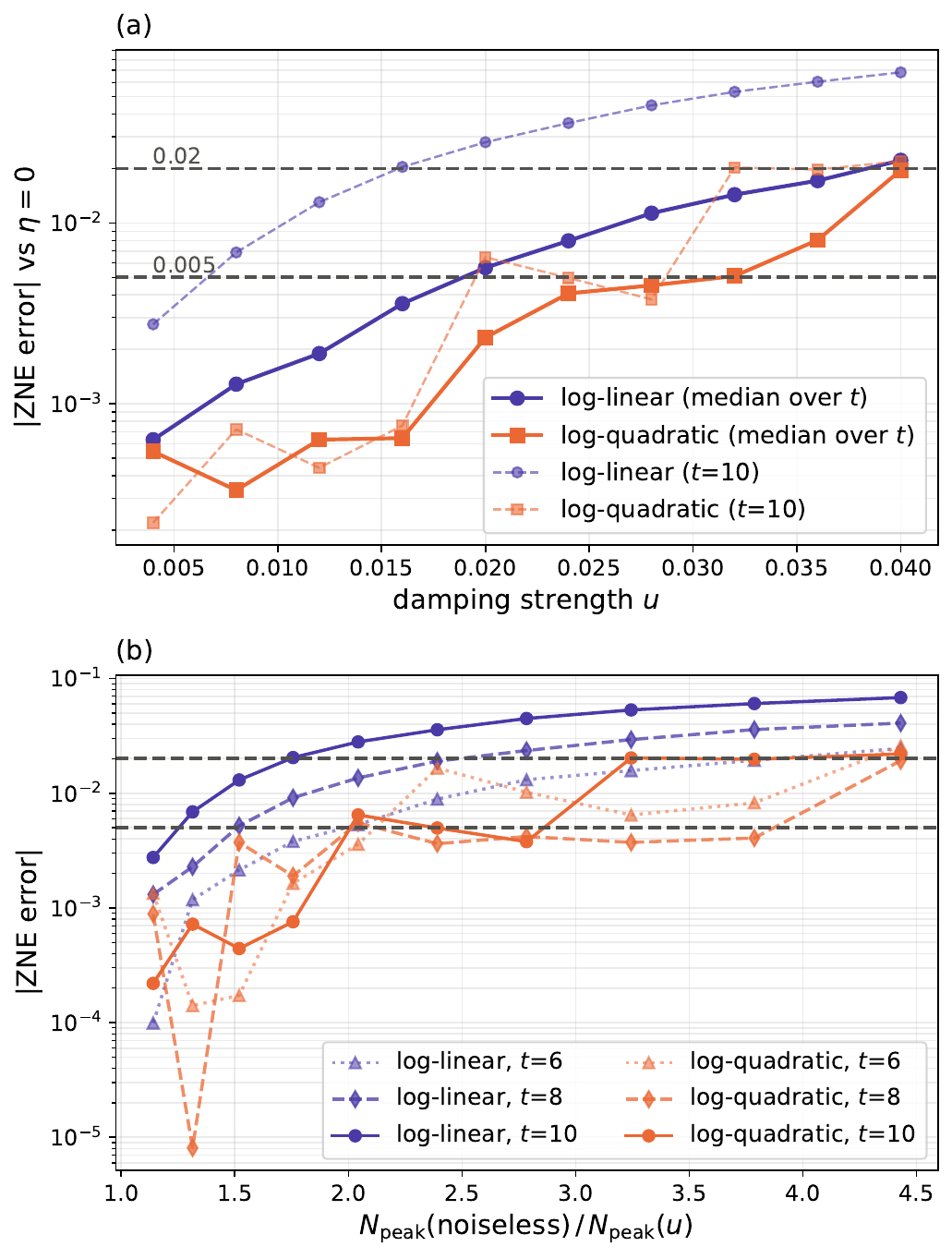}
    \caption{Memory advantage of Pauli-path ZNE along the depolarizing direction ($\langle Z_{\mathrm{avg}} \rangle$, 51-qubit patch, $T=10$ Trotter steps, relative truncation $\delta_{\mathrm{rel}} = 2\times10^{-5}$, no weight cutoff). The noise ladder $u \in \{0.004, \dots, 0.048\}$, $\Delta u = 0.004$, is scanned with a sliding three-point window: the family $\{u,\, u{+}\Delta u,\, u{+}2\Delta u\}$ is extrapolated to zero noise, log-linearly and log-quadratically, and compared against the measured $\eta = 0$ baseline.(a)~Absolute ZNE error versus the damping strength $u$ of the family's least-damped run; solid lines show the median over $t \le 10$, dashed lines the final step $t = 10$. (b)~The same errors at $t = 6, 8, 10$ versus the family's memory advantage
    $N_{\mathrm{peak}} (\eta{=}0)/N_{\mathrm{peak}}(u)$ with increasing $u$; the peak string count is attained by $t \le 6$ in every run, so the advantage is fixed before the plotted steps. Horizontal dashed lines mark absolute tolerances $0.005$ and $0.02$.
    }
    \label{fig:pps_zne_memory_advantage}
\end{figure}

Another aspect of our study stems from the observation that nothing in this method -- of injecting Pauli noise channels into circuits we want to simulate with SPP -- requires the engineered channel to be isotropic. We chose depolarizing noise (for experiments summarized in Fig.~\ref{fig:pps_zne_main}) for its simplicity and physical relevance. Relaxing this condition yields both a diagnostic (on which Pauli content the observable depends on) and a design freedom (which content to damp without loosing much in accuracy).

In Fig.~\ref{fig:pps_zne_probe} we show results from running Pauli-path ZNE for $Z_{\mathrm{avg}}$ along different noise directions $\eta = (p_x, p_y, p_z)$. We sweep six directions sharing one noiseless ($\eta = 0$) baseline: four physical ones  -- isotropic (depolarizing) noise and the three pure-rate channels, which by CPTP constraint damp \emph{pairs} of axes (pure-$p_x$ damps $Y,Z$; pure-$p_y$ damps $X,Z$; pure-$p_z$ damps $X,Y$) -- and two engineered single-axis directions, $Y$-only and $Z$-only. The latter correspond to no CPTP channel, but the simulator's noise step is simply a diagonal rescaling of stored coefficients, so it readily implements direct-$\lambda$ damping for arbitrary $\lambda = (\lambda_X, \lambda_Y, \lambda_Z)$.   

The left panel in Fig.~\ref{fig:pps_zne_probe} quantifies each direction's bias, which measures how much of the evolved observable's support lies on the damped axis. For $Z_{\mathrm{avg}}$, dominated by $Z$-content, pure-$p_z$ is the least physically biased direction, and $Y$-only the least biased of all, and $Z$-only the most biased single axis. Moreover, as a consistency check, we verified that the bias of pure-$p_x$ (damping $Y$ and $Z$) agrees with the sum of the $Y$-only and $Z$-only biases to less than a percent. The middle panel shows the growth of peak Pauli strings, $N_{\mathrm{peak}}$, as damping strength approaches $u = 0$. We see that the isotropic direction provides the maximum memory advantage. The right panel shows the accuracy of the per-direction three-point ZNE -- at each step $t$ the three noisy runs are fit log-linearly in $u$ and extrapolated to $u=0$ --  against the measured $\eta = 0$ baseline; all six families extrapolate to within a few times $10^{-3}$ of the baseline at every step.

Once a noise direction is fixed, a separate question is how far the noise can be dialed up along that direction before extrapolation fails -- that is, how large a memory advantage Pauli-path ZNE can actually provide. If Fig.~\ref{fig:pps_zne_memory_advantage} we extend the depolarizing noise ladder of Fig.~\ref{fig:pps_zne_probe} to $u = 0.048$ in uniform steps $\Delta u = 0.004$. We use a three-point sliding method to extrapolate to $u=0$, both log-linearly and log-quadratically. We find that dialing up the depolarizing noise buys a 3-4x memory reduction at percent-level absolute accuracy.    

The directional probe thus provides a soft-truncation lever: one chooses the direction that buys the most memory per unit of bias. The probe itself is cheap --- a dozen or so short runs --- and is rerun per truncation policy. 

A more systematic study of general Pauli noise channels acting on the space of Pauli paths may yield further tailored truncation strategies. We leave this to future work.

\section{Tensor-Network Simulations}
\label{app.tn}

This appendix summarizes the tensor-network methods used in Secs.~\ref{sec.results} and \ref{sec.classical}. We begin with the belief-propagation projected entangled pair state (PEPS-BP) method for Schr\"odinger-picture simulations \cite{alkabetz2021tensor,tindall2023gauging}, followed by the corresponding projected entangled pair operator (PEPO-BP) method for Heisenberg-picture evolution \cite{begusic2023fastconverged}. Next, we analyze the accuracy, convergence, and computational complexity of these methods, and conclude with a description of the Heuristic-Corrected TEBD-based extrapolation method, along with an analysis of its convergence.

\subsection {PEPS with Belief Propagation (PEPS-BP)}
\label{app.peps_bp}
We simulate the circuit dynamics considered in this work using
projected entangled pair states (PEPS)
\cite{verstraete2004renormalization,orus2014tensor}, which naturally
preserve the heavy-hex lattice geometry, evolved and contracted
using the Belief-Propagation (BP) algorithm
\cite{alkabetz2021tensor,tindall2023gauging}. Belief propagation is
a well-known message-passing algorithm from the fields of statistical
mechanics, classical error correction, and graphical models, which
can be used to find the marginal of a multivariate distribution
\cite{pearl1982proceedings,wainwright2008graphical,koller2009probabilistic,mezard2009information}. Recently, it has been adapted to the field of
tensor-networks~\cite{alkabetz2021tensor, robeva2019duality}, where it is
used as a fast and approximate method for contracting tensor-networks with very large bond dimensions. This, in turn, has
made PEPS-BP a powerful method for simulating real-time quantum
dynamics, where large bond dimension is necessary to capture the
high entanglement that is generated in such dynamics \cite{tindall2024efficient, begusic2023fastconverged,haghshenas2026digitalmagnetism, tindall2026dynamics}.
\par
In the PEPS-BP algorithm, we apply BP to the double-layered
tensor-network that describes $\braket{\psi}{\psi}$. The algorithm
passes messages between neighbouring nodes, which then converge to a
fixed point. From the converged messages, one can either construct
an approximation for the local environments in the network, which is
exact on tree tensor networks and approximate on loopy geometries
such as the heavy-hex lattice. Such environments can then be used to
calculate the expectation value of local observables. In addition,
the converged messages can be used to transform the TN to the
so-called ``Vidal gauge'', which defines for every bond a set of
``Vidal weights'' $\lambda_1\ge \lambda_2 \ge \lambda_3 \ge \ldots$.
When the underlying tensor network has topology of a tree, these
weights are exactly the Schmidt values of the corresponding Schmidt
decomposition. In such case, one can use the weights to perform an
optimal truncation of the TN bond dimension by keeping the $D$
largest weights and discarding the rest. When the underlying TN has
loops, such truncation is no longer optimal, but can still yield
excellent results, and this forms the basis for using BP for TN
truncation. 

As a heuristic proxy for the truncation quality, we
follow~\cite{tindall2023gauging} and define a local error
for truncation step $k$ by
\begin{equation}
  \epsilon_k = \frac{\sum_{i>D_{\max}}\lambda_i^2}
    {\sum_i \lambda_i^2},
    \label{eq:peps_err}
\end{equation}
This quantity measures the fraction of the Vidal weights discarded
during the truncation. The cumulative truncation error after $K$
truncation steps is then defined as
\begin{equation}
\label{eq:err_trunc}
  \epsilon_{\mathrm{tot}} = \sum_{k=1}^{K}\epsilon_k.
\end{equation}
We stress that $\epsilon_{\mathrm{tot}}$ is only a proxy for the actual truncation error in the PEPS-BP simulation. However, by analyzing small-scale simulations, where full statevector results are available, we find that it is an excellent approximation.
In addition for the truncation error, one can define a proxy for the
fidelity between PEPS-BP state and the exact state.  Specifically,
the fidelity between the state $\ket{\psi_k}$ before the $k$'th
truncation and the state $\ket{\phi_k}$ after the truncation can be
approximated by the retained Vidal weight,
\begin{equation} 
  f_k = |\langle\phi_k|\psi_k\rangle|^2 \approx
    \frac{\sum_{i\le D_{\max}}\lambda_i^2} {\sum_i\lambda_i^2}
    = 1-\epsilon_k.
\end{equation}
Within this approximation, the accumulated fidelity after $K$
truncation steps is given by
\begin{equation}
  F_K \approx \prod_{k=1}^{K} f_k.
  \label{eq:fidelity_peps}
\end{equation}

\subsection{PEPO with Belief Propagation (PEPO-BP)}
\label{app.pepo_bp}
In addition to evolving quantum states in the Schrödinger picture,
we also evolve the studied observables directly in the Heisenberg
picture using projected entangled pair operators (PEPOs), the
operator analogue of PEPS
\cite{liao2023simulation,begusic2023fastconverged}. Here, we focus
on the Liouville (vectorized) representation, where we represent the PEPO in the local Pauli basis, so that each physical index corresponds to one of the four single-qubit Pauli operators ${I,X,Y,Z}$. Starting from an initial observable $\Phi_0$, applying one Floquet
cycle evolves it to
\begin{equation}
\Phi_1 = U^\dagger \Phi_0 U,
\end{equation}
and after $N_c$ Floquet cycles,
\begin{equation}
\Phi_{N_c}=\left(U^\dagger\right)^{N_c}\Phi_0 U^{N_c}.
\end{equation}
The expectation value is then obtained by contracting the evolved
PEPO with the initial product-state $\rho_0 =
|000\cdots 0\rangle \langle 000\cdots 0|$
\begin{equation}
  \langle \Phi_{N_c} \rangle = \operatorname{Tr}\!\left(\rho_0 \,
    \Phi_{N_c}\right).
\end{equation}
As in the Schr\"odinger-picture PEPS evolution, applying two-qubit
gates increases the PEPO bond dimensions. We therefore use the same belief-propagation (BP) truncation scheme to compress the PEPO after each gate application, restricting the virtual bond dimension to a prescribed maximum value $D$. 

To evolve the total magnetization
\begin{equation}
  M=\frac{1}{N_q}\sum_{i\in V} Z_i,
\end{equation}
where $V$ denotes the set of all qubits (vertices) of the heavy-hex
graph, or a magnetization restricted to a subset $S\subseteq V$
(such as the coordination-two magnetization $M_{z=2}$ considered in
this work), we first construct the corresponding operator
\begin{equation}
  Z_S=\sum_{v\in S} Z_v
\end{equation}
as a PEPO in the vectorized Pauli basis. Such operators admit a PEPO representation with bond dimension $\sim2$. Here we construct this representation by choosing a rooted spanning tree of the underlying graph and using it to define the local PEPO tensors via
walks on the tree from the root to every node in $S$.

\subsection{PEPS-BP and PEPO-BP Accuracy and Computational Complexity}
Having introduced the PEPS-BP and PEPO-BP methods, we now benchmark their reliability and discuss their computational cost: we first assess the accuracy of the belief-propagation approximation on a smaller system for which exact statevector simulation remains feasible, then examine convergence with bond dimension, compare Schr\"odinger-picture PEPS evolution with Heisenberg-picture PEPO evolution, and finally discuss the associated computational-resource requirements.
\begin{figure}
\centering
\includegraphics[width=0.8\linewidth]{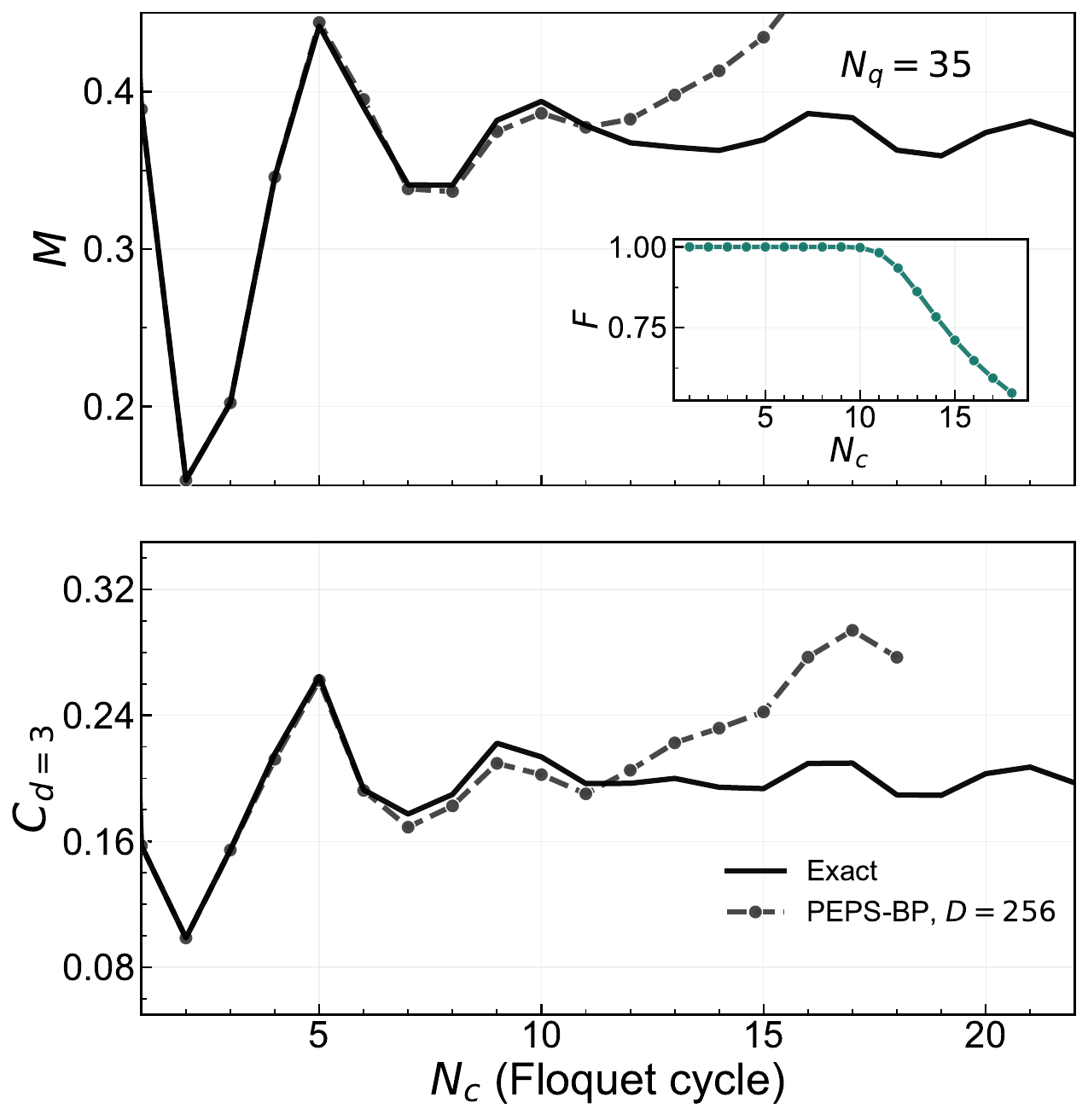}
\caption{Magnetization, $M$, (top) and the next-next-nearest-neighbour two-point correlation, $C_{d=3}$, (bottom) as a function of Floquet cycle number for the $N_q=35$ qubit patch (Fig.~\ref{fig:small_systems}), comparing PEPS-BP with the exact evolution to assess the accuracy of the belief propagation approximation. 
The inset in the top panel shows the PEPS simulation fidelity as a function of Floquet cycle.}
\label{fig:35q_peps_app}
\end{figure}

\begin{figure}[tb]
\centering
\begin{minipage}[t]{\linewidth}
\centering
\subfloat[]{
    \includegraphics[width=0.8\linewidth]{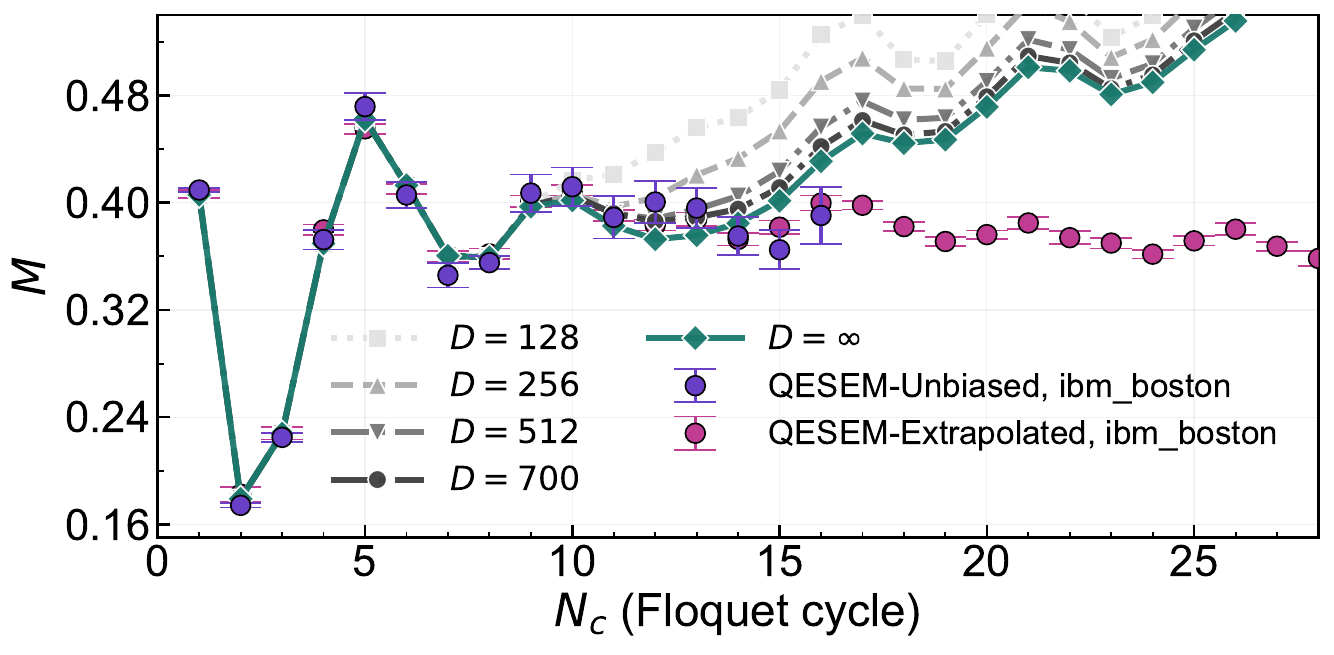}
    \label{fig:extrapolation_infinite_chi_plot_a}
}
\end{minipage}\\[0.5em]
\begin{minipage}[t]{\linewidth}
\centering
\subfloat[]{
    \includegraphics[width=0.8\linewidth]{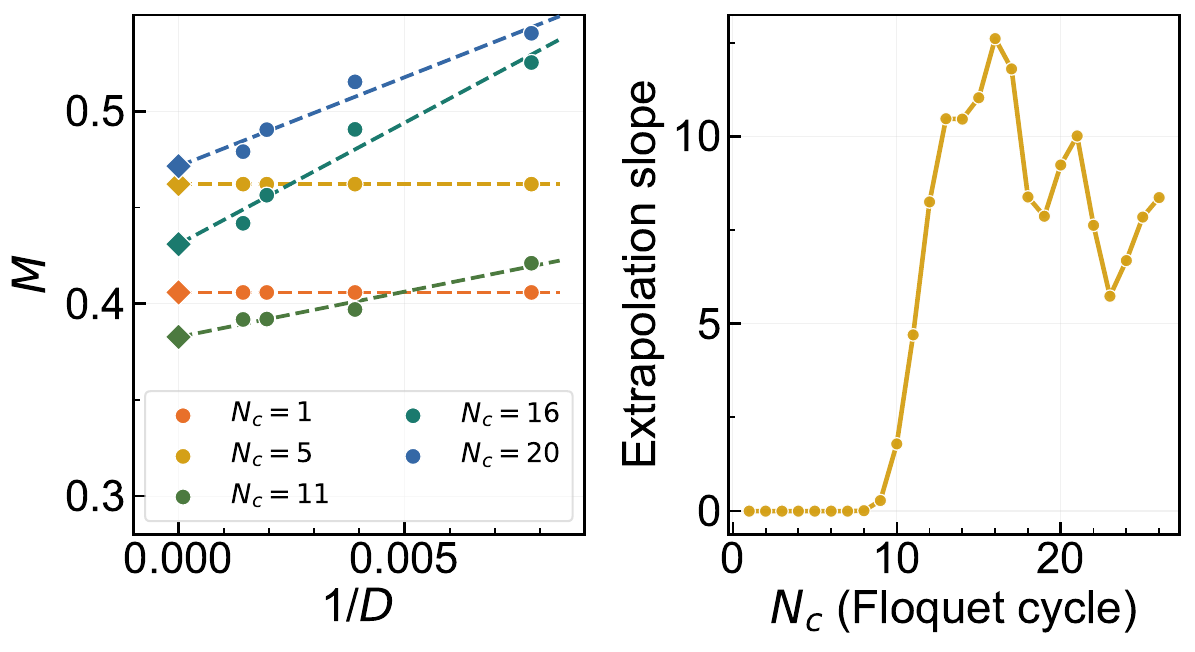}
    \label{fig:extrapolation_infinite_chi_fits_b}
}
\end{minipage}
\caption{
\textbf{\protect\subref{fig:extrapolation_infinite_chi_plot_a}}
Magnetization dynamics obtained by extrapolating the PEPS-BP results for the $N_q=51$-qubit system to the $D\rightarrow\infty$ limit using a linear fit in $1/D$ at each Floquet cycle.
\textbf{\protect\subref{fig:extrapolation_infinite_chi_fits_b}}
Top: Magnetization as a function of $1/D$ for selected Floquet cycles, illustrating the linear extrapolation used to obtain the $D\rightarrow\infty$ estimates. Bottom: Slopes of the linear fits as a function of Floquet cycle. Larger slopes indicate a stronger dependence on bond dimension and, consequently, poorer convergence of the PEPS-BP simulations at the largest $D$ values considered.
}
\label{fig:extrapolation_infinite_chi_fits}
\end{figure}
We begin by assessing the accuracy of the PEPS-BP approximation on a smaller system, without including loop-correction schemes as those presented in \cite{midha2026belief,park2025simulating,evenbly2026loop,midha2026beyond,gray2026tensor}. Fig.~\ref{fig:35q_peps_app} compares the magnetization dynamics obtained from PEPS-BP with bond dimension $D=256$ against exact statevector simulations of the $N_q=35$ heavy-hex patch, which contains four loops (Fig.~\ref{fig:small_systems}). Although beyond a certain Floquet cycle the PEPS-BP results cease to agree with the exact results and the fidelity begins to decrease, the magnetization remains in excellent agreement throughout the time window in which the fidelity is high. This suggests that, at least for this observable, neglecting loop corrections introduces only small errors compared to the compression errors due to the finite bond dimension over the relevant simulation regime. More generally, observables involving longer-range correlations, such as the other studied two-point correlation functions in this work, presented in Sec.~\ref{app.correlation}, are expected to be more sensitive to the lack of loop corrections in our BP-based contraction. 

Having established that the BP approximation itself captures the magnetization dynamics to a good approximation while the PEPS remains well converged, we next investigate whether an extrapolation to infinite bond dimension can systematically recover the exact dynamics at later Floquet cycles.
Fig.~\ref{fig:extrapolation_infinite_chi_fits} shows a simple linear extrapolation performed independently at each Floquet cycle by fitting the magnetization as a function of $1/D$, and then evaluating the resulting fit in the limit $1/D \rightarrow 0$. As shown in Fig.~\ref{fig:extrapolation_infinite_chi_plot_a}, the extrapolated dynamics exhibits the same unphysical upward drift from the expected prethermal value of the magnetization, remaining inconsistent with both physical expectations and quantum results. The quality of the linear fit deteriorates significantly at later Floquet cycles, as can be seen in Fig.~\ref{fig:extrapolation_infinite_chi_fits_b}, where the fitted slopes increase substantially and the linear fit becomes poorer. Together, these observations indicate that a straightforward linear extrapolation in $1/D$ for the large Floquet cycle dynamics results does not provide a reliable estimate of the limit of infinite bond-dimension. 

Besides evolving the quantum state as a PEPS, one may alternatively evolve observables, such as the total magnetization or a local $Z$ operator, directly in the Heisenberg picture using a PEPO representation as discussed in the previous sections. While these two approaches are formally equivalent, they generate tensor networks with different entanglement structure and, consequently, different computational complexity. As shown in Fig.~\ref{fig:param_scan_circ}, the Floquet circuit studied in this work generates relatively large operator entanglement, making PEPO-BP simulations challenging similar to how the relatively high state entanglement makes PEPS-BP challenging. The comparison between the PEPO-BP and PEPS-BP simulations presented in the main text, together with their cumulative truncation errors (Fig.~\ref{fig:classical_simulation_methods_vs_quantum}), shows that this is indeed the case, and that evolving an operator as a PEPO is substantially more challenging than evolving a quantum state as a PEPS. To further understand the difference, Fig.~\ref{fig:peps_pepo_comparison_appendix} compares the two simulations, for the $N_q=51$-qubit system where here evolving the central-qubit $Z$ operator, rather than the coordination-two magnetization considered in the main text with the same bond dimension $D=512$, and on the same CPU hardware.

In Fig.~\ref{fig:vidal_entropy_app}, we compare the maximum
Vidal bond entropy of the two simulations. The Vidal bond entropy of
an edge is the entropy of the corresponding Vidal weights that are
obtained after passing to the Vidal gauge. As mentioned in
Sec.~\ref{app.peps_bp}, when the underlying TN has the topology of a
tree, the Vidal weights are exactly the Schmidt coefficients of the
corresponding Schmidt decomposition, defined by partitioning the
system along that edge. Therefore, in such cases the von Neumann
Vidal bond entropy coincides with the entanglement entropy. In more
general cases, we can still view this entropy as a proxy for
how much entanglement is carried along that PEPS bond. Very loosely
speaking, once the Vidal bond entropy approaches $\log D$, the
simulation is expected to suffer from truncation errors. We see in
Fig.~\ref{fig:vidal_entropy_app} that the second-R\'enyi Vidal bond
entropy grows significantly faster for the PEPO than for the PEPS.
The PEPO reaches a peak around $N_c \approx 10$, after which it
begins to decrease. In contrast, the PEPS develops substantially
smaller bond entanglement throughout the evolution, reaching a
maximum only around $N_c\approx 14$. This demonstrates that
Heisenberg-picture operator evolution here generates considerably
richer virtual-bond structure than Schr\"odinger-picture state
evolution at the same bond dimension. The increased bond entanglement of the PEPO-BP simulation with
respect to the PEPS-BP simulation is also reflected in the proxy BP
truncation error and fidelity, which were introduced in
Sec.~\ref{app.peps_bp},
Eqs.~(\ref{eq:err_trunc},\ref{eq:fidelity_peps}). As shown in
Figs.~\ref{fig:cumulative_error_app}, \ref{fig:fidelity_app}, the
truncation error is consistently larger for the PEPO evolution than
for the PEPS evolution, while the drop in the fidelity happens much
quicker in the PEPO-BP simulation. Both quantities closely follow
the increase in the Vidal bond entropy in
Fig.~\ref{fig:vidal_entropy_app}, and indicate that the broader
virtual-bond spectra of the PEPO makes tensor compression
substantially more challenging.

Finally, in Fig.~\ref{fig:peak_ram_app} we compare the global
memory resources of the two simulations. The larger 
complexity of the PEPO is also reflected here, as the running peak RAM reaches
approximately twice that required for the PEPS evolution. Likewise,
after the initial transient, the runtime per Floquet cycle of the
PEPO evolution is approximately a factor of two larger than that of
the corresponding PEPS evolution.
\par
\begin{figure}[tb]
\centering

\subfloat[]{
\includegraphics[width=0.49\linewidth]{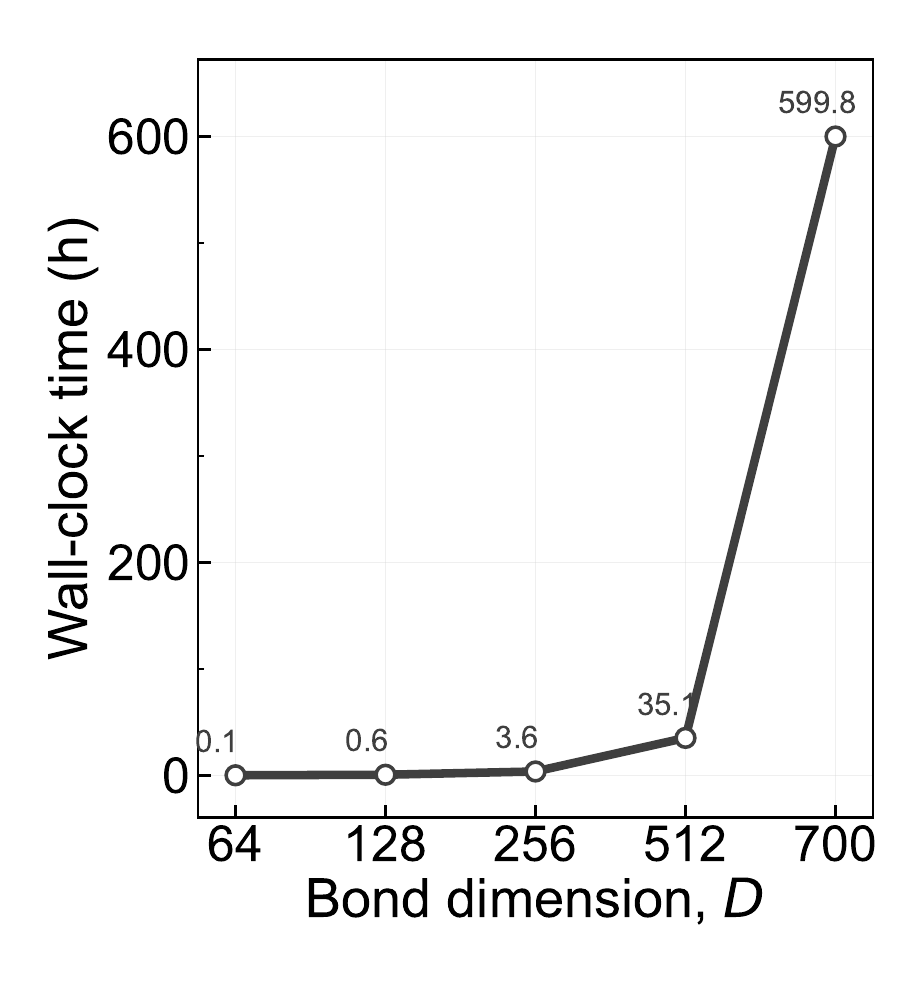}
\label{fig:walltime_scaling_app}
}
\subfloat[]{
\includegraphics[width=0.49\linewidth]{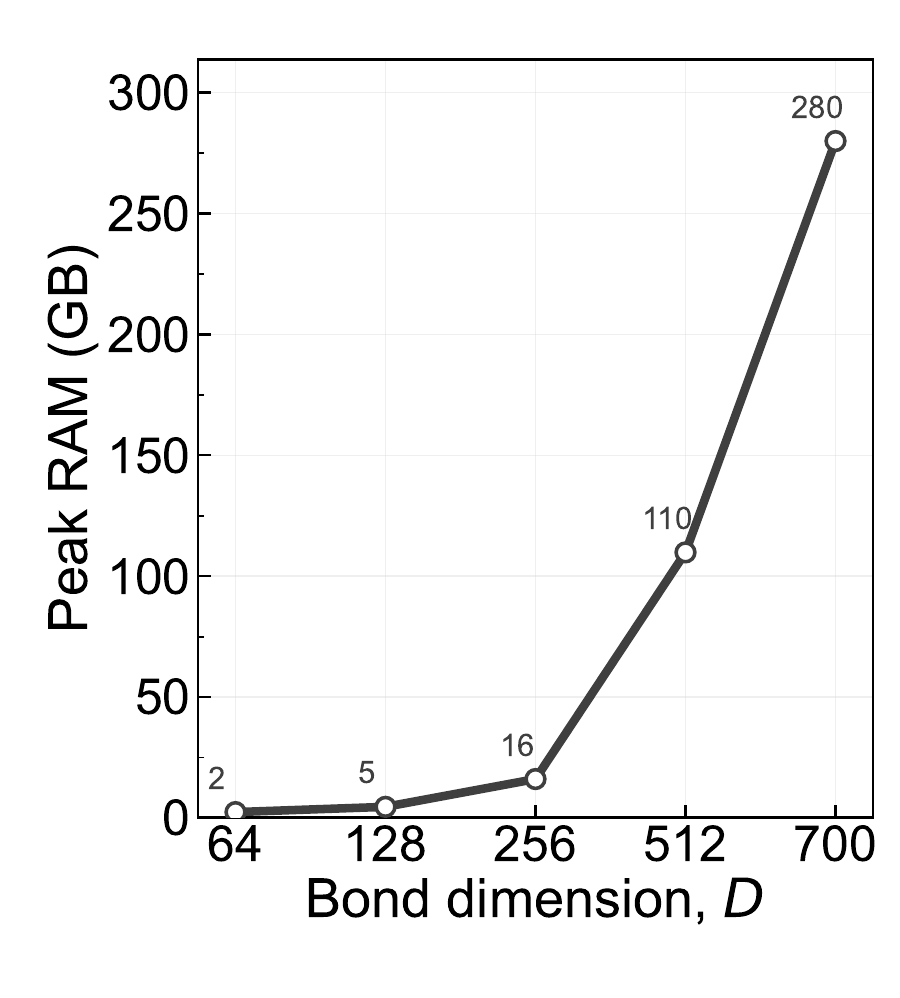}
\label{fig:ram_scaling_app}
}

\caption{
Computational resources required for the PEPS-BP simulations presented in Fig.~\ref{fig:classical_simulation_methods_vs_quantum}, as a function of bond dimension, \(D\).
\textbf{\protect\subref{fig:walltime_scaling_app}} Total wall-clock time.
\textbf{\protect\subref{fig:ram_scaling_app}} Peak CPU RAM usage.
}
\label{fig:peps_resource_scaling1}
\end{figure}

\begin{figure}[tb]
\centering

\begin{minipage}[t]{0.49\linewidth}
\centering
\subfloat[]{
    \includegraphics[width=\linewidth]
    {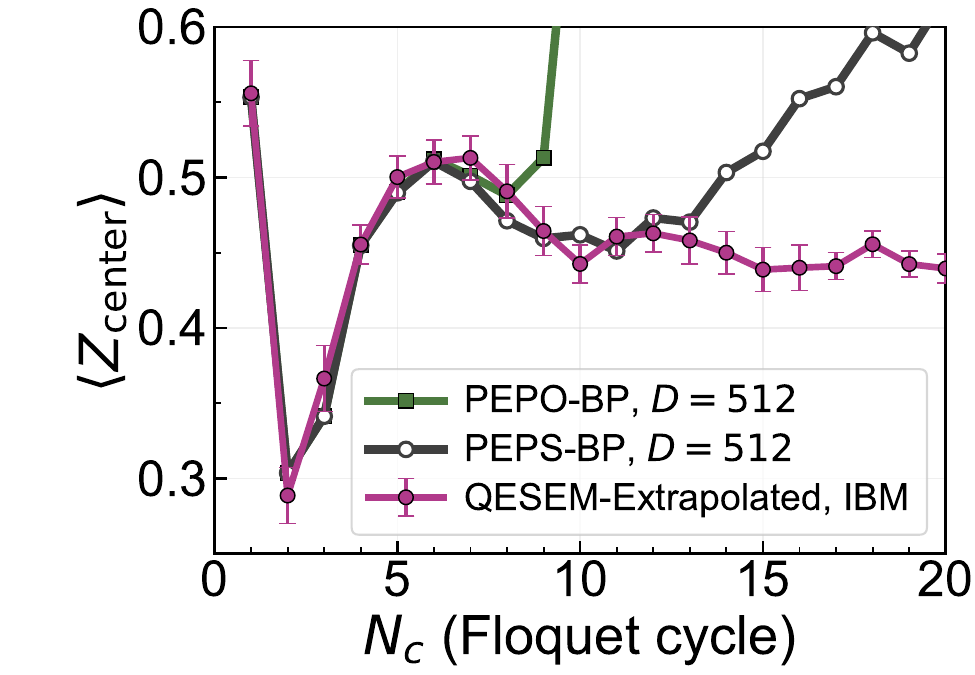}
    \label{fig:z23_app}
}
\end{minipage}
\hfill
\begin{minipage}[t]{0.49\linewidth}
\centering
\subfloat[]{
    \includegraphics[width=\linewidth]
    {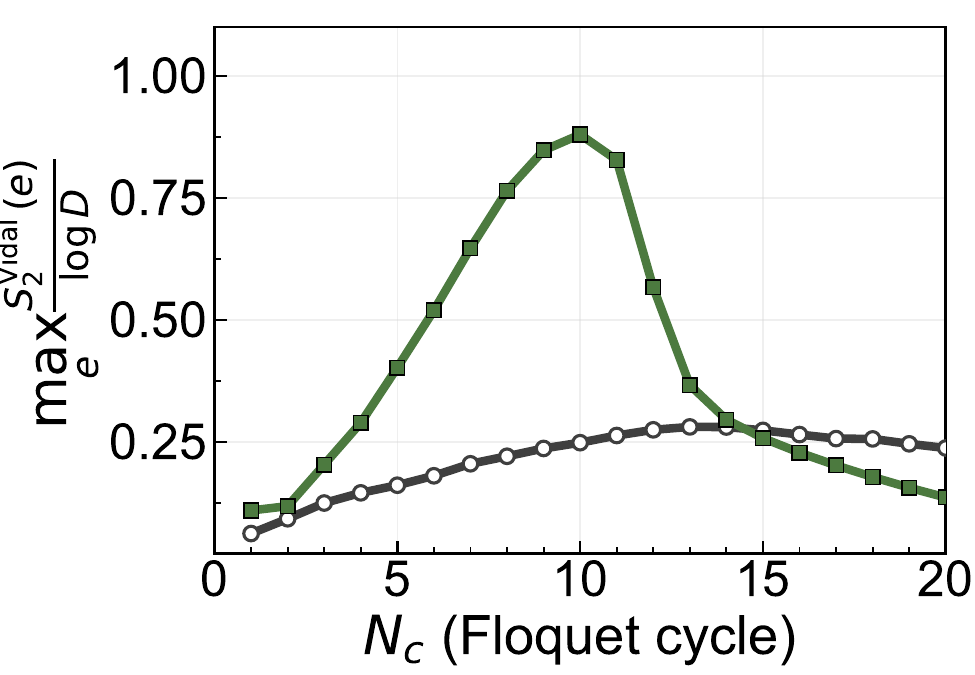}
    \label{fig:vidal_entropy_app}
}
\end{minipage}

\vspace{0.3em}

\begin{minipage}[t]{0.49\linewidth}
\centering
\subfloat[]{
    \includegraphics[width=\linewidth]
    {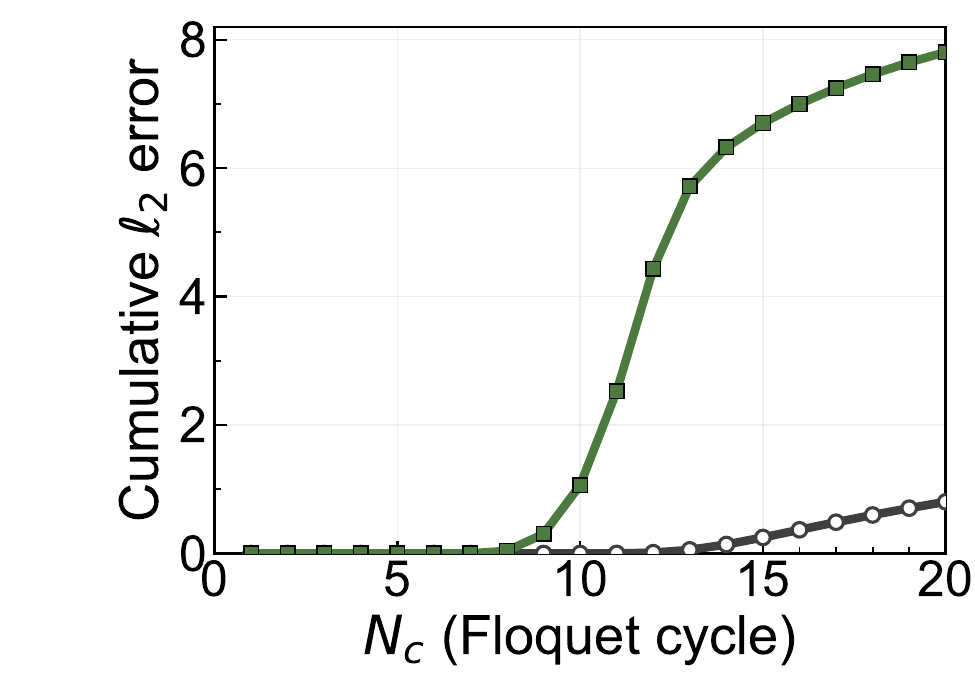}
    \label{fig:cumulative_error_app}
}
\end{minipage}
\hfill
\begin{minipage}[t]{0.49\linewidth}
\centering
\subfloat[]{
    \includegraphics[width=\linewidth]
    {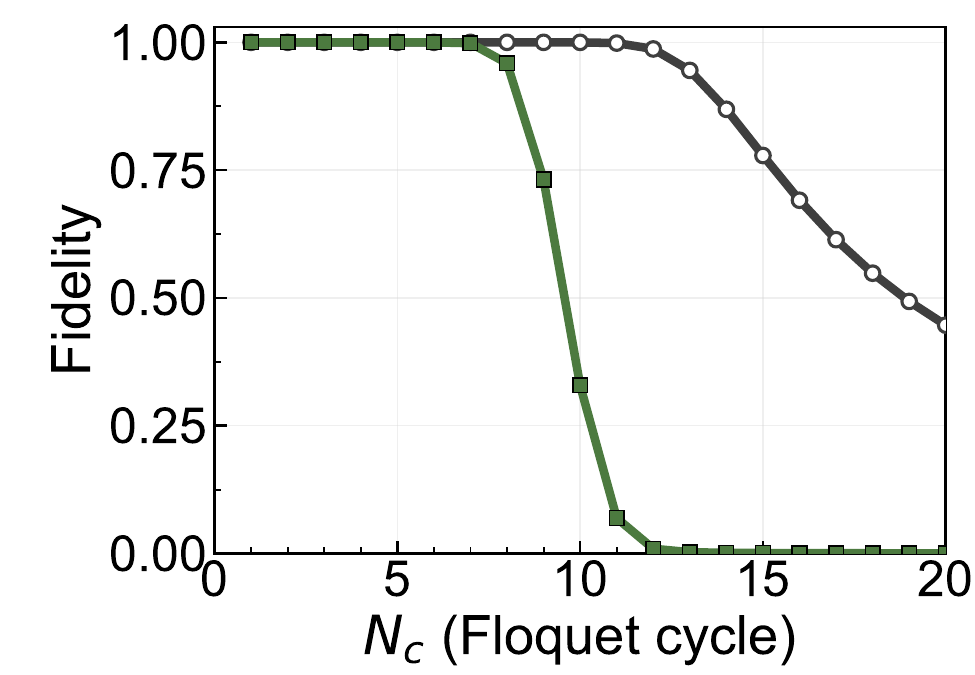}
    \label{fig:fidelity_app}
}
\end{minipage}

\vspace{0.3em}

\begin{minipage}[t]{0.62\linewidth}
\centering
\subfloat[]{
    \includegraphics[width=\linewidth]
    {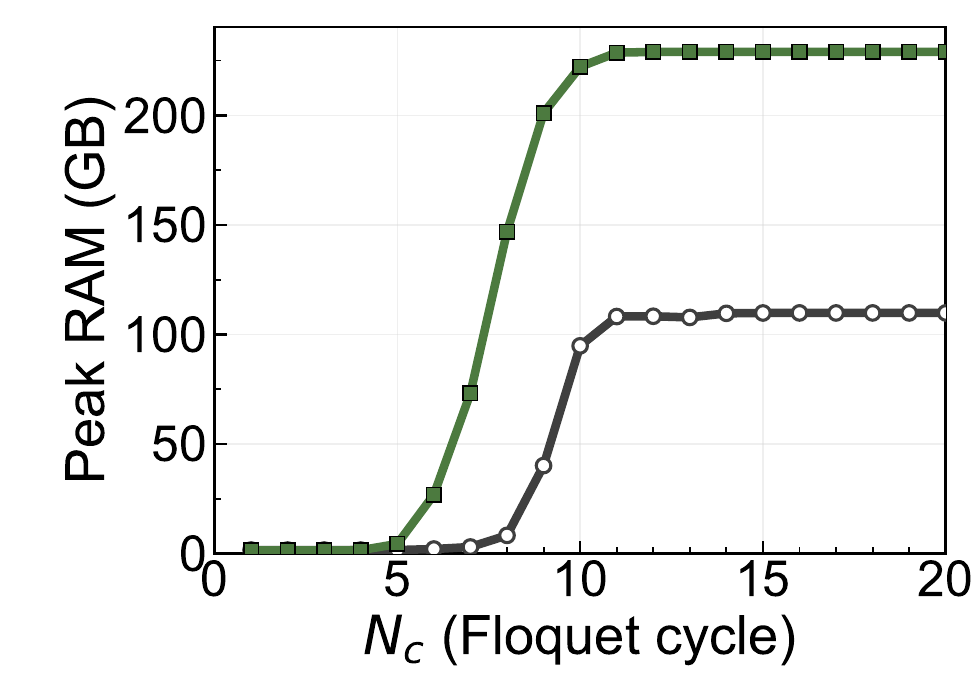}
    \label{fig:peak_ram_app}
}
\end{minipage}

\caption{
Comparison between PEPS-BP and PEPO-BP simulations for the evolution of the central-$Z$ operator at the same bond dimension, $D=512$.
\textbf{\protect\subref{fig:z23_app}}
Expectation value of the central-$Z$ operator as a function of Floquet cycle, comparing PEPO-BP, PEPS-BP, and the error-mitigated QESEM-Extrapolated results. The PEPO-BP simulation loses convergence earlier than the corresponding PEPS-BP simulation.
\textbf{\protect\subref{fig:vidal_entropy_app}}
Maximum normalized Vidal bond second-R\'enyi entropy,
$\max_e S^{\mathrm{Vidal}}_2(e)/\log D$.
\textbf{\protect\subref{fig:cumulative_error_app}}
Cumulative $\ell_2$ truncation error.
\textbf{\protect\subref{fig:fidelity_app}}
Simulation fidelity.
Panels \protect\subref{fig:vidal_entropy_app}--\protect\subref{fig:fidelity_app} show the faster growth of virtual-bond complexity and truncation errors, together with the earlier loss of fidelity, during PEPO-BP evolution.
\textbf{\protect\subref{fig:peak_ram_app}}
Running peak RAM usage, illustrating the larger memory requirements of PEPO-BP relative to PEPS-BP at the same bond dimension.
}
\label{fig:peps_pepo_comparison_appendix}
\end{figure}
The largest PEPS-BP and PEPO-BP simulations presented in this section and Sec.~\ref{sec.classical} were performed on a machine equipped with $128$ vCPUs (AMD EPYC 7R13), utilizing $90$ vCPUs for each simulation and an allocated RAM of $334$ GiB. All simulations were carried out using the \texttt{TensorNetworkQuantumSimulator.jl} package \cite{Tindall2025TensorNetworkQuantumSimulator}, where each simulation included both the tensor-network evolution under the Floquet gates and the subsequent evaluation of all observables and diagnostic quantities reported in this work. Fig.~\ref{fig:ram_scaling_app} summarizes the corresponding computational resources, including the peak RAM usage and wall-clock time as a function of the bond dimension. The peak RAM scales approximately as  $O(D^3)$. While all simulations used the same number of CPU threads, the average thread utilization decreases for the largest bond dimensions, indicating reduced parallel efficiency. Consequently, the measured wall-clock time exhibits a steeper increase at the largest bond dimensions, making straightforward extrapolation of the runtime to larger bond dimensions unreliable. It is worth noting that the simulations considered here could considerably benefit from GPU acceleration, however, these calculations exceeded the memory of a single GPU requiring more than $200\,\mathrm{GB}$ of memory, making it challenging to run on currently accessible single-GPU systems. Fig.~\ref{fig:peps_resource_scaling1} shows the corresponding gate-application runtimes measured on a single H100 GPU for the smaller bond dimensions.

\subsection{Matrix Product State Simulations}
\label{app.mps}

The Heuristic-Corrected TEBD method, discussed in Sec.~\ref{sec.classical}, uses a one-dimensional ordering of the heavy-hex circuit together with time-evolving block decimation (TEBD). Here we describe the implementation details together with the truncation diagnostics and computational resources.

\begin{figure}[t]
\centering
\includegraphics[width=\linewidth]{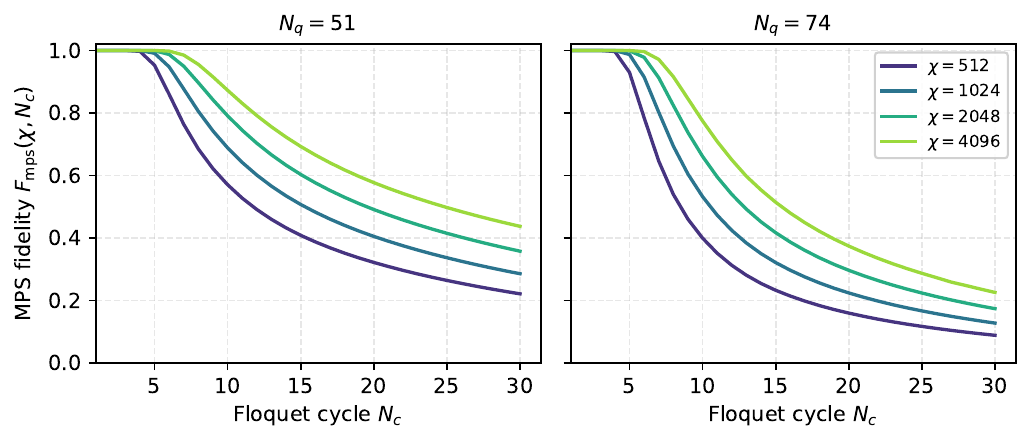}
\caption{
    Raw MPS fidelity $F_{\mathrm{mps}}(\chi,N_c)$ (Eq.~\eqref{eq:mps_fidelity}) as a function of Floquet cycle and bond dimensions $\chi\in\{512,1024,2048,4096\}$ and system sizes $N_q=51$ and $74$.
    }
\label{fig:mps_fidelity}
\end{figure}

We order the qubits according to their physical-qubit index on the device, yielding a snake-like path through the two-dimensional lattice. 
Each truncation step $k$, after the application of a physical gate or a SWAP operation, incurs a truncation error $\epsilon_k$ and an associated gate fidelity $f_k=1-\epsilon_k$, analogous to the PEPS case (Eq.~(\ref{eq:peps_err}) and Eq.~(\ref{eq:fidelity_peps})) with $D\!\rightarrow\!\chi$, but here evaluated exactly rather than approximately.
Leaving the state unnormalized, its squared norm accumulates these gate fidelities into the cumulative fidelity after $N_c$ Floquet cycles,
\begin{equation}
    F_{\mathrm{mps}}(\chi,N_c) = \prod_{k} f_k  = \langle \tilde\psi_\chi | \tilde\psi_\chi \rangle \le 1,
    \label{eq:mps_fidelity}
\end{equation}
where $|\tilde{\psi}_k\rangle$ denotes the unnormalized MPS.
Figure~\ref{fig:mps_fidelity} shows the evolution of $F_{\mathrm{mps}}$ as a function of the Floquet cycle for $N_q=51$ and $74$.
Although $F_{\mathrm{mps}}$ quantifies the accumulated truncation error, it does not directly determine the precision of a given observable, $O$.
Expectation values are computed from the normalized state, according to
$\langle O\rangle_\chi = \langle \tilde\psi_\chi|O|\tilde\psi_\chi\rangle / \langle \tilde\psi_\chi|\tilde\psi_\chi\rangle$.
\par
\begin{figure}[tb]
\centering
\includegraphics[width=0.8\linewidth]{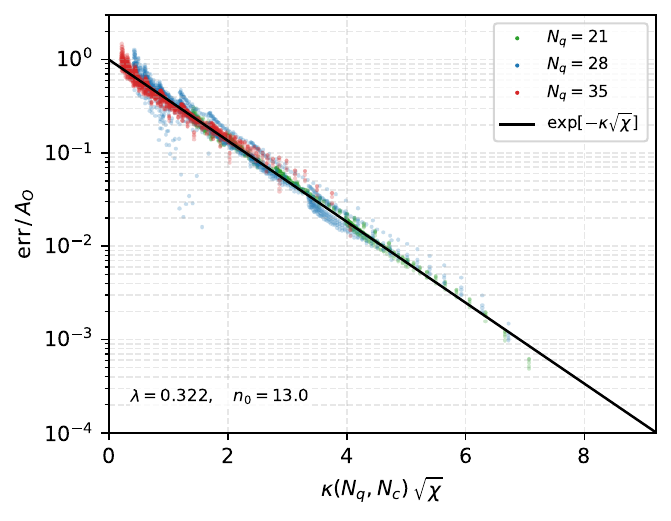}
\caption{
    Universal collapse of the empirical model, Eq.~\eqref{eq:err_model}: all $3870$ measured errors (three sizes, five
    observables, bond ladder $\chi=64$--$4096$, $N_c\ge8$), are rescaled by the fitted amplitude $A_O$ and
    shown as a function of $\kappa(N_q,N_c)\sqrt{\chi}$. The black line shows the model prediction.
    }
\label{fig:error_model}
\end{figure}

To estimate the computational resources required to simulate the magnetization dynamics studied in the main text, we next construct an empirical model for the bond dimension required to achieve a given accuracy. 
Using data obtained by exact statevector simulations of system sizes with $N_q\le35$, we find, as shown in Figure~\ref{fig:error_model}, that the relative error of a local observable
$O$ can be described using a stretched exponential in bond dimension,
\begin{equation}
    \begin{gathered}
    \mathrm{err}
    \equiv \frac{\lvert \langle O \rangle - \langle O \rangle_\chi \rvert}{\lvert\langle O \rangle\rvert}
    = A_O\, e^{-\kappa \sqrt{\chi}} \\
    \kappa = \kappa_\infty(N_q) + \tfrac{\alpha(N_q)}{N_c}.
    \end{gathered}
    \label{eq:err_model}
\end{equation}
Here, $A_O = O(1)$, is an observable-dependent amplitude $A_O = O(1)$ and $\kappa$ is an observable-independent decay rate, which relaxes with depth to a plateau $\kappa_\infty$ shrinking with size as
$\kappa_\infty = [\,\lambda\,(N_q-n_0)\,]^{-2}$.\footnote{
    Curiously, even though $n_0$ was treated as a continuous fit parameter, its fitted value lies within
    $10^{-4}$ of the integer $13$, one more than the $12$ qubits in a single heavy-hex plaquette.
}
Figure~\ref{fig:error_model} shows the resulting collapse.

\begin{figure*}[t]
\centering
\includegraphics[width=\textwidth]{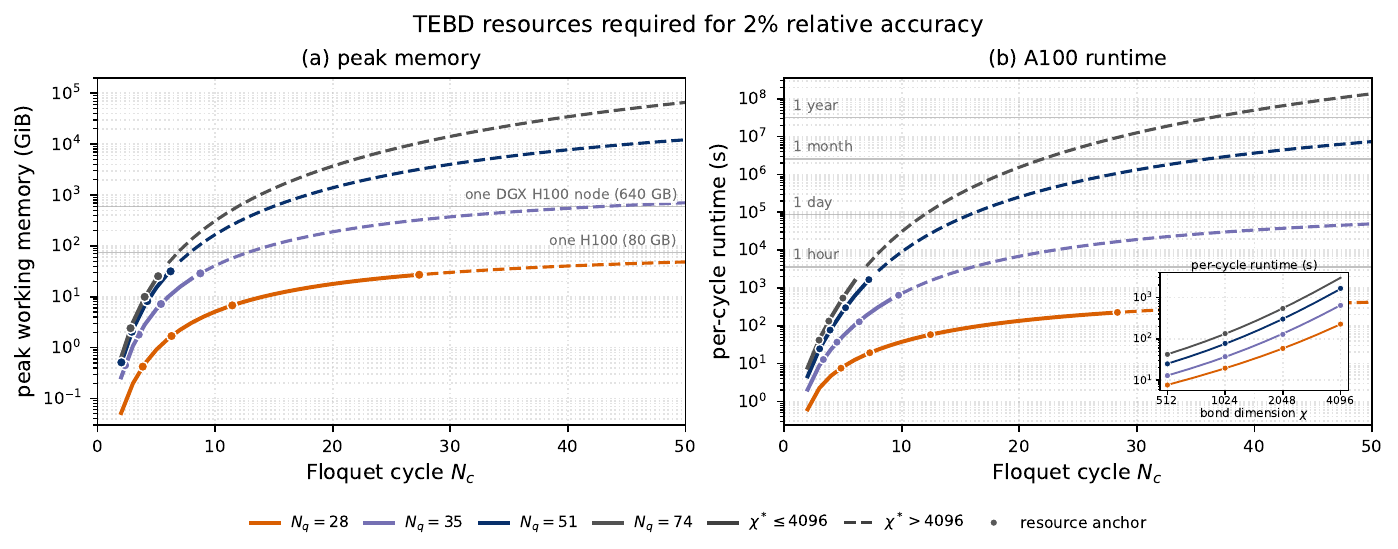}
\caption{
    TEBD resources required for $2\%$ relative accuracy in $z_{\mathrm{tot}}$.
    \textbf{(a)} Peak working memory, with gray guides marking the nominal HBM of one H100 and one
    8xH100 DGX node. \textbf{(b)} A100 per-cycle
    runtime; the inset compares the measured timings with the cubic runtime model. The main curves
    compose Eq.~\eqref{eq:err_model} with the resource models. Circles mark the resource anchors;
    memory markers without retained profiles are model values. Solid segments lie within the
    measured bond range $\chi\le4096$; dashed segments extrapolate beyond it.
    }
\label{fig:mps_resources}
\end{figure*}

Inverting Eq.~\eqref{eq:err_model} yields the bond dimension required to achieve a target relative accuracy $\varepsilon$:
\[
    \chi^\star(\varepsilon) = [\ln(A_O/\varepsilon)/\kappa]^2
\]
Combining $\chi^\star$ with measured TEBD
peak memory usage and a cubic fit to the per-cycle runtime yields an estimate to the runtime, shown in the inset
of Fig.~\ref{fig:mps_resources}(b) ($R^2=0.999$).

All MPS trajectories were generated using the \texttt{quimb} library~\cite{gray2018quimb} on
\texttt{cupy}/NVIDIA A100-80GB, in single precision with SVD cutoff $10^{-7}$. A double-precision replica (SVD cutoff $10^{-12}$) differed by at most
$2.6\times10^{-4}$ in the evaluated correlators.

\subsection{Heuristic-Corrected TEBD}
\label{app.ibe}

The Heuristic-Corrected TEBD scheme used in Sec.~\ref{sec.classical} is based on Ref.~\cite{mandra2025heuristic}. The method attempts to improve raw Matrix Product State (MPS) simulations by rescaling the expectation value of an observable using the fidelity of the MPS,
\begin{equation}
    \langle O\rangle \approx
    F_{\mathrm{MPS}}(\chi,N_c)^{\gamma(\chi,N_q)}
    \langle O\rangle_\chi,
    \label{eq:ibe}
\end{equation}
where $\langle O\rangle_\chi$ denotes the MPS estimate obtained with bond dimension $\chi$, $\langle O\rangle$ is the exact expectation value, and $\gamma(\chi,N_q)$ is an observable-dependent exponent that depends on the bond dimension and system size.

Following Ref.~\cite{mandra2025heuristic}, we calibrate $\gamma$ using system sizes for which exact statevector simulations are available. For each pair $(\chi,N_q)$, an optimal exponent $\gamma^*(\chi,N_q)$ is obtained by matching Eq.~\eqref{eq:ibe} to the exact dynamics. These values are then fitted using the empirical fitting model

\begin{figure}[tb]
\centering
\includegraphics[width=\linewidth]{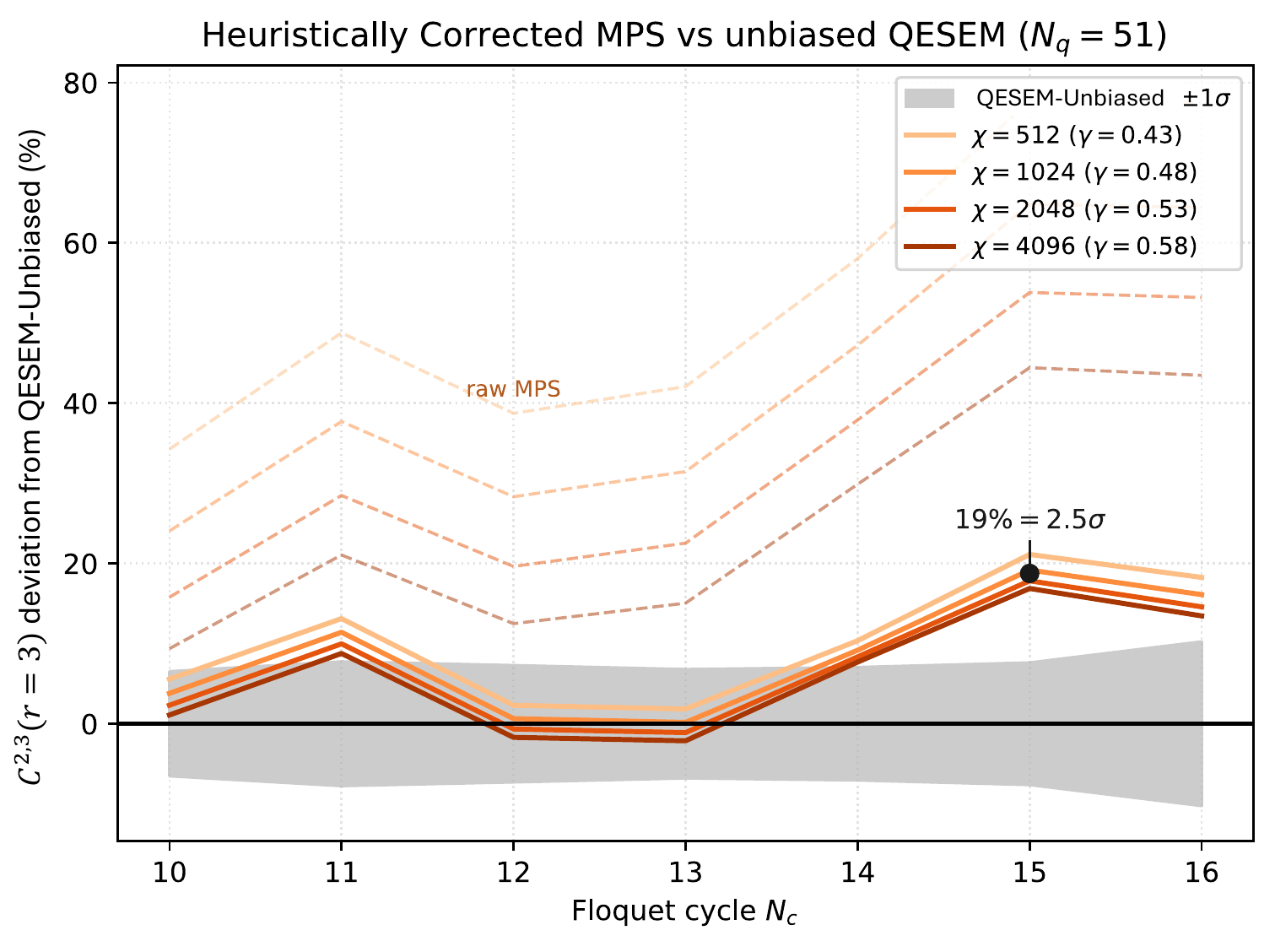}
\caption{
Heuristic-Corrected TEBD applied to the $C^{2,3}(r=3)$ two-point correlator on the $N_q=51$ heavy-hex patch, restricted to qubit pairs separated by graph distance $3$ with coordination numbers $2$ and $3$. Curves show the deviation from the QESEM-Unbiased result in percent, while the shaded band indicates its $\pm1\sigma$ uncertainty. Near Floquet cycle $15$, the spread among the corrected trajectories (solid) has narrowed to $0.2\sigma$, yet their common value remains $19\%$ ($2.5\sigma$) above the QESEM-Unbiased reference. The corresponding raw MPS results (dashed) are shown for comparison.
}
\label{fig:ibe_false_convergence}
\end{figure}

\begin{equation}
    \gamma(\chi,N_q)
    =
    a\frac{\log_2\chi}{N_q}
    +b,
    \label{eq:gamma_law}
\end{equation}
which can be then evaluated at the target system size.
In the main text, this procedure is applied to the coordination-two magnetization $M_{z=2}$ using exact simulations for $N_q\in\{28,35\}$ and bond dimensions $\chi\in\{512,1024,2048,4096\}$ (Fig.~\ref{fig:classical_simulation_methods_vs_quantum}). The resulting fitted model is then used to correct the raw MPS dynamics for the $N_q=51$ system, where no exact reference is available.
To assess the robustness of this extrapolation, we additionally consider the $C^{2,3}(r=3)$ correlator, which is similar to $C(r=3)$ (Eq.~(\ref{eq:corr_def})) but restricted to pairs of degree-two and degree-three qubits separated by graph distance three on the $51$-qubit heavy-hex lattice. As shown in Fig.~\ref{fig:ibe_false_convergence}, the corrected trajectories appear to converge with increasing bond dimension. However, they converge to a value that remains inconsistent with the QESEM-Unbiased result beyond the statistical uncertainty, demonstrating that the heuristic correction can exhibit false convergence in this regime.

\section{Quantum and Classical Runtime Estimates}
\label{app.runtimes}

\subsection{Quantum runtimes}
\label{app:runtime_estimates}

We estimate the resources required for QESEM-Unbiased, and QESEM-Extrapolated, as well as the limit of what is possible with unmitigated execution. All three estimates depend on the
noise accumulated in the portion of the circuit relevant to the
measured observable, the target precision, and the QPU execution
timescales.

\subsubsection{Unmitigated execution}

The sensitivity of \(O\) to noise within its active volume is described by 
\begin{equation}
    \frac{O_{\mathrm{noisy}}}{O_{\mathrm{ideal}}}
    \simeq
    \exp\!\left(-\rho I_FV_A\right),
    \label{eq:dilution_definition}
\end{equation}
where \(\rho<1\) describes dilution of the effect of errors \cite{QESEM,QuantinuumDilution2025} within the active volume $V_A$ on the measured observable and $I_F$ is the average two-qubit gate infidelity.

We model the accumulated bias of the unmitigated circuit as
\begin{equation}
    B(V_A)
    =
    1-\exp\!\left(-\rho I_FV_A\right).
\end{equation}
The unmitigated limit is defined by \(B(V_A^\star)=\epsilon\), giving
\begin{equation}
    V_A^\star
    =
    -\frac{\ln(1-\epsilon)}{\rho I_F}\approx\frac{\epsilon}{\rho I_F}.
    \label{eq:unmitigated_limit}
\end{equation}
This gives the largest active volume for which the modeled
unmitigated bias remains below the target error $\epsilon$.
For the parameters used in
\hyperref[fig:fig1b]{Fig.~\ref*{fig:fig1b}},
Eq.~\eqref{eq:unmitigated_limit} predicts a $0.02$ accuracy
limit at Floquet cycle $N_c=5$. Experimentally, however, the
unmitigated magnetization in Fig.~\ref{fig:magnetization}
deviates from the mitigated result by more than $0.02$ already
at $N_c=4$, likely reflecting state-preparation and measurement errors which were not included in the active-volume model.
We therefore use the empirical limit $N_c=4$ in
\hyperref[fig:fig1b]{Fig.~\ref*{fig:fig1b}}.

\subsubsection{QESEM-Unbiased}

We use the phenomenological variance model and optimized resource
allocation derived in Ref.~\cite{QESEM}. The quasiprobability norm is
approximated as
\begin{equation}
    W=\exp\!\left(2I_FV_A\right).
\end{equation}
Using the effective-volume model in
Eq.~\eqref{eq:dilution_definition}, Eqs.~(B26)--(B28) of
Ref.~\cite{QESEM} give
\begin{align}
    V_c
    &=
    \left(W^{2-\rho}-1\right)
    O_{\mathrm{ideal}}^2/n_{\mathrm{eff}},
    \label{eq:runtime_vc}
    \\
    V_s
    &=
    W^2-W^{2-\rho}O_{\mathrm{ideal}}^2/n_{\mathrm{eff}},
    \label{eq:runtime_vs}
\end{align}
where \(V_c\) is the variance between sampled quasiprobabilistic
circuits and \(V_s\) is the shot variance within a sampled circuit. $n_{\mathrm{eff}}$ is a measure of the intrinsic variance of the observable. In the absence of covariance between the spins, its value for the magnetization would simply equal the number of qubits. In practice, however, spin covariances significantly reduce this value. Optimizing the number of circuits and shots per circuit yields
\begin{equation}
    \tau_{\mathrm{unbiased}}
    =
   \left(
            \sqrt{V_ct_c}
            +
            \sqrt{V_st_s}
        \right)^2,
    \label{eq:unbiased_runtime}
\end{equation}
where \(t_c\) is the cost of loading a new circuit, \(t_s\) is the execution time per shot~\cite{QESEM}.

\subsubsection{QESEM-Extrapolated}

Without error mitigation, the quantum runtime is determined by the number of shots required to achieve the desired statistical precision for the observable. Specifically, if the standard deviation of the estimator after $n_{\mathrm{s}}$ shots is $\sqrt{V_{\mathrm{o}}/n_{\mathrm{s}}}$, then achieving a precision of $\epsilon$ requires a QPU time of $\tau=t_{\mathrm{s}}V_{\mathrm{o}}/\epsilon^2$, where $V_{\mathrm{o}}\equiv1/n_{\mathrm{eff}}$ is the intrinsic variance of the observable.

Error mitigation increases this runtime by a largely predictable overhead. For PEA-ZNE, assuming an optimal allocation of circuits and shots across the different noisy and noise-amplified computations, the runtime overhead is given by
\begin{equation}
    W^2 = \left[\frac{fe^{\gamma}+\left(1+\sqrt{\frac{t_{\mathrm{c}}V_{\mathrm{c,o}}}{t_{\mathrm{s}}V_{\mathrm{o}}}}\right)e^{f\gamma}}{f-1}\right]^2,
\end{equation}
with $V_{\mathrm{c,o}}$ being the circuit-to-circuit variance arising from the different Pauli insertions. At present, QESEM-Extrapolated uses a default amplification factor of ${f=2}$. Accordingly, its overall runtime is estimated as
\begin{equation}
    \tau_{\mathrm{QESEM-Ext}} = \frac{t_{\mathrm{s}}V_{\mathrm{o}}}{\epsilon^2}\left[2e^{\gamma}+\left(1+\sqrt{\frac{t_{\mathrm{c}}V_{\mathrm{c,o}}}{t_{\mathrm{s}}V_{\mathrm{o}}}}\right)e^{2\gamma}\right]^2.
\end{equation}

\begin{figure}[tb]
    \centering
    \includegraphics[width=\columnwidth]{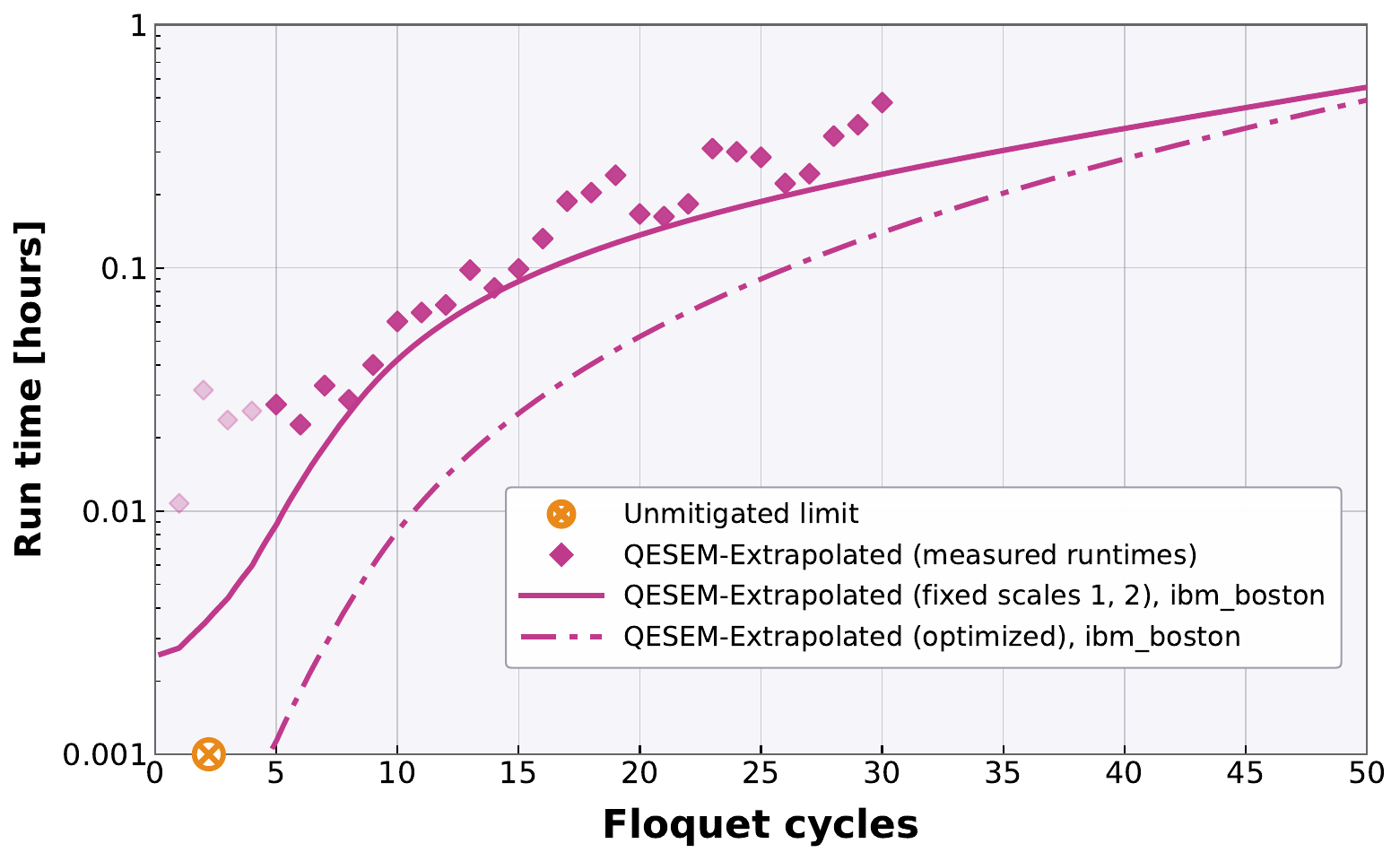}
    \caption{Estimated and measured runtimes for QESEM-Extrapolated to achieve a precision of $5\times10^{-3}$. The solid line corresponds to the current implementation of QESEM, while the dash-dotted line represents a forthcoming amplification-factor optimization feature. The unmitigated limit differs from that in \ref{fig:fig1b} because it is calibrated for the higher target precision.}
    \label{fig:optimal_pea}
\end{figure}

In the future, QESEM-Extrapolated will use an optimized amplification factor $f=1+\lambda/\gamma$, with $\lambda$ being the solution to
\begin{equation}
    \left(1+\sqrt{\frac{t_{\mathrm{c}}V_{\mathrm{c,o}}}{t_{\mathrm{s}}V_{\mathrm{o}}}}\right)(\lambda-1)e^{\lambda}=1.
\end{equation}
In this case, the estimated QPU time is reduced to
\begin{equation}
    \tau_{\mathrm{Opt-Ext}} = \frac{t_{\mathrm{s}}V_{\mathrm{o}}}{\epsilon^2}e^{2\gamma}\left(1+\frac{\gamma}{\lambda-1}\right)^2.
\end{equation}
Fig.~\ref{fig:optimal_pea} shows the estimated runtimes for the circuits considered in this work using the current implementation of QESEM-Extrapolated, as well as the projected runtimes obtained with the optimized amplification factor.

\subsubsection{Numerical inputs and uncertainty estimates}

To obtain numerical estimates from the above equations, we use a set of hardware and circuit-specific parameters.
For IBM QPUs, the shot time is dominated by the so-called \textit{rep-delay} parameter, which specifies the cooling time between consecutive shots. On Heron processors, this delay is $250\mu s$. Circuit execution times exhibit greater variability, but for our estimates we assume $t_{\mathrm{c}}=50ms$, which is typically a slight overestimate. The mean infidelity, $I_F$, naturally varies across experiments, but we use a representative average value of $2.4\times10^{-3}$.

For the circuit-specific parameters, at Floquet cycle numbers greater than 15, we find $0.07\lesssim\rho\lesssim0.11$, $6\lesssim n_{\mathrm{eff}}\lesssim15$ and $0.36\lesssim|O_{\mathrm{ideal}}|\lesssim0.41$.

These parameter ranges determine the central curves in Fig.~\ref{fig:fig1b}, as well as the surrounding uncertainty bands.

\subsection{Statevector simulation}

\begin{figure}
    \centering
    \includegraphics[width=\linewidth]{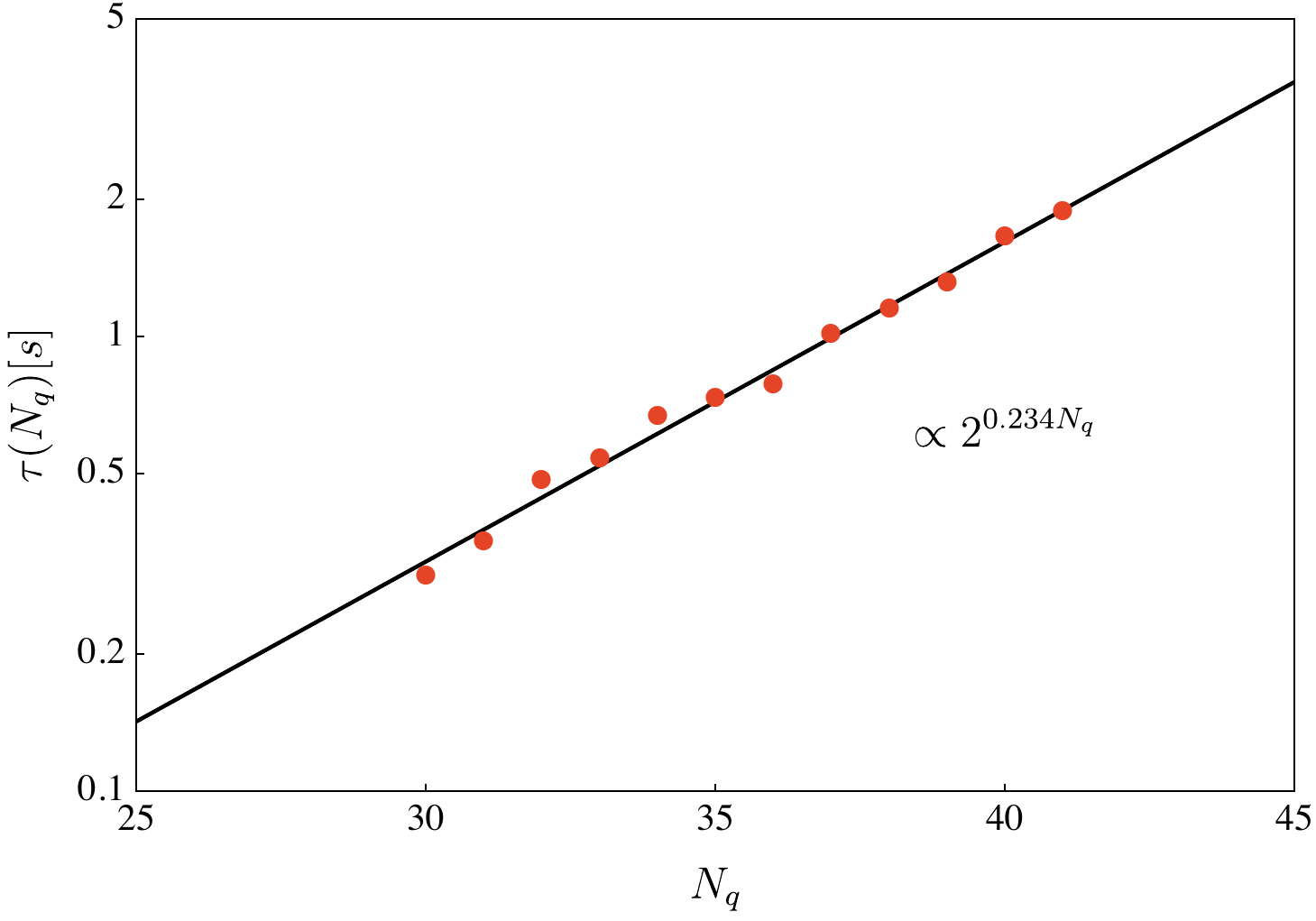}
    \caption{Estimates for the wall-time per quantum gate with respect to the number of qubits using the RIKEN-braket statevector simulator. These results were obtained by increasing the parallel resources at the same pace as the statevector dimension. }
    \label{fig:statevector-runtime-braket}
\end{figure}

We perform the runtime estimate for the statevector simulations in the weak-scaling sense, where the degree of parallelization is increased at the same pace as the problem size. Using the RIKEN-braket statevector simulator~\cite{Yoshioka2026jhpc}, we obtain an estimate of the average wall-time $\tau(N_q)$ per quantum gate as a function of the number of qubits $N_q$. We show these results in Fig.~\ref{fig:statevector-runtime-braket}. 
Taking into account the light-cone of a local magnetization in the 51-qubits setup, we obtain an effective number of qubits $N_q^{(\mathrm{eff})}(N_c)$ per Floquet cycle $N_c$. 
We multiply the number of gates $N_g(N_c)$ in the circuit up to a Floquet cycle $N_c$ with the wall-time per gate extrapolated to the corresponding effective number of qubits and obtain our total runtime-estimate as 
\begin{equation}
    \tau(N_c)=\tau(N_q^{(\mathrm{eff})}(N_c)) \times N_G(N_c).
\end{equation}

\subsection{ORQA}
The runtime of SPP simulations performed using ORQA scales close to linearly with the number of Pauli strings retained in the simulation, and benefits greatly from parallelization on the Fugaku supercomputer. While the details may vary depending on the dynamics of the simulation and the exact structure of the parallelization, we can empirically approximate the average wall-time per quantum gate as 
\begin{equation}
    \tau \approx \frac{|O|}{10^7 N_\mathrm{proc}}\mathrm{s},
\end{equation}
where $|O|$ is the number of Pauli strings at a given point in time, and $N_\mathrm{proc}$ is the number of computational processes across which the simulation is parallelized. Note that this wall-time is bounded from below by communication overhead that increases with the number of processes and eventually dominates the runtime. For a detailed scaling analysis from which we derive the above estimate, we refer to Ref.~\cite{broers2025scalablesimulationquantummanybody}.

Within the limitations of the supercomputer Fugaku, it is possible to retain a few trillion Pauli strings distributed over less than a hundred-thousand parallel compute cores. In App.~\ref{app.sppconvergence}, we have identified a conservative estimate of roughly $10^{30}$ Pauli strings that are necessary to obtain converged results and capture the characteristic oscillations in the case of 51 qubits. The computational requirements for a simulation at that scale are infeasible. Even a significantly more modest estimate for the number of Pauli strings would necessitate resources far beyond what exists today.

\subsection{PEPS-BP}

\begin{figure}[tb]
\centering
\includegraphics[width=1\linewidth]{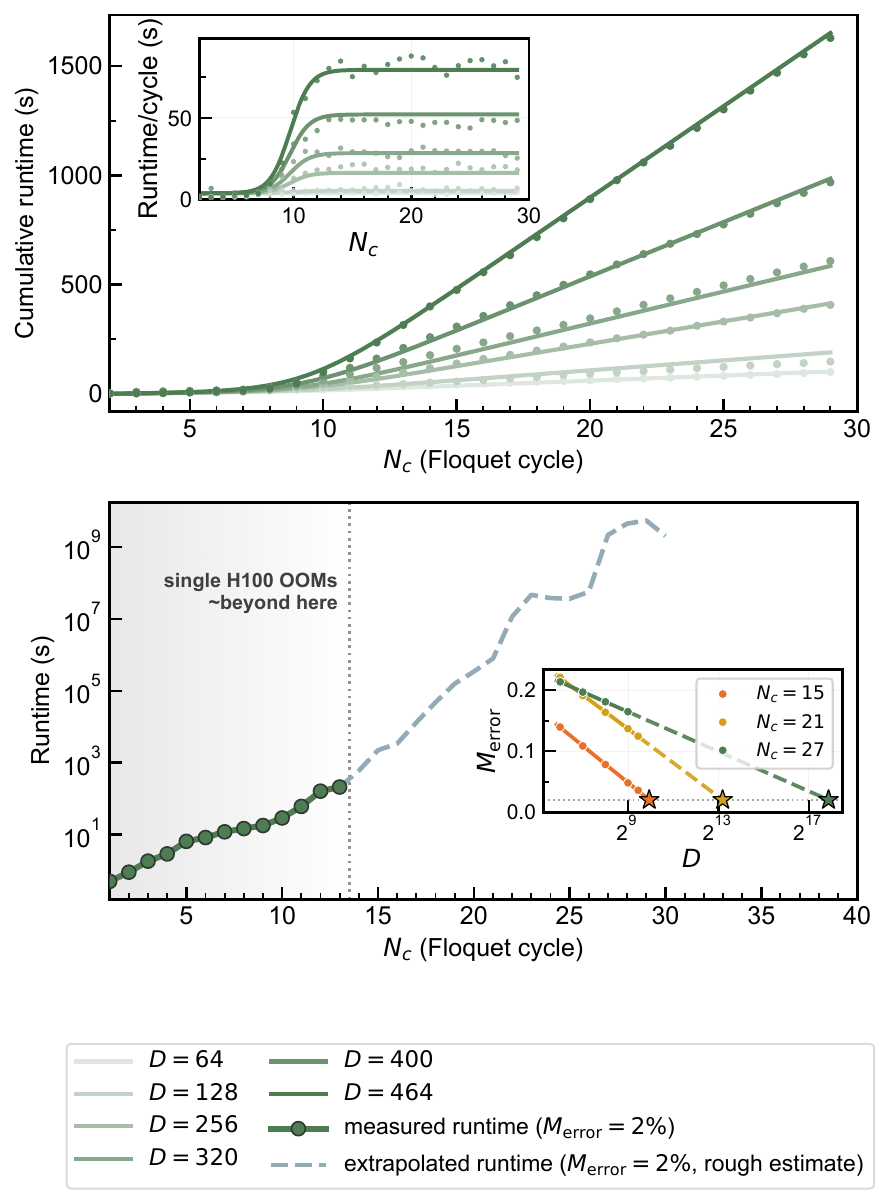}
\caption{
PEPS-BP runtime on a single NVIDIA H100 GPU ($80$~GB of memory).
Top: cumulative runtime as a function of Floquet cycle for several bond dimensions. The inset shows the runtime per cycle, with solid lines denoting fits to Eq.~\eqref{eq:runtime_model}.
Bottom: runtime required to reach a $2\%$ absolute error in the magnetization, with respect to the QESEM-Extrapolated quantum mitigated data. The shaded region marks the regime in which the required bond dimension exceeds the memory capacity of a single H100 GPU. Beyond this point, the runtime is extrapolated using the procedure described in the text.}
\label{fig:peps_resource_scaling2}
\end{figure}

To support the runtime comparison presented in Fig.~\ref{fig:fig1b}, we provide here additional details on the PEPS-BP runtime scaling and we present a rough extrapolation to Floquet cycles beyond those directly accessible on a single H100 GPU presented in Fig.~\ref{fig:fig1b}.
The runtime of the PEPS-BP method per Floquet cycle can be modeled using the form
\begin{equation}
t(D, N_c) =
b_0 +
CD^{\alpha}\,
\sigma\!\left(\frac{N_c - N_c^{(0)}}{\gamma}\right),
\label{eq:runtime_model}
\end{equation}
where $\sigma(x)=\frac{1}{1+e^{-x}}$
is the sigmoid function, $N_c^{(0)}$ denotes the Floquet cycle at which the runtime approaches its steady-state plateau, and $\gamma$ sets the width of this crossover. The factor $CD^\alpha$, captures the expected power-law dependence of the plateau runtime on the bond dimension, $D$.
The accumulated runtime is fitted independently using the cumulative functional form, $T(D, N_c)$, corresponding to the per-cycle sigmoid model. Both the per-cycle and accumulated runtime models are fitted to data obtained from different simulations performed on a single H100 GPU using the \texttt{TensorNetworkQuantumSimulator.jl} package \cite{Tindall2025TensorNetworkQuantumSimulator} for a range of different bond dimensions. The resulting fits are shown in the top panel of Fig.~\ref{fig:peps_resource_scaling2}. 

To estimate the runtime required to reach a fixed target accuracy as the circuit depth grows, we combine this runtime model with the bond-dimension requirement $D_\mathrm{needed}(N_c)$ obtained by fitting the PEPS-BP error with respect to the QESEM-Extrapolated data, $M_\mathrm{error}(D)=|M_{\mathrm{PEPS}}(D)-M_{\mathrm{QESEM\text{-}Ext.}}|$ linearly in $\log_2 D$, and determining the required bond dimension, $D _\mathrm{needed}$, at which the error reaches $2\%$ at each $N_c$ (inset in the top panel of Fig.~\ref{fig:peps_resource_scaling2}). Substituting $D_\mathrm{needed}(N_c)$ into $T(D, N_c)$ gives the projected runtime needed to achieve $2\%$ accuracy as a function of circuit depth, shown in the bottom panel of Fig.~\ref{fig:peps_resource_scaling2}. Beyond $N_c\approx 13$, the required bond dimension exceeds what fits in the single H100 memory budget, and the curve is extrapolated past this point using the approach above. This extrapolation is optimistic as there are currently no GPUs with the required memory and thus achieving most of these values of $D$ would require distributing the tensor network across multiple GPUs or compute nodes. Such distributed execution, if feasible, would necessarily incur communication overhead and reduced parallel efficiency that are not captured by the single-device fit.

\end{document}